\documentclass[preprint,authoryear]{elsarticle}

\usepackage{hyperref}
\hypersetup{hidelinks}
\journal{Neural Networks}

\usepackage{amssymb}
\usepackage{amsmath}
\usepackage{amsfonts}
\usepackage{graphicx}
\usepackage{siunitx}
\usepackage{booktabs}
\usepackage{afterpage}\usepackage{longtable}\newcommand{\needspace}[1]{\par\ifdim\dimexpr\pagegoal-\pagetotal\relax<#1\relax\newpage\fi}
\makeatletter\RequirePackage{etoolbox}\patchcmd{\pprintMaketitle}{\par\vskip36pt}{\par\vskip10pt}{}{}%
\def\LT@makecaption#1#2#3{%
  \LT@mcol\LT@cols c{\hbox to\z@{\hss\parbox[t]\LTcapwidth{%
    \footnotesize
    \sbox\@tempboxa{#1{#2: }#3}%
    \ifdim\wd\@tempboxa>\hsize
      #1{#2: }#3%
    \else
      \hbox to\hsize{\hfil\box\@tempboxa\hfil}%
    \fi
    \endgraf\vskip\baselineskip}%
  \hss}}}
\renewcommand{\floatpagefraction}{0.8}\makeatother
\usepackage{xcolor}
\usepackage{url}
\usepackage{placeins}

\graphicspath{{./}}
\DeclareGraphicsExtensions{.pdf,.png,.jpg,.eps}
\usepackage{epstopdf}\DeclareUnicodeCharacter{00B1}{\ensuremath{\pm}}\DeclareUnicodeCharacter{2212}{\ensuremath{-}}\DeclareUnicodeCharacter{03B1}{\ensuremath{\alpha}}\DeclareUnicodeCharacter{1D63}{\ensuremath{_{r}}}%
\newcommand{\SB}{S_{\mathrm{B}}}
\newcommand{\SW}{S_{\mathrm{W}}}
\newcommand{\SR}{\mathcal{S}}

\newcommand{\LayoutKeepFourLines}{\widowpenalties 4 10000 10000 10000 0\relax}\input{preamble_supplement.tex}\usepackage{latexml}\iflatexml\makeatletter
\renewenvironment{composegraphic}{}{}\renewenvironment{composetable}{}{}\renewenvironment{composerow}{}{}\renewenvironment{composecell}[1][]{}{}%
\renewenvironment{composepage}[1][]{\ComposePageMeta#1\relax\ifnum\ComposePageIndex>1 \addtocounter{figure}{-1}\fi}{}%
\renewcommand{\afterpage}[1]{#1}%
\let\ComposeOrigRef\ref\def\ComposeSplitRef#1:part#2:part#3\relax{\ComposeOrigRef{#1}#2}\renewcommand{\ref}[1]{\ComposeSplitRef#1:part:part\relax}%
\makeatother\fi
\begin{document}
\begin{frontmatter}
\title{Short-Term Synaptic Plasticity Stabilizes\texorpdfstring{\\}{ }Goal-Conditioned Dynamics\texorpdfstring{\\}{ }in a \texorpdfstring{Prefrontal Cortex--Inspired}{Prefrontal Cortex\textendash Inspired} Reservoir Model\texorpdfstring{\\}{ }for Multistep Goal-Directed Action Planning}

\author[osaka]{Jin Nakamura\corref{cor1}}
\ead{jin.nakamura@ist.osaka-u.ac.jp}

\author[hakodate]{Yuichi Katori}
\ead{katori@fun.ac.jp}

\cortext[cor1]{Corresponding author.}

\affiliation[osaka]{
    organization={Graduate School of Information Science and Technology, The University of Osaka},
    addressline={1-5 Yamadaoka},
    city={Suita},
    postcode={565-0871},
    state={Osaka},
    country={Japan}
}
\affiliation[hakodate]{
    organization={School of Systems Information Science, Future University Hakodate},
    addressline={116-2 Kamedanakano-cho},
    city={Hakodate},
    postcode={041-8655},
    state={Hokkaido},
    country={Japan}
}

\begin{abstract}
Goal-directed behavior requires remembered goals to guide later actions. We tested whether short-term synaptic plasticity (STP) improves this function in a reservoir model inspired by recurrent prefrontal circuitry. We trained the action-value readout using temporal-difference learning, with the structural recurrent connections held fixed. We evaluated 100 paired random networks on a grid task requiring sequential moves after a brief goal cue and delay. Success required reaching the remembered goal. State noise was applied during training and evaluation. At the same recurrent connection strength, mean success was 81.9\% with STP versus 43.7\% without STP ($p=\ensuremath{1.2\times10^{-19}}$). After separate selection of connection strength, success was 86.3\% versus 71.6\%. The advantage persisted when selection and evaluation used separate network seeds. Goal identity was highly decodable during the delay in both conditions. At separately selected strengths, later decoding remained high in both conditions, while STP strengthened action-value preferences for goal-directed moves before the third action. Synaptic efficacy showed goal-specific patterns during the delay. A separately trained control used the steady-state STP equations to recompute synaptic efficacy, and hence effective connectivity, from current activity. At a unit time constant of 1{,}000\,ms and the same connection strength, this control reached 83.9\% success, similar to 81.9\% with STP. This similarity also held at unit time constants of 500 and 750\,ms. These results support STP-derived activity-dependent recurrent modulation as a mechanism that stabilizes goal-conditioned dynamics, keeping remembered goals usable for successive actions under noise in this model.
\end{abstract}

\end{frontmatter}

\section{Introduction}\label{sec:01}

Goal-directed behavior often requires maintaining a goal across a delay and converting it into a sequence of actions. How recurrent neural circuits keep goal information in a form usable for later action selection remains unclear.

The prefrontal cortex (PFC) plays a central role in future action planning and supports the maintenance of goal information during delay periods. To examine the involvement of the PFC in multistep action, \citet{mushiake2006activity} recorded neuronal activity in the PFC of monkeys performing a route-search task. In that task, the monkey had to move a cursor on a grid to a designated goal, which required working memory (WM) to retain the final goal briefly and transform it into a sequence of immediate subgoals. Analysis of the recorded activity revealed coding for future actions, including goal position and movement direction. Preparatory activity represented the forthcoming moves on the grid rather than the arm movements used to produce them. These findings link PFC activity to prospective representations of action outcomes.

Prominent features of the PFC relevant to WM include recurrent network architecture and short-term synaptic plasticity (STP). The PFC contains large, highly branched pyramidal cells \citep{elston2011pyramidal}. Its nonrandom local connectivity includes an overrepresentation of bidirectional connections (\citealp{wang2006heterogeneity}; see also \citealp{song2005highly} for visual cortex). Its heterogeneous synaptic dynamics include both facilitating and depressing synapses \citep{wang2006heterogeneity}. At a facilitating synapse, repeated presynaptic activity transiently increases the efficacy of each transmission event; at a depressing synapse, it transiently decreases that efficacy. Which of the two effects dominates at a given synapse depends on the synaptic parameters and on the presynaptic firing rate. Classically, WM has been explained by persistent firing stabilized by recurrent excitation \citep{goldman1995cellular,wang2001synaptic} in the context of the integrative control functions of the PFC \citep{miller2001integrative,fuster2015prefrontal,funahashi2017working}. At the population level, stable mnemonic coding can coexist with heterogeneous single-neuron dynamics \citep{murray2017stable}. STP transiently modulates synaptic transmission efficacy over time scales ranging from tens to thousands of milliseconds through presynaptic facilitation and depression \citep{tsodyks1997neural,tsodyks1998neural,markram1998differential,varela1997differential}. Models of synaptic working memory propose that STP can retain information in synaptic efficacy when elevated firing is absent, with subsequent activity enabling readout \citep{mongillo2008synaptic,barak2014working,jackman2017mechanisms}. Synapses with STP, referred to as dynamic synapses, are formalized by the Tsodyks--Markram model of resource and utilization dynamics \citep{tsodyks1998neural} and influence both network dynamics and function \citep{torres2013emerging,katori2013stability}. Cortical STP time scales vary across connection types and experimental conditions. Facilitating connections between layer-5 pyramidal cells in ferret medial PFC have facilitation and recovery time constants of hundreds of milliseconds \citep{wang2006heterogeneity}. Rat medial PFC also exhibits sustained augmentation lasting seconds to tens of seconds \citep{hempel2000multiple}. In a related modeling study using a similar goal-oriented action-planning paradigm, \citet{katori2011representational} proposed a PFC network model with STP. In that model, STP dynamically reorganizes attractor structure and qualitatively reproduces representational switching during such tasks. Here, we examine how STP contributes to goal-directed action selection, using random recurrent connectivity and temporal-difference (TD)-based readout learning under state noise. In the present model, a single set of STP parameters is shared by all units, a simplification of the heterogeneity of synaptic dynamics described above (\S\ref{sec:05}; see also \S\ref{sec:34}).

Action planning is thought to rely on optimization based on reward prediction within cortico--basal ganglia circuits that include the PFC and striatum. The striatum receives excitatory cortical input and dopaminergic modulatory signals from the midbrain, and plays a key role in decision making. \citet{schultz1997neural} described phasic dopamine responses that depended on reward predictability, consistent with a reward-prediction-error signal. This discovery was interpreted in relation to TD error in reinforcement learning (RL) \citep{sutton1988learning,montague1996framework}, and led to the hypothesis that cortico--basal ganglia circuits implement RL based on TD error \citep{daw2006computational,nakahara2001parallel}.

Dynamical-systems approaches relate neural activity to information processing \citep{rabinovich2006dynamical}, and learning can produce stable computational outputs from chaotic networks \citep{sussillo2009generating,abbott2016building,aljadeff2016transition}. We use reservoir computing (RC) to study how STP modulates recurrent dynamics while task-dependent learning is restricted to the readout. RC is a class of recurrent neural network (RNN) in which time-varying inputs are processed by the high-dimensional and nonlinear dynamics of an intermediate layer (the reservoir) whose random and sparse recurrent connections are fixed. Learning is restricted to the connections from the reservoir to the output layer, while the internal reservoir connections remain fixed. Representative examples are the liquid state machine proposed by \citet{maass2002real} and the echo state network (ESN) proposed by \citet{jaeger2001echo}, which showed that randomly connected recurrent networks provide diverse temporal representations \citep{lukovsevivcius2009reservoir}. RC has also been used as a framework for modeling computational principles of the brain, for example in liquid-state-machine models of the cerebellar granular layer \citep{yamazaki2007cerebellum}, analyses of chaos and computational expressivity in the cerebellar granular layer \citep{tokuda2021chaos}, and demonstrations of physical reservoirs using cultured neural circuits \citep{sumi2023biological}. High-dimensional mixed selectivity in monkey PFC supports flexible linear readout \citep{rigotti2013importance}, and reservoir models reproduce features of the dynamic mixed selectivity observed in dorsal anterior cingulate cortex \citep{enel2016reservoir}. These findings motivate RC as a functional model of recurrent cortical computation. The recurrent connectivity of the PFC is structurally similar to that of an RNN, and the RC property that fixed recurrent connections maintain temporal information from input streams as internal states over a finite interval can serve as a computational substrate for short-term information retention in the PFC. \citet{yoshino2022short} combined RC with TD learning and showed that a transient directional cue could guide a later choice in a T-maze navigation task. Here, we use RC and TD-based RL to build a PFC-inspired RC--RL model of multistep goal-directed action planning: the reservoir includes STP at recurrent synapses, the task is inspired by the multistep route-search task of \citet{mushiake2006activity}, and the Q-value feedback to the reservoir abstracts re-entrant cortico--basal ganglia--thalamo--cortical loops.

 Whereas standard RNNs such as long short-term memory networks (LSTMs) and gated recurrent units (GRUs) typically update recurrent parameters by backpropagation through time (BPTT), the present RC-based model avoids backpropagation through recurrent connections and restricts task-dependent learning to the readout weights, which are updated using a TD-error-based rule. Because the structural recurrent weights are fixed, within-trial changes in the efficacy of recurrent synapses arise from STP alone.

We use \emph{goal representation} to denote goal information retained in the network after cue removal. The term \emph{goal-conditioned dynamics} describes evolving network states that depend on the previously cued goal and accompany subsequent goal-directed action selection. Here, stabilization means that the cued goal remains usable for successive goal-directed actions after cue removal; \S\ref{sec:17} gives the operational criteria. \emph{Effective connectivity} denotes the fixed recurrent matrix multiplied by the current STP scaling factors (\S\ref{sec:17}).

We test whether STP keeps the delay-period goal representation usable for successive actions under state noise. We also ask whether its contribution is associated with goal-dependent effective connectivity and differs from simply lengthening the unit time constant. Fading-memory dynamics motivate this question because the influence of an earlier input decreases over a design-dependent time scale. Dynamic synapses can improve short-term memory and influence recurrent network function \citep{mori2015dynamic,katori2013stability}. Synaptic working-memory and representational-switching models address storage, readout, and goal-to-action transformations \citep{mongillo2008synaptic,barak2014working,katori2011representational}, while mixed-selectivity studies and reservoir models address population coding and flexible readout \citep{enel2016reservoir,rigotti2013importance}. \citet{kozachkov2022robust} found greater robustness to synaptic ablation with STP in BPTT-trained networks, whereas fixed-synapse networks tended to be more robust to process noise. \citet{masse2019circuit} distinguished decodable information from information used for performance by shuffling neuronal and synaptic states across trials. Our replay intervention reassigns STP scaling trajectories across units (\S\ref{sec:19}). Here, fixed random recurrent connections and a TD-trained readout support delayed goal use across successive GO opportunities, each permitting a single move (\S\ref{sec:07}).

Our earlier conference reports introduced the RC--RL model with STP and an initial analysis of goal-dependent activity separation \citep{nakamura2024nolta,nakamura2024wcci}. Here, we extend the evaluation to 100 paired random networks under state noise and analyze time-resolved goal representations, action-value readout, and STP modulation of effective connectivity and its spectrum.

\subsection{Aims and contributions}\label{sec:02}
We examine delayed goal use in a $5{\times}5$ grid task in which a goal is cued briefly and, after a delay, must be reached by sequential moves, one per GO opportunity. We address four questions:
\begin{description}
\item[\textbf{RQ1} (task learning)] Can an RC--RL model combining a PFC-inspired recurrent circuit with basal-ganglia-inspired reward-based readout learning appropriately learn the multistep goal-directed action-selection task with sequential GO opportunities?
\item[\textbf{RQ2} (functional contribution of STP)] Does STP keep the delay-period goal representation usable for action selection later in the trial under state noise? Does this improve the robustness of task performance, and which properties of STP-derived modulation support this contribution?
\item[\textbf{RQ3} (internal representation and dynamics)] How does STP reorganize, in a goal-specific manner, the state space and effective-connectivity structure that support goal-conditioned dynamics?
\item[\textbf{RQ4} (parameter dependence)] How does task success depend on STP parameters and the rate scale that converts unit activity to a presynaptic rate (\S\ref{sec:05}), and how do these settings relate to physiological time scales reported in cortex?
\end{description}
In this study, we evaluate the goal-conditioned dynamics described above using four operational indices: state-space separability, dynamic modulation of the effective spectral radius, goal specificity of effective connectivity, and an operational action-value difference at GO opportunities that tests whether the maintained goal-conditioned structure is expressed at the level of action-value readout (details are given in \S\ref{sec:03}).

The remainder of this paper is organized as follows. \S\ref{sec:03} describes the model architecture, task design, and analytical methods. \S\ref{sec:21} reports the experimental results, \S\ref{sec:26} discusses their implications, and \S\ref{sec:36} concludes the paper.

\section{Methods}\label{sec:03}

\subsection{Model architecture and overview}\label{sec:04}

\paragraph{Design motivation}
The model combines fixed recurrent connectivity, activity-dependent STP modulation, and a TD-trained action-value readout to study delayed goal use. The TD-trained readout provides a functional abstraction of reward-dependent learning in cortico--basal ganglia circuits. Because of the leak term $(-m_i)$ and the small update ratio $\Delta t / \tau = 0.01$ in the discretized leaky rate model, it operates under conditions different from the usual design criterion $\rho < 1$ for discrete-time ESNs. In the parameter regime used here ($\alpha_r{=}3.0$), numerically stable behavior was obtained. Goal retention was assessed with time-resolved decoding (\S\ref{sec:14}). The correspondence to the PFC and basal ganglia is functional rather than anatomical, and the model is not intended to reproduce individual anatomical pathways one-to-one.

\begin{figure}[tp]
\centering
\includegraphics[width=1\textwidth]{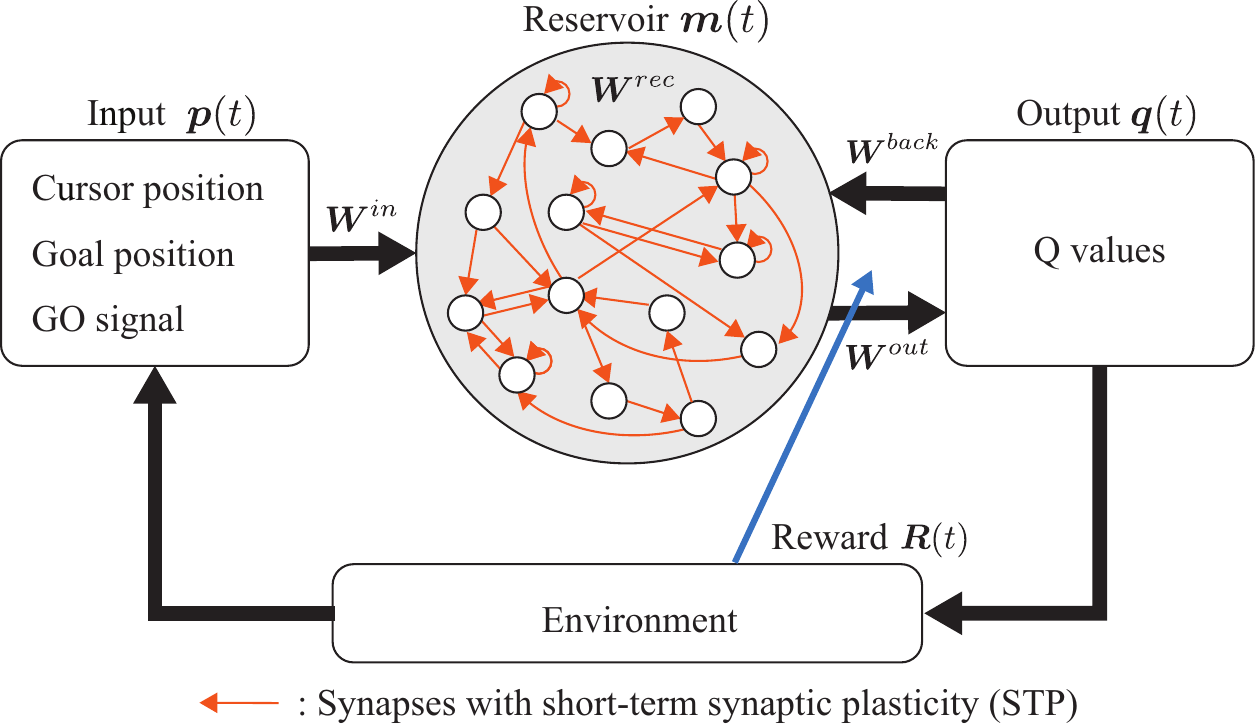}
\caption{\textbf{Network architecture of the proposed model.} The model consists of an input layer, a reservoir layer (with STP), and an output layer. Environmental information is processed through the reservoir, and Q values for action selection are generated. Only the output weights are trained by TD learning. The orange arrow denotes STP-modulated recurrent synapses, and the blue arrow denotes the reward signal $R(t)$.}
\label{fig:01}
\end{figure}

\paragraph{Overall architecture}
The model consists of an input layer, a recurrent reservoir with STP-equipped synapses, and an output layer that computes action values (Q values) (Fig.~\ref{fig:01}). Only the output weights are trained; the recurrent and feedback weights are fixed. A sparse, fixed feedback pathway ($W^{\mathrm{back}}$) conveys the Q values from the output layer back to the reservoir (Q-value feedback), and is intended as a functional abstraction of re-entrant pathways in cortico--basal ganglia--thalamo--cortical loops. This Q-value feedback is distinct from the reward signal $R(t)$, which enters only the learning rule for the output weights (\S\ref{sec:08}). The input is a 34-dimensional vector $p(t)$ representing sensory input to the PFC. It concatenates a 25-dimensional one-hot position vector for the $5 \times 5$ grid and an 8-dimensional one-hot goal vector for the eight target locations. It also includes a 1-dimensional binary GO channel indicating when action is permitted. The action selected from the output-layer Q values is interpreted as a functional abstraction of downstream action-selection signals in cortico--basal ganglia--premotor/motor pathways.

\begingroup Table~\ref{tab:01} defines the model and analysis symbols; Table~\ref{tab:03} lists parameter values, including the integration and synaptic time constants, and their sources.

\needspace{55mm}\renewcommand{\arraystretch}{0.93}\begin{longtable}{@{}p{0.17\textwidth}@{\hspace{1em}}p{0.62\textwidth}@{\hspace{1em}}p{0.10\textwidth}@{}}
\caption{\textbf{List of symbols.} Meanings and units of the symbols used in the Methods. Table~\ref{tab:03} gives parameter values and their sources. ``--'' denotes a dimensionless quantity.}\label{tab:01}\\
\toprule
Symbol & Meaning & Unit \\ \midrule
\endfirsthead
\multicolumn{3}{@{}l}{\footnotesize\tablename~\thetable{} (continued)}\\
\toprule
Symbol & Meaning & Unit \\ \midrule
\endhead
\bottomrule
\endfoot
\multicolumn{3}{@{}l}{\emph{State variables}} \\
$m_i(t)$ & Internal state of reservoir unit $i$ & -- \\
$\bar r_j(t)$ & Normalized activity of unit $j$, $\bar r_j\in[0,1]$ & -- \\
$r^{\mathrm{STP}}_j(t)$ & Presynaptic activity expressed as a rate in the STP update & Hz \\
$x_j(t)$ & Fraction of available synaptic resources of presynaptic unit $j$ & -- \\
$u_j(t)$ & Utilization variable of presynaptic unit $j$ & -- \\
$s_j(t)$ & Scaling factor for recurrent transmission, $s_j=2x_ju_j/U_{\mathrm{SE}}$ with STP & -- \\
$p_k(t)$ & Component $k$ of the input vector (position, goal, and GO signal; 34 elements) & -- \\
$q_a(t)$ & Action value (Q value) of action $a$ & -- \\
$\xi_i(t)$ & State noise added to $m_i$ at each integration step & -- \\
\multicolumn{3}{@{}l}{\emph{Connectivity}} \\*
$W^{\mathrm{rec}}$ & Recurrent weights (fixed) & -- \\
$W^{\mathrm{in}}$ & Input weights (fixed) & -- \\
$W^{\mathrm{back}}$ & Q-value feedback weights from the output layer to the reservoir (fixed) & -- \\
$W^{\mathrm{out}}$ & Output (readout) weights (trained) & -- \\
$W^{\mathrm{rec}}_{\mathrm{eff}}(t)$ & Effective recurrent connectivity, $W^{\mathrm{rec}}\,\mathrm{diag}(\mathbf{s}(t))$ & -- \\
\multicolumn{3}{@{}l}{\emph{Parameters}} \\
$N_m$ & Number of reservoir units & -- \\
$\Delta t$ & Integration step & ms \\
$\tau$ & Effective integration time constant of a unit & ms \\
$T$ & Temperature of the activation function ($1/T$: inverse temperature) & -- \\
$U_{\mathrm{SE}}$ & Utilization of synaptic efficacy (baseline of $u_j$) & -- \\
$\tau_{\mathrm{R}}$, $\tau_{\mathrm{F}}$ & Time constants of resource recovery and of facilitation & ms \\
$\kappa$ & Rate scale of the STP update ($\bar r_j{=}1$ corresponds to $\kappa/\Delta t$) & -- \\
$\alpha_r$, $\alpha_i$, $\alpha_b$, $\alpha_o$ & Scaling factors of $W^{\mathrm{rec}}$, $W^{\mathrm{in}}$, $W^{\mathrm{back}}$, and the initial $W^{\mathrm{out}}$ & -- \\
$p_r$, $\beta_i$, $\beta_b$ & Fractions of nonzero elements of $W^{\mathrm{rec}}$, $W^{\mathrm{in}}$, and $W^{\mathrm{back}}$ & -- \\
$\sigma$ & Standard deviation of the state noise per integration step & -- \\
$\gamma$ & Discount factor per integration step & -- \\
$\eta$ & Learning rate of the output weights & -- \\
$R$ & Reward & -- \\
\multicolumn{3}{@{}l}{\emph{Operators and analysis quantities}} \\
$\odot$ & Element-wise (Hadamard) product of two vectors & \\
$\mathrm{diag}(\mathbf{s})$ & Diagonal matrix with the elements of $\mathbf{s}$ on its diagonal & \\
$\SR(t)$ & Scatter ratio (between-class over within-class scatter) & -- \\
$\rho_{\mathrm{eff}}$, $\rho_{\mathrm{rel}}$ & Effective spectral radius and its value relative to the Without-STP condition & -- \\
$\Delta_{\mathrm{sim}}(t)$, $\Delta_s(t)$ & Goal specificity of $W^{\mathrm{rec}}_{\mathrm{eff}}$ and of $\mathbf{s}$ & -- \\
$D_Q$ & Action-value difference at a GO opportunity & -- \\
\end{longtable}

\endgroup \subsection{Reservoir dynamics with short-term synaptic plasticity}\label{sec:05}
Each reservoir unit represents collective neural activity at a phenomenological rate level. Its dimensionless state $m_i(t)$ integrates recurrent, external and Q-value feedback inputs, with recurrent transmission modulated by STP:
\begin{equation}
\begin{split}
m_i(t{+}\Delta t) ={} & \left(1-\frac{\Delta t}{\tau}\right)m_i(t) \\
& + \frac{\Delta t}{\tau}\biggl[
  \sum_{j} W^{\mathrm{rec}}_{ij}
  \Bigl(\frac{2\,\bar r_j(t)\,x_j(t)\,u_j(t)}{U_{\mathrm{SE}}}-1\Bigr) \\
& \quad + \sum_{k} W^{\mathrm{in}}_{ik}\,p_k
  + \sum_{\ell} W^{\mathrm{back}}_{i\ell}\,q_\ell
\biggr] + \xi_i(t),
\end{split}
\label{eq:01}
\end{equation}
where the dimensionless reservoir activity is $\bar r_j(t)=\tfrac{1}{2}\!\left[1+\tanh(m_j(t)/T)\right]\in[0,1]$ ($T{=}0.4$, i.e., $1/T{=}2.5$). Here $T$ is the temperature parameter of the sigmoid function, and smaller $T$ yields a steeper transition in the activation function. We set $\Delta t=\SI{10}{ms}$ and $\tau=1{,}000\,\mathrm{ms}$, and additive noise is given by $\xi_i\!\sim\!\mathcal{N}(0,\sigma^2)$. The parameter $\tau$ is the effective integration time constant of a unit that represents collective activity rather than a single neuron; it is therefore longer than a single-neuron membrane time constant, and its value is motivated in \S\ref{sec:06}. The temporal dynamics of the coupled circuit depend jointly on this unit integration, recurrent interactions, STP and Q-value feedback. $U_{\mathrm{SE}}$ is the utilization of synaptic efficacy of the Tsodyks--Markram model, that is, the baseline value to which the utilization variable $u_j$ returns in the absence of presynaptic activity (Eq.~\eqref{eq:04}). Normalizing the synaptic efficacy by $U_{\mathrm{SE}}$ makes $x_j u_j/U_{\mathrm{SE}}$ equal to 1 at baseline, matching the baseline recurrent drive between conditions.

\paragraph{STP (Tsodyks--Markram model)}
\begingroup\setlength{\emergencystretch}{2em}The synaptic transmission efficacy of recurrent connections is modeled using a discrete-time rate-based adaptation of the Tsodyks--Markram model of short-term synaptic plasticity \citep{tsodyks1997neural,tsodyks1998neural}. For clarity, we distinguish the dimensionless reservoir activity from the effective presynaptic activation rate used in the STP update. The normalized reservoir activity $\bar r_j(t)\in[0,1]$ defined above (Eq.~\eqref{eq:01}) is used as the reservoir output for recurrent input and readout. To make the rate dimension explicit, in the STP update we express the same normalized presynaptic activation as an effective rate over one integration interval,\par\endgroup
\begin{equation}
r_j^{\mathrm{STP}}(t)=\frac{\bar r_j(t)}{\Delta t}.
\label{eq:02}
\end{equation}
The presynaptic resource variable $x_j$ and utilization variable $u_j$ then evolve as
\begin{align}
x_j(t{+}\Delta t)
&= x_j(t)
+ \Delta t\!\left[\frac{1 - x_j(t)}{\tau_{\mathrm{R}}}
- r_j^{\mathrm{STP}}(t)\,x_j(t)\,u_j(t)\right],
\label{eq:03}\\
u_j(t{+}\Delta t)
&= u_j(t)
+ \Delta t\!\left[\frac{U_{\mathrm{SE}} - u_j(t)}{\tau_{\mathrm{F}}}
+ U_{\mathrm{SE}}\bigl(1 - u_j(t)\bigr)r_j^{\mathrm{STP}}(t)\right].
\label{eq:04}
\end{align}
Here $\tau_{\mathrm{R}}$ is the time constant with which the resource variable $x_j$ recovers toward 1 (recovery from depression), and $\tau_{\mathrm{F}}$ is the time constant with which the utilization variable $u_j$ relaxes back to its baseline $U_{\mathrm{SE}}$ (decay of facilitation). The main comparison used $\tau_{\mathrm{R}}{=}\SI{100}{ms}$ and $\tau_{\mathrm{F}}{=}\SI{500}{ms}$.
Because $r_j^{\mathrm{STP}}(t)\,\Delta t = \bar r_j(t)$, Eqs.~\eqref{eq:03}--\eqref{eq:04} are algebraically equivalent to the discrete-time update used in the simulations. This notation makes the dimensional structure explicit while preserving the implemented dynamics, and writes the STP update in an Euler-difference form for the fixed integration step used here ($\Delta t{=}\SI{10}{ms}$).  Thus, $r_j^{\mathrm{STP}}(t)$ should be read as the rate-scale representation of normalized reservoir activity in this rate-based STP model, not as an additional independently fitted biophysical parameter. To examine STP parameter values taken from the literature (\S\ref{sec:18}), we additionally used a lower rate scale, $r_j^{\mathrm{STP}}(t)=\kappa\,\bar r_j(t)/\Delta t$ with $\kappa{=}0.2$. This assigns 20\,Hz to $\bar r_j{=}1$ instead of $1/\Delta t=100$\,Hz. All other analyses used $\kappa{=}1$. The classical Tsodyks--Markram model is driven by spike trains; here, $r_j^{\mathrm{STP}}(t)$, a rate-level (mean-field) representation of the activity of unit $j$, drives representative resource and utilization states associated with that unit, and all outgoing recurrent connections from the unit share these states. Population-rate formulations with dynamic synapses provide precedents for coupling collective activity to synaptic states \citep{tsodyks1998neural,mongillo2008synaptic}; here utilization relaxes toward $U_{\mathrm{SE}}$, as in \citet{mongillo2008synaptic}. The constraints imposed by this unit-wise representation are discussed in \S\ref{sec:34}. In the main comparison, $U_{\mathrm{SE}}{=}0.1$ was fixed (a value inherited from preliminary simulations and the conference-version model, with published precedents in Table~\ref{tab:03}), and parameter exploration was performed over $(\tau_{\mathrm{R}},\tau_{\mathrm{F}})$ and $U_{\mathrm{SE}}$ (\S\ref{sec:18}).\par No clipping or other boundary constraint was applied to $x_j$ or $u_j$. For the update in Eqs.~\eqref{eq:03}--\eqref{eq:04}, the admissible ranges $x_j \in [0,1]$ and $u_j \in [U_{\mathrm{SE}},1]$ are preserved for any activity sequence $\bar r_j\in[0,1]$ provided that $\Delta t/\tau_{\mathrm{R}}\le 1$, $\Delta t/\tau_{\mathrm{F}}+U_{\mathrm{SE}}\le 1$, and $\kappa\le 1$, so that the depletion term $\kappa\bar r_j u_j\le 1$ cannot make $x_j$ negative. These conditions hold for all STP parameter values used in this study ($\kappa{=}1$ or $0.2$); the left-hand sides of the first two conditions are 0.1 and 0.12 for the main values. For the main values, the same update equations further give $x_j\ge 0.105$ and $u_j\le 0.85$ along every trajectory that starts from the reset state $(x_j,u_j)=(1,U_{\mathrm{SE}})$; these two bounds are limits on such trajectories, not on the whole range $[0,1]\times[U_{\mathrm{SE}},1]$. These inequalities establish range preservation for the implemented discrete-time system. Trial resets prevent direct carry-over of the reservoir and STP state variables between episodes.

\paragraph{Biophysical interpretation of the STP variables}
The variable $x_j$ represents the fraction of available synaptic resources associated with presynaptic unit $j$, whereas $u_j$ is a utilization variable that controls the fraction of those resources used for transmission \citep{tsodyks1998neural}. In the Tsodyks--Markram formalism, $u_j$ is often interpreted as reflecting facilitation-related presynaptic processes, such as residual calcium effects, but it is not itself a direct measurement of calcium concentration. Presynaptic activity contributes a resource-depletion term to the update of $x_j$ and a facilitation term to the update of $u_j$; recovery and relaxation act concurrently. In the absence of presynaptic activity, $x_j$ recovers toward its steady state ($x_j{=}1$) with time constant $\tau_{\mathrm{R}}$, and $u_j$ recovers toward its steady state ($u_j{=}U_{\mathrm{SE}}$) with time constant $\tau_{\mathrm{F}}$. The effective synaptic transmission efficacy $u_j(t)\,x_j(t)$ is given by the product of these two competing processes. We use ``facilitation-dominant'' for the selected regime with slower utilization relaxation than resource recovery ($\tau_{\mathrm{F}} > \tau_{\mathrm{R}}$) and low baseline utilization. 

\paragraph{Without STP}
In the Without-STP condition, the recurrent drive in Eq.~\eqref{eq:01} reduces to $\sum_j W^{\mathrm{rec}}_{ij}(2\bar r_j(t)-1)$. The Without-STP condition therefore corresponds to a standard leaky-integrator rate-based reservoir with a fixed recurrent scaling and a TD-trained linear readout, and serves as the baseline for evaluating the contribution of STP-induced effective-connectivity modulation. In the comparison at the same recurrent scaling (\S\ref{sec:12}), the recurrent connectivity $W^{\mathrm{rec}}$, input weights $W^{\mathrm{in}}$, feedback weights $W^{\mathrm{back}}$, unit integration time constant $\tau$, noise level, TD-learning rule, reward structure, and all other hyperparameters were identical across the With-STP and Without-STP conditions; only the activity-dependent STP modulation was added or removed. For analyses that extract the coefficient multiplying $\bar r_j(t)$, the corresponding operational effective recurrent matrix is $W^{\mathrm{rec}}_{\mathrm{eff}}=2W^{\mathrm{rec}}$; the constant offset term is not included in spectral analyses.

\subsection{Connectivity structure, scaling, and initialization}\label{sec:06}
For each network (random seed), the weight matrices are drawn at random as specified below; the hyperparameter values themselves are not random, and their sources are given in Table~\ref{tab:03}.
$W^{\mathrm{rec}}$ is a sparse matrix with a fixed fraction $p_r{=}0.1$ of its entries nonzero, and its nonzero elements are drawn from $\mathcal{N}(0,\,1)$ and scaled by $\alpha_r / \sqrt{p_r N_m}$ ($\alpha_r{=}3.0$ in the comparison at the same $\alpha_r$; \S\ref{sec:12}). Under this construction, the expected spectral radius of $W^{\mathrm{rec}}$ is about $\alpha_r$. The present model is implemented as a discretized leaky-rate system $m(t{+}\Delta t) = (1-\Delta t/\tau)m(t) + (\Delta t/\tau) F(m(t))$ with $\Delta t / \tau = 0.01$. The discrete-time ESN criterion $\rho(W){<}1$ therefore does not directly provide a design rule for this system. Each update forms a weighted average of the previous state and the total drive $F(m(t))$, with weights $1-\Delta t/\tau$ and $\Delta t/\tau$, and adds state noise. The drive is finite because $\bar r_j\in[0,1]$, the input vector, including any added input noise, has finite entries, the Q values are linear in $\bar r_j$, and the STP variables remain within their admissible ranges at every step (\S\ref{sec:05}). All state variables therefore remain finite throughout each trial. The input weights $W^{\mathrm{in}}$ have a fraction $\beta_i{=}0.1$ of their entries nonzero; these entries are drawn from $\mathcal{U}(-1,1)$ and scaled by the input-scaling factor $\alpha_i{=}18.112$. The feedback weights $W^{\mathrm{back}}$ have a fraction $\beta_b{=}0.45$ of their entries nonzero; these entries are drawn from $\mathcal{U}(-1,1)$ and scaled by $\alpha_b{=}1.0$. The output weights $W^{\mathrm{out}}$ were initialized with Gaussian random values drawn from $\mathcal{N}(0,1)$ and scaled by $\alpha_o/\sqrt{N_m}$ ($\alpha_o{=}0.02$). The values of $\alpha_i$, $\alpha_b$, $\alpha_o$, $\beta_i$, $\beta_b$, and $p_r$ were inherited from preliminary simulations and the conference-version model and were kept at those values in all comparisons reported here, except that $\alpha_b$ was varied in the Q-value feedback controls (\S\ref{sec:12}; Supplementary Fig.~\ref{fig:30}). Variance-based scaling specifies the recurrent scaling $\alpha_r$, and the input and feedback scalings $(\alpha_i,\alpha_b)$ as well as the unit time constant $\tau$ are reported explicitly for reproducibility. Additive Gaussian noise is applied during both training and evaluation. The default noise level is $\sigma{=}0.001$, a value likewise inherited (Table~\ref{tab:03}). The main robustness comparison used $\sigma{=}0$ (without noise) and $\sigma{=}0.001$; a noise axis extending to $\sigma{=}0.064$, independent training and evaluation noise levels, and other forms of noise are described in \S\ref{sec:10}.

\paragraph{Hyperparameter selection}
The With-STP and Without-STP conditions were compared in two ways (\S\ref{sec:12}): at the same recurrent scaling ($\alpha_r{=}3.0$ in both conditions), and with the recurrent scaling selected for each condition from the same grid of 16 values. Both conditions used the same input and feedback scaling factors $(\alpha_i,\alpha_b)$. The unit time constant $\tau$ was also fixed in these comparisons. The main STP parameter pair was selected in an earlier grid search and was confirmed within the high-success-rate, facilitation-dominant band of the grid search with 100 seeds per grid point (\S\ref{sec:18}); the pair was fixed across seeds.

\begingroup\setlength{\emergencystretch}{3em}We chose $\tau=1{,}000$\,ms because it lies within the range of reward-memory time scales reported in primate PFC \citep{bernacchia2011reservoir,murray2014hierarchy,spitmaan2020multiple}. The primary comparison used this phenomenological unit integration time constant in both conditions to examine whether STP acting on recurrent dynamics keeps goal information usable for later action selection. Recurrent interactions in a cortical network model can also sustain prefrontal responses for seconds \citep{chaudhuri2015large}. We also evaluated $\tau{=}250$--$3{,}000$\,ms with the same recurrent scaling in both conditions and with the recurrent scaling selected for each condition, together with a passive control (\S\ref{sec:12}; \S\ref{sec:22}; Fig.~\ref{fig:05}).\par\endgroup

\subsection{Task design and input}\label{sec:07}
The present task is inspired by the multistep action-planning experiment of \citet{mushiake2006activity}. The grid layout is shown in Fig.~\ref{fig:07}A. In our implementation, a goal cue is followed by a delay and sequential GO windows on a $5{\times}5$ grid. The binary \emph{GO signal} is active during each 1-s \emph{GO window}. At the end of the window, the agent executes the action with the largest Q value as a single one-cell move of the cursor (\emph{action execution}). We call this window and its concluding move a \emph{GO opportunity}; each trial contains at most four opportunities, GO1--GO4. In the original experiment, monkeys planned a multistep action sequence to reach a goal and executed one step at a time in response to each GO signal; the GO signal of the present task plays the same role. The model task differs from the monkey experiment in grid size, action discretization, and noise sources. Accordingly, we do not treat monkey behavioral measures as a quantitative benchmark in this study. The agent starts from the central position $(3,3)$ and aims to reach one of eight target locations. The four corner cells $(1,1)$, $(1,5)$, $(5,1)$, and $(5,5)$ were treated as blocked cells and were not available for movement; the eight goal locations were non-corner perimeter cells. This design provided eight symmetric goal options from the central start position while keeping the required action horizon within three GO opportunities. For reproducibility, the four action indices in the implementation were fixed as $a{=}0,1,2,3$ for left, right, up, and down moves, respectively, and the eight goal indices $g{=}0,\ldots,7$ corresponded to the locations (in row, column with $1$-indexing) $(2,1), (2,5), (4,1), (4,5), (1,2), (1,4), (5,2), (5,4)$, in that order. Trial timing was as follows:
\[\begin{aligned}
&\text{Position input: } t\ge 1~\text{s},\\
&\text{Goal cue: } 2\le t\le 3~\text{s},\\
&\text{Delay period: } 3<t<4~\text{s},
\end{aligned}\]
\[
\text{GO windows: } 4\le t<5,\ 6\le t<7,\ 8\le t<9,\ 10\le t<11~\text{s}.
\]
Actions were executed at $t = 5, 7, 9, 11$\,s (GO1--GO4). The 34-dimensional input $p$ consisted of a 25-dimensional one-hot position vector, an 8-dimensional one-hot goal vector, and a binary GO channel. No input was presented during the first second of a trial, and the position input was presented from $t{=}1$\,s until the end of the trial. The goal input was set to zero after $t{=}3$\,s. The integration step was $\Delta t=\SI{10}{ms}$. No additional interior obstacles were placed beyond the four blocked corner cells. The timeline of a single trial is shown in Table~\ref{tab:02}.

\begin{table}[t]
\centering
\caption{\textbf{Timeline of a single trial in the grid task.} Times are in seconds from trial onset. The goal cue is presented and then removed, and actions are executed at the end of each GO window.}
\label{tab:02}
\vspace{8pt}\renewcommand{\arraystretch}{0.91}\begin{tabular}{@{}ll@{}}
\toprule
Interval & Event \\ \midrule
$2$ to $3$ & Goal cue (one-hot goal input on) \\
$>3$ to $<4$ & Delay period (goal input off; internal retention only) \\
$4$ to $<5$ & GO window 1 \\
$6$ to $<7$ & GO window 2 \\
$8$ to $<9$ & GO window 3 \\
$10$ to $<11$ & GO window 4 \\
Action execution & Actions executed at $t = 5, 7, 9, 11$\,s \\ \bottomrule
\end{tabular}
\end{table}\noindent The action sequence required to reach each non-corner perimeter goal was therefore relatively simple without noise. Under state noise, however, decisions varied across GO windows, motivating an analysis of progress toward the goal at the GO opportunity following a distance-increasing or invalid move (\S\ref{sec:23}). Because every goal is three moves away from the start position, a trial can reach the goal at GO3 at the earliest, and it proceeds to GO4 ($t{=}10$--$11$\,s) only if at least one of the first three executed actions failed to make progress toward the goal. Accordingly, the analyses below focus mainly on the interval from $t{=}0$ to $9$\,s.

\subsection{Reinforcement learning and policy}\label{sec:08}
We next specify the learning rule that maps reservoir states to action selection. The output layer linearly reads out Q values for each action from the normalized reservoir activities, and only the output weights are updated on the basis of TD error.
The Q value for action $a\in\{\mathrm{LEFT},\mathrm{RIGHT},\mathrm{UP},\mathrm{DOWN}\}$ is defined as a linear readout:
\begin{equation}
q_a(t)=\sum_{j} W^{\mathrm{out}}_{a,j}(t)\, \bar r_j(t).
\end{equation}
Action selection uses a greedy policy ($\arg\max_a q_a(t)$); when multiple actions share the maximum Q value, ties are broken deterministically by the smallest action index, and this convention was fixed across all conditions. The model does not include an explicit exploration mechanism such as epsilon-greedy. Because the output weights were initialized with small random values and action selection followed a deterministic greedy policy with fixed tie-breaking, early action-value differences arose from the random readout initialization and were then shaped by reward-dependent TD updates. Additive state noise introduced further variability during training and evaluation. In this four-action task, shaped rewards provided immediate negative feedback on nonprogressive moves, allowing TD updates to change action values and subsequent greedy choices (\S\ref{sec:34}). Let $t_n=n\,\Delta t$ denote the $n$th integration step of a trial and $a_n=\arg\max_a q_a(t_n)$ the greedy action at that step. The Q values and the greedy action are computed at every integration step, but an action is executed by the environment only at the end of each GO window. We write $k\in\{1,2,3,4\}$ for the GO opportunities, $t_k\in\{5,7,9,11\}\,\mathrm{s}$ for the GO-window end times, and $t_k^{-}$ for the time immediately before the $k$th action execution, so that the executed action is the greedy action at $t_k^{-}$. During training, the TD error is computed at every integration step as
\begin{equation}
\delta_n = R_{n-1} + \gamma\, q_{a_n}(t_n) - q_{a_{n-1}}(t_{n-1}),\quad \gamma=0.985\ \text{per integration step},
\label{eq:05}
\end{equation}
where $q_{a_n}(t_n)=\max_{a'} q_{a'}(t_n)$ and $R_{n-1}$ is the reward returned by the environment after step $n-1$. Reservoir states, Q values, and, during training, the output weights are thus all updated at every integration step.
State transitions of the cursor and nonzero rewards occur only for one-step moves at the GO-window end times $t_k$. At all other steps, $R_{n-1}{=}0$. The update then bootstraps the preceding step's selected action value from the current maximum action value, discounted by $\gamma$.

The reward entering the TD update was defined as follows. When an action was executed at the end of a GO window and the trial continued, a move that decreased the squared Euclidean distance to the goal received $+0.1$, whereas all other moves received $-0.3$. Because actions changed only one coordinate by one grid cell, decreasing the squared Euclidean distance to the goal selected the same progress-making moves as decreasing the Manhattan distance among valid moves in this task; the squared Euclidean form was used because it matches the implementation. A trial ended when the goal was reached or at the 12-s time limit. The per-step TD(0) update was identical in the With-STP and Without-STP conditions. Like the values marked ``Inherited'' in Table~\ref{tab:03}, the reference reward values were inherited from preliminary simulations and the conference-version model; they encourage progress along the shortest path while suppressing nonprogressive actions. To assess sensitivity to their overall magnitude and relative weighting, we trained and evaluated networks with nine pairs of positive progress and negative nonprogress rewards while retaining the circuit settings of both main comparisons (Supplementary Fig.~\ref{fig:28}). Only the row corresponding to the action selected at the preceding step is updated:
\begin{equation}
W^{\mathrm{out}}_{a_{n-1},j}\leftarrow W^{\mathrm{out}}_{a_{n-1},j}+\eta\,\delta_n\,\bar r_j(t_{n-1}),
\end{equation}
where the learning rate was set to $\eta=3\times10^{-3}$ during the initial training phase and was reduced to $\eta=7\times10^{-4}$ after the cumulative number of successful training episodes reached $633$ (the integer part of $1{,}900/3$, one third of the training horizon). This success-count threshold was inherited from preliminary simulations and the conference-version model; the threshold was not tuned and was kept fixed in the comparisons reported here, apart from the control with fixed learning-rate schedules (\S\ref{sec:12}).

\subsection{Training, evaluation, and randomization}\label{sec:09}
We used $n{=}100$ different random seeds. NumPy's random-number stream was initialized with seeds 0--99. Each sparse matrix had a fixed number of nonzero entries, placed at positions drawn by random permutation of all matrix positions (for $W^{\mathrm{rec}}$, including the diagonal). For each seed, the recurrent connectivity $W^{\mathrm{rec}}$, input connectivity $W^{\mathrm{in}}$, and feedback connectivity $W^{\mathrm{back}}$ were initialized independently, so the 100 seeds correspond to 100 distinct network structures. The With-STP and Without-STP conditions shared the same connectivity structure $(W^{\mathrm{rec}}, W^{\mathrm{in}}, W^{\mathrm{back}})$ for a given seed and were evaluated as paired conditions that differed only in the presence or absence of STP (in the comparison with selected $\alpha_r$, the two conditions shared the same unscaled recurrent matrix, each multiplied by its selected $\alpha_r$). The initial output weights were also identical within a pair. For both conditions of a pair, the random-number generator was initialized with the same seed; goals and state noise were drawn as each simulation proceeded. Because episode lengths depended on behavior, the goal sequences and noise realizations could diverge between the two conditions.\par At the beginning of each episode, the cursor was reset to the central cell, and the reservoir and STP variables were initialized as $m_i(0)=\bar r_j(0)=0$, $x_j(0)=1$, and $u_j(0)=U_{\mathrm{SE}}$. The initial activity was set directly to zero. From the first update onward, activity was obtained by applying the activation function to the updated reservoir state; the STP update used the activity entering that update. Saved reservoir and Q-value trajectories, and the mid-trial STP snapshots (\S\ref{sec:17}), represented updated states before the corresponding environment step. Actions were executed only at the end of each GO window; if a selected action would have moved the cursor outside the grid or into a blocked corner cell, the cursor remained at the current position and that GO opportunity was consumed; this case received the same negative reward as a move away from the goal. An episode terminated when the goal was reached or when the simulation reached the trial time limit of 12\,s without reaching the goal.\par Each network was trained for 1,900 episodes (Fig.~\ref{fig:02}); final performance was evaluated over 100 test episodes with learning disabled ($\eta{=}0$), in which the eight goals were presented randomly. We report the mean $\pm$ standard deviation (SD) across seeds. Unless otherwise noted, the representational analyses below used all evaluation trials in each condition.\par As a supplementary analysis, we recalculated scatter ratio and decoding accuracy, in both the delay-period and the later-trial windows, using only successful trials (those in which the goal was reached) in each condition. We included 75 paired seeds that met three criteria in both conditions: at least 10 successful trials, successful trials for at least four of the eight goals, and at least as many successful trials for the most frequent goal as cross-validation folds. Decoding in this supplementary analysis used activity averaged within each window. For each seed, decoding used only the goals with at least two successful trials in both conditions (the same goals in both conditions), so that every goal in a test fold was also present in the corresponding training set. After excluding seeds with fewer than two such goals or with too few trials for five-fold stratification, 73 of the 75 seeds remained (Supplementary Fig.~\ref{fig:14}). The main hyperparameters are summarized in Table~\ref{tab:03}.

\subsection{Noise conditions}\label{sec:10}
To test robustness, we injected additive noise into the reservoir state and evaluated the sensitivity of success rate to noise intensity.
During both training and testing, independently and identically distributed Gaussian noise $\xi_i\!\sim\!\mathcal{N}(0,\sigma^2)$ was injected in Eq.~\eqref{eq:01}. The default value was $\sigma{=}0.001$, and the main robustness analysis used two training/evaluation-matched conditions: $\sigma{=}0$ (without noise) and $\sigma{=}0.001$. The noise term was implemented as discrete-time state noise added to $m_i$ once per integration step, outside the factor $\Delta t/\tau$ in Eq.~\eqref{eq:01}; $\sigma$ is therefore the standard deviation per 10-ms step, and the term was not intended as a continuous-time diffusion term. In addition to these two conditions, we examined three extensions. (i) \emph{Noise axis.} Networks were trained and evaluated at the same noise level for 8 levels, $\sigma\in\{0,\,0.001,\,0.002,\allowbreak\,0.004,\,0.008,\allowbreak\,0.016,\,0.032,\allowbreak\,0.064\}$ (Fig.~\ref{fig:06}). (ii) \emph{Independent training and evaluation noise.} Networks trained at each of 4 noise levels ($\sigma_{\mathrm{train}}\in\{0,\,0.001,\,0.002,\,0.004\}$) were evaluated, with learning disabled, at each of 7 noise levels ($\sigma_{\mathrm{eval}}\in\{0,\,0.001,\,0.002,\allowbreak\,0.004,\,0.008,\allowbreak\,0.016,\,0.032\}$); this analysis is reported in Supplementary Fig.~\ref{fig:19}. (iii) \emph{Other forms of noise.} Trained networks were evaluated under two further forms of noise, with learning disabled and the additive state noise kept at its training level. For multiplicative noise on synaptic transmission, the contribution from unit $j$ to the recurrent drive in Eq.~\eqref{eq:01} was replaced by $W^{\mathrm{rec}}_{ij}\bigl(2\,\bar r_j x_j u_j(1+\varepsilon_j)/U_{\mathrm{SE}}-1\bigr)$ with STP and by $W^{\mathrm{rec}}_{ij}\bigl(2\,\bar r_j(1+\varepsilon_j)-1\bigr)$ without STP, with $\varepsilon_j\sim\mathcal{N}(0,s^2)$ drawn independently for each unit and step. For input noise, independent Gaussian noise $\mathcal{N}(0,\sigma_{\mathrm{in}}^2)$ was added at each step to every element of the input vector $p$ before the input weights were applied, so that the resulting perturbation is correlated across reservoir units through $W^{\mathrm{in}}$. The levels were $s\in\{0.05,\,0.25\}$ and $\sigma_{\mathrm{in}}\in\{0.1,\,0.5\}$, applied to the networks trained at $\sigma{=}0.001$ (Supplementary Fig.~\ref{fig:25}).

\subsection{Ablation experiment}\label{sec:11}
We compared the With-STP condition with the Without-STP condition, the standard fixed-scaling reservoir defined in \S\ref{sec:05}. Both conditions were trained separately using the paired network initialization described in \S\ref{sec:09}. Recurrent-scaling selection and the separately trained controls are specified in \S\ref{sec:12}; the gain-matched controls and the perturbations of the STP state (STP reset, frozen STP, and replay) are specified in \S\ref{sec:19}.

\begingroup\subsection{Design of the comparisons and controls}\label{sec:12}
Separately trained controls apply the specified recurrent rule, learning schedule, or pathway change during readout learning. Evaluation-time perturbations retain the trained readout and change a specified state, modulation pattern, or pathway. Unless otherwise noted, each condition in each comparison or control comprised the same 100 seeds, 1{,}900 training episodes, and 100 evaluation episodes as the main comparison, with $\tau=1{,}000$\,ms, $\sigma{=}0.001$, and the default STP, input, and feedback parameters.

\paragraph{The two comparisons and the grid of recurrent scalings}
Both conditions were trained and evaluated at each of 16 values of the recurrent scaling, $\alpha_r\in\{0.25,\allowbreak\,0.5,\allowbreak\,\ldots,\allowbreak\,4.0\}$ (steps of 0.25). The \emph{comparison at the same recurrent scaling} contrasts the two conditions at the same value, $\alpha_r{=}3.0$, so that the two conditions differ in the presence of STP alone. The \emph{comparison with the recurrent scaling selected for each condition} contrasts the two conditions at the values of $\alpha_r$ that gave each condition its highest mean success rate across the 100 networks at $\sigma{=}0.001$: $\alpha_r{=}2$ with STP and $\alpha_r{=}1.25$ without STP. For brevity, we refer to these as the comparisons at the same $\alpha_r$ and with selected $\alpha_r$. In the noise axis (\S\ref{sec:10}), the selected values were kept fixed across noise levels; values re-selected at each noise level are reported separately in Supplementary Fig.~\ref{fig:34}. The main mechanistic analyses of state representation, action values, and effective connectivity (\S\ref{sec:13}--\S\ref{sec:17}) were carried out on the networks of the comparison at the same $\alpha_r$. State-representation and action-value measures with the recurrent scaling selected for each condition are also reported in Supplementary Fig.~\ref{fig:35}A--C. To evaluate recurrent-scaling selection on separate networks, we also selected $\alpha_r$ by mean success in one half of the seeds and evaluated it in the other, reversing the halves for a second comparison (Supplementary Fig.~\ref{fig:37:parta}A).

\paragraph{Axis of unit time constants}
The unit time constant $\tau$ was varied over 12 values, from 250 to 3{,}000\,ms in steps of 250\,ms, which include the main value (1{,}000\,ms). At each value of $\tau$, both conditions were trained and evaluated on the same grid of 16 recurrent scalings, and the comparisons at the same $\alpha_r$ and with selected $\alpha_r$ were formed as above, with the selection repeated at each value of $\tau$. To measure how much of the task can be solved by the slow integration of the units alone, we trained a \emph{passive control}: Without-STP networks in which the recurrent connections were almost removed and the Q-value feedback was removed ($\alpha_r{=}0.01$, $\alpha_b{=}0$; the implementation requires $\alpha_r>0$), at each of the 12 values of $\tau$.

\paragraph{Learning without Q-value feedback}
To examine the contribution of the Q-value feedback pathway, we trained both conditions from scratch with the feedback pathway removed ($\alpha_b{=}0$) on the same grid of recurrent scalings at $\sigma{=}0$ and $\sigma{=}0.001$ (Supplementary Fig.~\ref{fig:24}). Within these networks, we formed the comparisons at the same $\alpha_r$ and with selected $\alpha_r$. The recurrent scalings selected at $\sigma{=}0.001$ were held fixed across these noise levels; the values re-selected at $\sigma{=}0$ are shown separately in that figure. In a separate analysis, the feedback pathway was removed only at evaluation in networks trained with feedback ($W^{\mathrm{back}}$ set to zero from trial onset, or after the first action execution). Because the executed actions still change the position input, we refer to these conditions as conditions without Q-value feedback rather than as open-loop conditions. We also varied the Q-value feedback scaling during both training and evaluation over $\alpha_b\in\{0,0.25,0.5,1,2\}$, with $\alpha_r{=}3.0$, $\tau{=}1{,}000$\,ms, and $\sigma\in\{0,0.001\}$ (Supplementary section ``Q-value feedback strength''; Fig.~\ref{fig:30}).

\paragraph{Fixed learning-rate schedules}
The comparisons at the same $\alpha_r$ and with selected $\alpha_r$ were repeated, at $\sigma{=}0$ and $\sigma{=}0.001$, under 4 learning-rate schedules that were fixed identically for both conditions and all networks: a switch from $3\times10^{-3}$ to $7\times10^{-4}$ after 633 training episodes, a switch after 1{,}267 training episodes, no switch ($3\times10^{-3}$ throughout), and the lower rate from the start ($7\times10^{-4}$ throughout).

\paragraph{Prescribed time-varying scaling}
To examine whether a prescribed time-varying but goal-independent recurrent scaling can account for the With-STP advantage, we trained Without-STP networks from scratch with a prescribed time course replacing the constant factor multiplying $\bar r_j$ in the recurrent drive. The recurrent drive to unit $i$ became $\sum_j W^{\mathrm{rec}}_{ij}\bigl(2\,\bar g_j(t)\,\bar r_j(t)-1\bigr)$. For each seed and integration step, $\bar g_j(t)$ is the average of the scaling $x_j(t)u_j(t)/U_{\mathrm{SE}}$ entering the recurrent update over still-active evaluation trials of the With-STP network, pooling goals. At integration steps with no active trials, the most recent average from active trials was retained. Two variants were used: Unit-wise prescribed scaling with $\bar g_j(t)$ as defined above, and Common prescribed scaling in which $\bar g_j(t)$ was further averaged over units. The prescribed time course was applied from trial onset in both training and evaluation. These controls were trained at each of the same 16 recurrent scalings, using the time courses recorded from the With-STP networks at $\alpha_r{=}3.0$.\par \paragraph{Activity-dependent modulation rules}\mbox{}\\ \begingroup\setlength{\emergencystretch}{2em}We additionally trained controls with single-variable scaling driven by population-mean activity (Population-mean scaling) or own-unit activity (Own-unit scaling), and STP driven by a different unit fixed within each network (Other-unit STP). These comparisons test the modulation rule and its driving activity under the same training and evaluation settings, at the same recurrent scaling ($\alpha_r{=}3.0$) and at each of the 16 recurrent scalings; see Supplementary section ``Comparisons with simpler activity-dependent rules'' and Figs.~\ref{fig:29}, \ref{fig:33:partb}D and~\ref{fig:37:partc}D, E. We also trained a steady-state STP control in which synaptic efficacy at each integration step was set by the steady-state solutions of Eqs.~\eqref{eq:03}--\eqref{eq:04}, evaluated at that unit's current activity, at unit time constants of 500, 750 and 1{,}000\,ms. This control retained the original STP parameters. History remained in the recurrent activity rather than in separately evolving resource and utilization states. The frozen-STP control (\S\ref{sec:19}) instead holds a reference efficacy vector fixed throughout a trial. See Supplementary section ``Steady-state STP modulation'' (Figs.~\ref{fig:33:parta}A--C, \ref{fig:37:partb}B, C and~\ref{fig:38}).\par\endgroup

\endgroup\subsection{\texorpdfstring{Analysis overview and dimensionality reduction}{Analysis overview and dimensionality reduction}}\label{sec:13}
\paragraph{Analysis overview}  The primary analyses assessed behavioral robustness, time-resolved goal decoding, the action-value difference $D_Q$, and effective-connectivity similarity $\Delta_{\mathrm{sim}}(t)$. The scatter ratio, relative effective spectral radius $\rho_{\mathrm{rel}}(t)$, and eigenvalue-distribution summaries were auxiliary dynamical indices. The STP-parameter grid search assessed dependence on the synaptic time constants (\S\ref{sec:18}).

\paragraph{PCA visualization} In the following analyses (PCA, decoding, and scatter ratio), we denote the vector of normalized reservoir activities $\bar{\mathbf{r}}(t)$ by $\mathbf{r}(t) \in \mathbb{R}^{N_m}$ for notational simplicity. Dimensionality reduction was performed on this normalized-activity vector.
We used principal component analysis (PCA) to visualize goal-conditioned normalized-activity trajectories in two dimensions. PCA was performed independently for each seed, with principal axes estimated from the normalized-activity time series obtained from the evaluation trials of that seed. For the representative visualization in \S\ref{sec:23}, PCA axes were estimated separately for each condition within the displayed seed; PCA provides a qualitative visualization of trajectory structure. Statistical comparisons rely instead on indices that do not depend on dimensionality reduction: the scatter ratio $\SR(t)$, goal-decoding accuracy, effective-connectivity similarity, the effective spectral radius, and the action-value difference $D_Q$.

\subsection{Linear decodability of goal information}\label{sec:14}
To quantify whether goal identity can be read out from the reservoir state, we performed goal decoding with a linear classifier. For each seed, the normalized reservoir activity $\bar r_j(t)=\tfrac{1}{2}\!\left[1+\tanh(m_j(t)/T)\right]$ ($T{=}0.4$) was averaged across the delay period ($t=3$--$4$\,s) to form an $N_m$-dimensional feature vector. Using the eight goal identities as class labels, we trained and evaluated an $L_2$-regularized multinomial logistic regression model (solver: limited-memory Broyden--Fletcher--Goldfarb--Shanno, L-BFGS, a quasi-Newton optimization method) with five-fold cross-validation. The inverse regularization strength was set to 1.0. Splits were generated with StratifiedKFold so that label proportions were preserved across folds, without shuffling. Cross-validation splits were defined at the episode level so that samples from the same trial did not span both training and test sets. Every goal represented in a test fold was also present in that fold's training set. Standardization parameters were estimated using only the training set of each fold, and the same transformation was applied to the test set. The mean cross-validation accuracy for each seed was taken as that seed's decoding accuracy, and conditions were then compared across the 100 seeds. Chance level was $1/8 = 12.5\%$.

In addition, to examine how the maintenance of goal representation changes over time, we also performed time-resolved decoding. At each sampled time point $t$ (every 100\,ms), the normalized-activity vector $\mathbf{r}(t)\in\mathbb{R}^{N_m}$ was used as the feature vector, and an $L_2$-regularized multinomial logistic regression model (solver: L-BFGS) was trained and evaluated independently at each time point using the eight goal identities as labels. Evaluation used five-fold cross-validation, with standardization parameters estimated only from the training set of each fold and applied unchanged to the test set. No information was shared across time points; each time point was classified independently. Time-specific cross-validation accuracy was computed for each seed and then averaged across the 100 seeds. Chance level was $12.5\%$. For the trial-by-trial analysis at GO3 (\S\ref{sec:23}), we additionally trained and evaluated the same decoder on the instantaneous normalized-activity vector at $t_3^{-}$, using the same episode-level folds. Each trial's held-out goal prediction was paired with the action selected by the trained readout at that integration step.
Delay-period decoding used window-averaged activity, whereas time-resolved decoding used activity at each sampled time point.

\subsection{Action-value difference at GO opportunities}\label{sec:15}
\begingroup\setlength{\emergencystretch}{2em}To examine whether goal-conditioned state structure was reflected in the action-value readout, we quantified the preference for target-consistent actions at GO opportunities ($t{=}5,7,9$\,s; corresponding to the steps at which environment-executed actions are determined). Because multiple shortest-path actions can be valid in the grid, the target-consistent action set $\mathcal{A}_{\mathrm{target}}(t_k)$ was defined as the valid actions that reduced the squared Euclidean distance to the goal (matching the reward shaping used in the environment) and did not move the cursor off the grid or into a blocked corner cell. Because the policy executes $\arg\max_a q_a(t)$ over all four action outputs, and off-grid or blocked-corner actions still consume a GO opportunity and incur negative reward, the competitor set was defined as $\mathcal{A}_{\mathrm{other}}(t_k) = \mathcal{A}_{\mathrm{all}} \setminus \mathcal{A}_{\mathrm{target}}(t_k)$ with $\mathcal{A}_{\mathrm{all}} = \{\mathrm{LEFT},\mathrm{RIGHT},\mathrm{UP},\mathrm{DOWN}\}$, so that invalid actions are also competitors at the readout. We then computed the difference between the largest Q value among target-consistent actions and the largest Q value among the remaining actions:\par\endgroup
\begin{equation}
D_Q(t_k) = \max_{a\in\mathcal{A}_{\mathrm{target}}(t_k)} q_a(t_k^{-})
         - \max_{a\in\mathcal{A}_{\mathrm{other}}(t_k)}  q_a(t_k^{-}).
\label{eq:06}
\end{equation}
Positive $D_Q$ values indicate that the readout assigned a larger action value to at least one target-consistent action than to every other action index, including invalid ones. This quantity was used only as an operational readout-level index aligned with the all-action argmax of the policy, and is not intended to denote the reinforcement-learning advantage function. GO opportunities after terminal goal reaching, and any entries corresponding to padded Q-value arrays after episode termination, were excluded. For each seed and GO opportunity, we averaged $D_Q$ over evaluable trials and calculated the fraction with $D_Q{>}0$. GO2--GO3 summaries then averaged the two GO-specific means or fractions with equal weight; group summaries used these seed-level values. At $D_Q{=}0$, action selection followed the deterministic tie-breaking rule described in \S\ref{sec:08}.

\subsection{Temporal changes in goal separability: between-class and within-class scatter ratio}\label{sec:16}

PCA and scatter-ratio calculations were based on the normalized-activity vector $\mathbf{r}(t)$ at each time point. Below, in the formal definition of the scatter ratio, we denote the activity vector by $\mathbf{z}(t)$ to avoid confusion with the STP resource variable $x_j(t)$. We measured separation among goal-conditioned activities at each time point $t$ as the ratio of between-class to within-class scatter in the full $N_m$-dimensional reservoir activity space.
Let $\{\mathbf{z}^{(i)}_g(t)\}_{i=1}^{n_g}\subset\mathbb{R}^{N_m}$ denote the activity vectors for goal $g\in\{1,\ldots,G\}$ at time $t$. The class mean and global mean are
\[
\mu_g(t)=\frac{1}{n_g}\sum_{i=1}^{n_g}\mathbf{z}^{(i)}_g(t),\qquad
\mu(t)=\frac{1}{\sum_h n_h}\sum_{g=1}^{G} n_g\,\mu_g(t)
\]
The total within-class scatter and total between-class scatter are
\[
\SW(t)=\sum_{g=1}^{G}\sum_{i=1}^{n_g}\bigl\|\mathbf{z}^{(i)}_g(t)-\mu_g(t)\bigr\|^2,\qquad
\SB(t)=\sum_{g=1}^{G} n_g\,\bigl\|\mu_g(t)-\mu(t)\bigr\|^2
\]
The separability score is defined as the scatter ratio
\begin{equation}
\SR(t)=\frac{\SB(t)}{\SW(t)}
\label{eq:07}
\end{equation}
Targets were presented randomly during evaluation; therefore, the number of trials $n_g$ could differ slightly across goals. Each goal's contribution to the between-class scatter was weighted by its number of trials, $n_g$. The main analyses used all evaluation trials without balancing or resampling of goal classes. As a sensitivity analysis, we additionally performed balanced-goal resampling in which the number of trials was matched across goals within each seed and condition (see Supplementary Fig.~\ref{fig:15}). Each resample selected distinct trials for each goal. In this balanced-goal resampling analysis, decoding used shuffled stratified episode-level folds, with the number of folds set to the smaller of five and the minimum per-goal trial count. Every goal represented in a test fold was also present in the corresponding training set. For the scatter-ratio calculations, states were used at their original scale, without additional standardization. The same preprocessing was applied to the With-STP and Without-STP conditions. We did not standardize because changes in effective scaling induced by STP could themselves contribute to goal separation, and we wished to preserve that effect. The quantity $\SR(t)$ was calculated at every simulation step for each seed. Because the absolute value of the scatter ratio depends on the geometry of the internal state space of each network, group statistics were based on seed-wise averages within predefined time windows, while the time-series plot in the main text shows a qualitative visualization from one representative network. A small positive denominator was used where within-class scatter was nonpositive.

\subsection{Dynamical indices}\label{sec:17}

\paragraph{Effective spectral radius}
To quantify how strongly STP modulates the effective dynamics of the reservoir, we computed the spectral radius of the effective recurrent connectivity matrix. In the recurrent drive term of Eq.~\eqref{eq:01}, the effective coefficient multiplying the normalized activity $\bar r_j(t)$ is $2 x_j(t) u_j(t) / U_{\mathrm{SE}}$. We define this as the scaling factor for unit $j$, $s_j = 2 x_j u_j / U_{\mathrm{SE}}$, and construct the effective recurrent connectivity matrix as $W^{\mathrm{rec}}_{\mathrm{eff}} = W^{\mathrm{rec}}\,\mathrm{diag}(\mathbf{s})$ (column scaling). Because the STP variables $(x_j, u_j)$ vary according to each unit's activity history, $s_j$ takes different values across units and time points. We define $\rho_{\mathrm{eff}} = \max_i |\lambda_i(W^{\mathrm{rec}}_{\mathrm{eff}})|$. In the Without-STP condition, $x_j = 1$ and $u_j = U_{\mathrm{SE}}$ always hold, so $s_j = 2$ (constant) and $W^{\mathrm{rec}}_{\mathrm{eff}} = 2 W^{\mathrm{rec}}$. Analyses of the goal specificity of $W^{\mathrm{rec}}_{\mathrm{eff}}$ are based on the time-resolved analysis described below. Note that $\rho_{\mathrm{eff}}$ is not the spectral radius of the full Jacobian of the nonlinear system, but is used as an operational index of the dynamic modulation of effective recurrent connectivity by STP. The Jacobian of the full system additionally involves the slope of the activation function, the leak and the integration step, the dynamics of the STP variables themselves, and the Q-value feedback. In this paper, the term ``stabilize'' is used in an operational sense: after the goal cue has been removed, goal-dependent state separation, goal decodability, and goal-consistent action selection are preserved; this usage is distinct from Lyapunov, fixed-point, or attractor stability of the full system. STP does not alter the structural recurrent connectivity $W^{\mathrm{rec}}$ itself, but forms a time-varying effective recurrent connectivity matrix $W^{\mathrm{rec}}_{\mathrm{eff}}(t)$ through the multiplicative scaling vector $\mathbf{s}(t)$. To track the temporal evolution of STP-induced scaling modulation, we calculated $\rho_{\mathrm{eff}}(t)$ at each time step. The per-network time course averaged the episode-specific spectral radii at each sampled step. For comparisons across conditions, we used the relative effective spectral radius $\rho_{\mathrm{rel}}(t) = \rho_{\mathrm{eff}}^{\mathrm{STP}}(t) / \rho_{\mathrm{eff}}^{\mathrm{Without\,STP}}$, normalized by the Without-STP value for each seed. This ratio shows how much STP changes the effective spectral radius of each network relative to its Without-STP value. The ratio still depends on the recurrent connectivity of that network. We also analyzed the distribution of eigenvalue magnitudes to characterize spectral changes beyond the largest magnitude. The Supplementary section ``Eigenvalue-magnitude distributions of effective recurrent connectivity'' gives the analysis of episode-averaged matrices (Fig.~\ref{fig:41}).

\paragraph{Time-resolved analysis of goal specificity in effective connectivity}
To quantify the temporal evolution of goal-specific reorganization of effective connectivity, for each evaluation episode $e$ and time point $t$ we computed the multiplicative scaling vector derived from the STP state variables,
\[
\mathbf{s}_e(t)=2\,\mathbf{x}_e(t)\odot\mathbf{u}_e(t)/U_{\mathrm{SE}}
\]
and constructed the effective recurrent connectivity
\[
W^{\mathrm{rec}}_{\mathrm{eff},e}(t)=W^{\mathrm{rec}}\,\mathrm{diag}(\mathbf{s}_e(t))
\]
Here $\odot$ denotes the element-wise (Hadamard) product, so that the $j$th element of $\mathbf{s}_e(t)$ is $2x_j(t)u_j(t)/U_{\mathrm{SE}}$, whereas $W^{\mathrm{rec}}\,\mathrm{diag}(\mathbf{s}_e(t))$ is the ordinary matrix product, which scales the $j$th column of $W^{\mathrm{rec}}$ by the $j$th element of $\mathbf{s}_e(t)$. The STP variables $(x,u)$ were reconstructed offline from normalized activity derived from the saved reservoir-state trajectory using Eqs.~\eqref{eq:03}--\eqref{eq:04} with initial conditions $x_j(0){=}1$, $u_j(0){=}U_{\mathrm{SE}}$, and $\bar r_j(0){=}0$, and with the same update equations, parameter values, and time indexing as in the simulator. Reconstruction was checked against the stored mid-trial $(x_j,u_j)$ snapshots taken after zero-based integration step $\lfloor L_e/2\rfloor$, where $L_e$ was the episode length in integration steps. The STP states reconstructed for the spectral-radius, eigenvalue, and both goal-specificity analyses agreed with the stored STP states at the same indices to numerical precision. Each $W^{\mathrm{rec}}_{\mathrm{eff},e}(t)$ was vectorized, and cosine similarity was computed between evaluation episodes within the same seed. Similarities were separated into same-goal and different-goal pairs (an episode paired with itself, whose similarity is 1, was excluded) and averaged with equal weighting across goals and goal pairs. We defined the time-resolved trajectory of goal specificity as
\[
\Delta_{\mathrm{sim}}(t)=\mathrm{sim}_{\mathrm{same}}(t)-\mathrm{sim}_{\mathrm{diff}}(t)
\]
We evaluated this quantity every 100\,ms over the interval $t=0$--$9$\,s and computed the group mean and 95\% confidence interval across seeds ($n{=}100$). In the Without-STP condition, because $x_j(t)=1$ and $u_j(t)=U_{\mathrm{SE}}$ always hold, $W^{\mathrm{rec}}_{\mathrm{eff}}(t)=2W^{\mathrm{rec}}$ is time-invariant, and therefore $\Delta_{\mathrm{sim}}(t)=0$. Inferential statistics rely on comparisons within predefined summary windows based on the task schedule. Because cosine similarity evaluates directional similarity between vectorized $W^{\mathrm{rec}}_{\mathrm{eff},e}(t)$, $\Delta_{\mathrm{sim}}(t)$ primarily reflects goal specificity in effective-connectivity patterns rather than directly measuring differences in the absolute Frobenius norm. To relate the effective-connectivity index to the STP scaling vectors, consider the matrix cosine for fixed $W^{\mathrm{rec}}$. Writing $A(\mathbf{v})=W^{\mathrm{rec}}\mathrm{diag}(\mathbf{v})$ and $\Omega=\mathrm{diag}(\omega_j)$ with $\omega_j=\sum_i(W^{\mathrm{rec}}_{ij})^2$, the matrix cosine satisfies \[\begin{aligned} &\cos(\mathrm{vec}A(\mathbf{s}),\mathrm{vec}A(\mathbf{v}))\\ &\quad=\frac{\mathbf{s}^{\mathsf T}\Omega\mathbf{v}}{\sqrt{(\mathbf{s}^{\mathsf T}\Omega\mathbf{s})(\mathbf{v}^{\mathsf T}\Omega\mathbf{v})}}\\ &\quad=\cos(\Omega^{1/2}\mathbf{s},\Omega^{1/2}\mathbf{v}) \end{aligned}\] for nonzero norms. We additionally defined the unweighted companion quantity $\Delta_s(t)$ directly on $\mathbf{s}(t)$:
\begin{equation*}
\Delta_s(t) = \mathrm{sim}^{(\mathbf{s})}_{\mathrm{same}}(t) - \mathrm{sim}^{(\mathbf{s})}_{\mathrm{diff}}(t),
\end{equation*}
where $\mathrm{sim}^{(\mathbf{s})}$ denotes cosine similarity between the $\mathbf{s}_e(t)$ vectors of the evaluation episodes within each seed, contrasting same-goal pairs against different-goal pairs. The two quantities describe the same STP scaling patterns with different unit weights: $\Delta_{\mathrm{sim}}(t)$ uses the squared recurrent-column norms, while $\Delta_s(t)$ weights units equally. By construction, $\Delta_s(t){=}0$ in the Without-STP condition. We use $\Delta_{\mathrm{sim}}(t)$ as the main effective-connectivity index (\S\ref{sec:24}) and report $\Delta_s(t)$ in Supplementary Fig.~\ref{fig:18}.

\subsection{Parameter exploration and biological correspondence}\label{sec:18}
To examine how the STP time constants affect success rate and to explore whether the high-success-rate region is consistent with the physiological range of cortical synapses, we carried out an exploratory grid search over $(\tau_{\mathrm{R}}, \tau_{\mathrm{F}})$. The grid range covers cortical STP time constants and parameter settings used in a synaptic working-memory model \citep{markram1998differential,wang2006heterogeneity,mongillo2008synaptic}. This range therefore includes both the facilitation time constant of a facilitating cortical connection modeled by \citet{tsodyks1998neural} ($\tau_{\mathrm{F}}\approx 530$\,ms) and the slower facilitation time constant used in a prefrontal working-memory model \citep{mongillo2008synaptic}.
All analyses in this subsection used With-STP networks at $\tau=1{,}000$\,ms and $\sigma{=}0.001$, with 100 networks (the same seeds as in the main comparison) per parameter setting; the Without-STP values used as references are those of the main comparisons.
We sampled $\tau_{\mathrm{R}}$ over 50--450\,ms in 50-ms increments (9 values) and $\tau_{\mathrm{F}}$ over 250--2{,}500\,ms in 250-ms increments (10 values), and evaluated success rate on this grid (90 grid points) with 100 seeds per grid point and $U_{\mathrm{SE}}{=}0.1$ fixed, at $\alpha_r{=}3.0$. We compared the resulting high-success-rate band with the time scales of facilitating synapses in prefrontal cortex (results are reported in \S\ref{sec:25}). Based on an earlier grid search with 10 seeds per grid point, we selected $(\tau_{\mathrm{R}}, \tau_{\mathrm{F}})=(100,500)$\,ms as a representative fixed parameter set from the high-success-rate band. This choice aligns with STP parameters reported for facilitating cortical synapses (e.g., the facilitating connection modeled by \citet{tsodyks1998neural}, with $\tau_{\mathrm{R}}\approx 130$\,ms and $\tau_{\mathrm{F}}\approx 530$\,ms). The parameter pairs $(\tau_{\mathrm{R}}, \tau_{\mathrm{F}}) = (150, 1{,}500)$~ms (near the slower PFC working-memory setting of \citet{mongillo2008synaptic}) and $(150, 1{,}000)$~ms are settings in this grid search (\S\ref{sec:25}). The subsequent 100-seed comparison between the With-STP and Without-STP conditions was conducted using this fixed parameter set, and the STP parameters were not individually optimized for each seed or condition.

\afterpage{{\small \begin{longtable}{@{}p{0.09\textwidth}@{\hspace{0.8em}}p{0.18\textwidth}@{\hspace{0.8em}}p{0.30\textwidth}@{\hspace{0.8em}}>{\leavevmode}p{0.35\textwidth}@{}} \caption{\textbf{Main hyperparameters and their sources.} ``Inherited'' denotes a default value carried over from preliminary simulations and the conference-version model. Symbols are defined in Table~\ref{tab:01}.}\label{tab:03}\\ \toprule Symbol & Value & Description & Source / rationale \\ \midrule \endfirsthead \multicolumn{4}{@{}l}{\footnotesize\tablename~\thetable{} (continued)}\\ \toprule Symbol & Value & Description & Source / rationale \\ \midrule \endhead \bottomrule \endfoot $\Delta t$ & \SI{10}{ms} & Integration step & Inherited; the same step was used in a recurrent network model with STP for working-memory tasks \citep{masse2019circuit} \\ $\tau$ & 1{,}000\,ms & Unit integration time constant & Main condition (\S\ref{sec:06}); 250--3{,}000\,ms examined in \S\ref{sec:12} \\ $N_m$ & 100 & Reservoir size & Inherited; the same number of recurrent units was used in \citet{masse2019circuit} \\ $p_r$ & 0.1 & Connection probability (fraction of nonzero entries) & Inherited; the same connection probability was used in a reservoir model of prefrontal dynamics \citep{enel2016reservoir}, and it is of the same order as local connection probabilities between pyramidal cells in ferret medial PFC \citep{wang2006heterogeneity} \\ $\alpha_r$ & 3.0{} (same $\alpha_r$); 2 with STP and 1.25 without STP (selected $\alpha_r$) & Recurrent-connection scaling & Same value in both conditions, or selected for each condition from the same grid of 16 values (\S\ref{sec:12}) \\ $\alpha_i$ & 18.112 & Input scaling factor & Inherited \\ $\alpha_b$ & 1.0 & Feedback scaling factor & Inherited; varied in the Q-value feedback controls (\S\ref{sec:12}; Supplementary Fig.~\ref{fig:30}) \\ $\eta$ & $3{\times}10^{-3}$, then $7{\times}10^{-4}$ & Output-weight learning rate (reduced after early successful training) & Inherited; reduced after 633 successful training episodes, approximately one third of the training horizon (\S\ref{sec:08}) \\ $\alpha_o$ & $0.02$ & Output-weight initialization scaling & Inherited \\ $T$ & $0.4$ & Activation-function temperature & Defined in \S\ref{sec:05} ($1/T{=}2.5$) \\ $\kappa$ & $1$ & Rate scale in the STP update & Defined in \S\ref{sec:05}; normalized activity 1 corresponds to 100\,Hz at $\Delta t{=}10$\,ms; $\kappa{=}0.2$ maps it to 20\,Hz in \S\ref{sec:18} \\ $\beta_i$ & $0.1$ & Fraction of nonzero input weights & Inherited \\ $\beta_b$ & $0.45$ & Fraction of nonzero feedback weights & Inherited \\ $\gamma$ & 0.985 & Discount factor (per integration step) & Inherited \\ $\sigma$ & 0.001 & State-noise standard deviation (per integration step) & Inherited; 0--0.064 examined in \S\ref{sec:10} \\ $U_{\mathrm{SE}}$ & 0.1 & Baseline STP utilization & Inherited; a value of 0.1 was obtained by fitting a facilitating pyramidal-to-interneuron connection in rat neocortex \citep{markram1998differential} and lies within the range estimated for pyramidal synapses in medial PFC \citep{barri2016quantifying}; five levels examined in \S\ref{sec:18} \\ $\tau_{\mathrm{R}}$ & \SI{100}{ms} & Resource recovery time constant (main comparison) & Selected as a pair with $\tau_{\mathrm{F}}$ in an earlier grid search with 10 seeds per grid point; confirmed in the grid search with 100 seeds per grid point (\S\ref{sec:18}) \\ $\tau_{\mathrm{F}}$ & \SI{500}{ms} & Facilitation time constant (main comparison) & Selected and confirmed as a pair with $\tau_{\mathrm{R}}$ \\ \end{longtable}}}\emph{Recovery time constant and recurrent scaling.} Because the recurrent scaling that suits a given STP setting may differ, $\tau_{\mathrm{R}}\in\{50,\,100,\,150,\,200,\,250\}$\,ms was additionally crossed with $\alpha_r\in\{0.5,\,1.0,\,\ldots,\,4.0\}$ (steps of 0.5) at $\tau_{\mathrm{F}}{=}\SI{500}{ms}$.

\emph{Utilization.} $U_{\mathrm{SE}}$ was varied over five levels, $U_{\mathrm{SE}}\in\{0.025,\allowbreak\,0.05,\allowbreak\,0.1,\allowbreak\,0.2,\allowbreak\,0.4\}$, at $(\tau_{\mathrm{R}}, \tau_{\mathrm{F}})=(100,500)$\,ms, for the same eight values of $\alpha_r$.

\emph{Parameter sets from the literature.} We tested the parameter set of the working-memory model of \citet{mongillo2008synaptic}, $(U_{\mathrm{SE}},\tau_{\mathrm{R}},\tau_{\mathrm{F}})\allowbreak=(0.2,\,\SI{200}{ms},\allowbreak\,1{,}500\,\mathrm{ms})$, and the group averages measured for facilitating connections between pyramidal cells in ferret medial PFC by \citet{wang2006heterogeneity}, $(0.28,\allowbreak\,\SI{194}{ms},\allowbreak\,\SI{507}{ms})$. Each set, together with the main parameter set, was evaluated under the default rate scale (100\,Hz for $\bar r_j{=}1$) and under the lower rate scale (20\,Hz; Eq.~\eqref{eq:02} and \S\ref{sec:05}), for $\alpha_r\in\{1,\,2,\,3,\,4\}$.

\subsection{Fixed recurrent-scaling and STP-state perturbation controls}\label{sec:19}
To test whether the With-STP advantage could be explained by a fixed increase in recurrent input strength, we trained a gain-matched Without-STP condition. The matched value was computed as
\[
\alpha_r^{\mathrm{matched}}
=
\rho_{\mathrm{rel}}^{\mathrm{delay}}\,\alpha_r^{\mathrm{default}},
\]
where $\rho_{\mathrm{rel}}^{\mathrm{delay}}\approx 1.55$ was the group mean of the delay-period-averaged effective spectral radius in the original With-STP condition divided by the corresponding group mean in the standard Without-STP condition, and $\alpha_r^{\mathrm{default}}=3.0$. This gave $\alpha_r^{\mathrm{matched}}\approx 4.65$. The same unscaled recurrent matrix and the same input and feedback weights were used as in the standard Without-STP condition; only the recurrent scaling factor $\alpha_r$ was changed. All other hyperparameters and the random seed were identical. The output weights were initialized from the same Gaussian distribution as in the standard Without-STP condition and trained from scratch using the same training procedure, so that the gain-matched condition serves as a baseline trained at the matched recurrent scaling. Writing $c=\alpha_r^{\mathrm{matched}}/\alpha_r^{\mathrm{default}}$ and $W^{\mathrm{rec}}$ for the recurrent matrix at $\alpha_r^{\mathrm{default}}$, the rescaled drive is $cW^{\mathrm{rec}}(2\bar{\mathbf r}-\mathbf 1)$, where $\mathbf 1$ is the all-ones vector. This operation scales both the activity coefficient and the constant component of the recurrent drive. We also trained a Gain-matched (constant component preserved) control, using $W^{\mathrm{rec}}(2c\bar{\mathbf r}-\mathbf 1)$ with the same calibrated $c=1.5506$. Its readout was trained from scratch using the same procedure (Supplementary Fig.~\ref{fig:16}).

We also performed three post-training analyses on the trained With-STP networks. In each case, the recurrent, input, feedback, and output weights $(W^{\mathrm{rec}}, W^{\mathrm{in}}, W^{\mathrm{back}}, W^{\mathrm{out}})$ were loaded from the original With-STP condition and only the evaluation episodes were re-run. In the \emph{intact re-evaluation} condition, the trained With-STP model was re-evaluated through the same post-training re-evaluation procedure without modifying the STP variables. In the \emph{STP reset} condition, $x_j$ and $u_j$ were reset to their baseline values, $x_j{=}1$ and $u_j{=}U_{\mathrm{SE}}$, once at the end of the update with the final goal-cue input, so that the first recurrent update after cue removal used the baseline STP state; the variables then evolved according to the standard STP equations. In the \emph{frozen-STP} condition, the dynamic STP-derived scaling vector $s_j(t){=}2x_j(t)u_j(t)/U_{\mathrm{SE}}$ was replaced throughout each trial by a static, time-invariant, goal-independent per-unit reference vector
\[
\bar s_j = \frac{1}{|\mathcal{T}|}\sum_{(e,t)\in\mathcal{T}} s_{j,e}(t),
\]
where $\mathcal{T}$ denotes the middle-of-trial snapshots from the first 10 original With-STP evaluation episodes within each seed (regardless of trial outcome) used to estimate the reference scaling vector. For an episode of length $L_e$ integration steps, the stored snapshot was taken after zero-based integration step $\lfloor L_e/2\rfloor$. This preserves the seed-specific average level of STP-derived scaling while removing both its time-specific and goal-specific patterning. The intact, reset, and frozen-STP conditions all used the output weights from the trained With-STP model; the latter two were acute post-training perturbations (Supplementary Figs.~\ref{fig:16} and~\ref{fig:17}).\par \begingroup\setlength{\emergencystretch}{2em}We additionally examined sensitivity to the assignment of replayed STP-derived scaling across units in the 100 trained With-STP networks at $\tau{=}1{,}000$\,ms, $\alpha_r{=}3.0$, and $\sigma{=}0.001$. From integration steps 301 through 899, we replaced the scaling $s_j(t)=2x_j(t)u_j(t)/U_{\mathrm{SE}}$ in the recurrent drive with a replayed trajectory. Integration steps were indexed from zero at trial onset; replay replaced the scaling entering the recurrent update at each indicated step. All weights and the input and feedback rules were held fixed. The variables $x_j$ and $u_j$ continued to evolve with the recipient network's activity, and the network's own scaling was restored from integration step 900. A separate bank of 96 donor trials (12 per goal) provided the mean trajectory $M_{g,j}(t)$ for each goal $g$; each donor block comprised one trial per goal, with the same noise realization across those trials. Intact used the network's own scaling, and Same-goal replayed the mean trajectory of the cued goal. Unit ranks were estimated by sorting the mean scaling over integration steps 301--399 in decreasing order, with ties broken by the smaller unit index. We write $\operatorname{rank}_{g}(j)$ for unit $j$'s rank for goal $g$. Other-goal rank reassignment (Other-goal) used a fixed permutation $\pi_{g,h}$ satisfying $\operatorname{rank}_{g}(\pi_{g,h}(j))=\operatorname{rank}_{h}(j)$ and replayed $M_{g,\pi_{g,h}(j)}(t)$ for another goal $h$. The cued-goal trajectories thus followed goal $h$'s unit ranking. For each network and goal pair, Shuffled used $\phi\circ\pi_{g,h}\circ \phi^{-1}$, where $\phi$ was a uniformly sampled permutation of unit labels, preserving the cycle structure of the corresponding Other-goal permutation. The three replay conditions preserved the same multiset of scaling values at each integration step and the same collection of scaling trajectories, while changing their assignment to units (see \S\ref{sec:22}). Unit reassignment changed how these trajectories interacted with the fixed recurrent weights and the recipient activities. For each network, the same 100 recipient trials and noise realizations were used across conditions, with donor noise generated separately. Results for the 7 other goals and their corresponding 7 shuffled permutations were averaged within each network. We measured trial success and first-move accuracy, defined as the proportion of all recipient trials in which the first move reduced the Manhattan distance to the cued goal by one grid step. Paired condition differences across the 100 networks were summarized with 95\% bias-corrected and accelerated (BCa) bootstrap confidence intervals (CIs) from 2{,}000 resamples.\par\endgroup

\subsection{Statistical tests and reproducibility}\label{sec:20}
 The primary comparisons were the success rate under noise, decoding accuracy in the later part of the trial ($t{=}8$--$9$\,s), comparisons of $\Delta_{\mathrm{sim}}(t)$ within predefined time windows, and the seed-wise mean action-value difference $D_Q$ averaged over GO2--GO3 (\S\ref{sec:15}). The success-rate comparison was made at the same $\alpha_r$ and, in addition, with selected $\alpha_r$ (\S\ref{sec:12}). For the comparisons along the axes of recurrent scaling, unit time constant, noise level, and STP parameters, and for the control comparisons, we report condition means and mean paired differences with 95\% confidence intervals for the displayed contrasts. For $D_Q$, conditions were compared using paired tests across the same 100 seeds, and we additionally summarized the fraction of evaluable GO decisions with $D_Q{>}0$; gain-matched, reset, and frozen-STP comparisons of $D_Q$ were treated as secondary supporting analyses for the readout-level interpretation. The scatter ratio $\SR(t)$ and $\rho_{\mathrm{rel}}(t)$ were treated as auxiliary indices showing consistent dynamical modulation, and the parameter search was treated as exploratory.
Because comparisons between the With-STP and Without-STP conditions were based on paired runs from the same seeds, we used paired $t$-tests and Cohen's $d_z$ (paired effect size). The $t$-tests, paired sign-flip tests, and Wilcoxon signed-rank tests were two-sided. Reported $p$ values and 95\% confidence intervals apply to individual comparisons and are unadjusted for multiple comparisons. Major time-series comparisons were restricted to predefined time windows based on the task schedule; exhaustive sequential testing at each time point was not performed. For similarity analyses, same-goal and different-goal similarities were first averaged within each seed; statistical tests used the resulting seed-level contrasts. Group comparisons for time-series indices (scatter ratio $\SR(t)$, decoding accuracy, goal specificity of effective connectivity $\Delta_{\mathrm{sim}}(t)$, and relative effective spectral radius $\rho_{\mathrm{rel}}(t)$) were performed on the seed-wise mean value within each predefined time window, treating the seed value as the statistical unit. The two windows were the delay period ($t{=}3$--$4$\,s) and the later interval ($t{=}8$--$9$\,s). To summarize condition differences in time-resolved decoding, we used the mean accuracy over $t{=}8$--$9$\,s, corresponding to the third GO window. For $\Delta_{\mathrm{sim}}(t)$, the predefined windows were the delay-period mean ($t{=}3$--$4$\,s) and the mean over the later interval ($t{=}8$--$9$\,s); we used a one-sample $t$-test to assess whether the delay-period mean exceeded the zero baseline, and a paired $t$-test within seeds to assess whether the mean over the later interval exceeded the delay-period mean. For $\Delta_{\mathrm{sim}}(t)$ (\S\ref{sec:24}), the delay-period mean used the 11 samples from 3.0 to 4.0\,s at 100-ms spacing, and for $\Delta_s(t)$ (Supplementary Fig.~\ref{fig:18}) the 10 samples from 3.0 to 3.9\,s; both used the 10 samples from 8.0 to 8.9\,s for the later interval. The same predefined windows ($t{=}3$--$4$\,s and $t{=}8$--$9$\,s) were also used for the relative effective spectral radius $\rho_{\mathrm{rel}}(t)$. For the summary statistics of the eigenvalue distribution, $Q_{95}(t)$ and $Q_{50}(t)$ (the 95th percentile and the median of the eigenvalue magnitudes of the episode-averaged $W^{\mathrm{rec}}_{\mathrm{eff}}(t)$; Supplementary Fig.~\ref{fig:41}), the delay-period mean ($t{=}3$--$4$\,s) was used as an auxiliary predefined window. For the scatter ratio and for decoding from window-averaged activity, these windows comprised the 10-ms samples with $3.0\le t<4.0$\,s and $8.0\le t<9.0$\,s; window averages of time-resolved decoding used the 11 samples at 100-ms spacing from 3.0 through 4.0\,s for the delay-period mean and the 10 samples from 8.0 to 8.9\,s for the later mean. For $\rho_{\mathrm{rel}}(t)$ and the eigenvalue-distribution summaries, the delay-period mean used the 10-ms samples from 3.0 through 4.0\,s inclusive. The 3.0-s sample included the final cue input and the 4.0-s sample included GO onset; all later-interval samples preceded the third action execution. We report Cohen's $d$ or $d_z$ as the effect size, as appropriate. As sensitivity checks, paired sign-flip permutation tests ($n_{\mathrm{resamples}}{=}10{,}000$), Wilcoxon signed-rank tests, and BCa bootstrap confidence intervals of the paired differences (2{,}000 resamples) were used to check whether the primary conclusions depended on the assumptions of the paired $t$-test; bootstrap CIs for the control-comparison differences are reported in Supplementary Figs.~\ref{fig:16} and~\ref{fig:17}.

BCa (bias-corrected and accelerated) bootstrap confidence intervals have end points adjusted for the bias and skewness of the resampled distribution. We chose this type of interval because success rate is bounded between 0 and 100\% and the seed-wise paired differences cannot be assumed to be normally distributed. The same 100 paired seeds shared between the With-STP and Without-STP conditions were used throughout, except where trial-inclusion criteria restricted an analysis to a subset of seed pairs (\S\ref{sec:23}) or recurrent-scaling selection and evaluation used separate halves of the seeds (Supplementary Fig.~\ref{fig:37:parta}A). The main and successful-trial decoders used an iteration limit of 1{,}000, and the decoder used for balanced-goal resampling had an iteration limit of 2{,}000. In time-resolved scatter-ratio calculations, nonpositive within-class scatter was replaced by $10^{-10}$ in the denominator.

\FloatBarrier
\section{Results}\label{sec:21}

We report task learning and noise robustness (RQ1--RQ2), state representation and action-value readout (RQ2), effective connectivity (RQ3), and STP parameter dependence (RQ4).

\subsection{Experiment 1: learning process and robustness under noise}\label{sec:22}

We evaluated task learning (RQ1) and the robustness of task performance under state noise (RQ2) at $\tau{=}1{,}000$\,ms. The With-STP and Without-STP conditions were compared in two ways (\S\ref{sec:12}): the comparison at the same recurrent scaling ($\alpha_r{=}3.0$ in both conditions) and the comparison with the recurrent scaling selected for each condition from the same grid. Unless otherwise noted, all analyses below are based on the condition in which additive state noise ($\sigma{=}0.001$) was applied during both training and evaluation. Comparisons with the condition without noise are presented separately in the noise-robustness analysis.

\begin{figure}[htbp]\centering\includegraphics{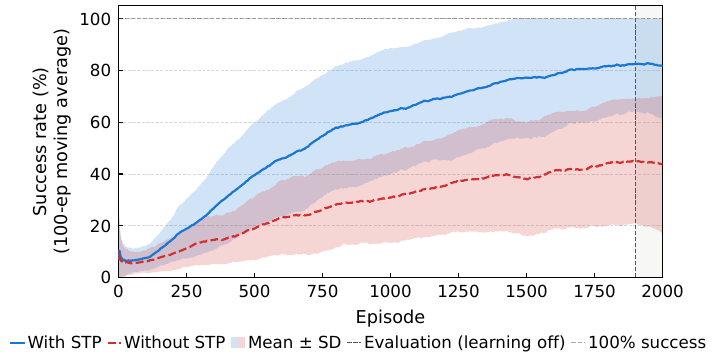}\caption{\textbf{Success rate of the With-STP and Without-STP conditions over the 1{,}900 training episodes and the subsequent 100 evaluation episodes.} Each condition used $n{=}100$ seeds, and shaded bands indicate the standard deviation (100-episode moving average). The moving average includes the current and preceding episodes; initial windows use all available episodes. The horizontal axis uses zero-based episode indices. Blue solid lines denote the With-STP condition and red dashed lines the Without-STP condition; the vertical dashed line marks the first evaluation episode (index 1{,}900). Readout learning was disabled for the final 100 evaluation episodes. The curves are shown for the condition in which the default state noise ($\sigma{=}0.001$) was applied during both training and evaluation ($\tau{=}1{,}000$\,ms; $\alpha_r{=}3.0$ in both conditions). The With-STP condition showed a higher mean success rate in the later training episodes and in the evaluation episodes. Post-training success rates are shown in Fig.~\ref{fig:03}. The standard-deviation band is clipped to the 0--100\% range; all individual seed values were within this range.}\label{fig:02}\end{figure}

\paragraph{Learning phase}
Regarding RQ1, Fig.~\ref{fig:02} shows the success rate of the two conditions during training (100-episode moving average). In the last 100 training episodes, the success rate was 82.6\% in the With-STP condition and 45.1\% in the Without-STP condition.

\begin{figure}[htbp]\ifdefined\OnepageFigureBox\else\newbox\OnepageFigureBox\fi\setbox\OnepageFigureBox=\vbox{\hsize=\textwidth\centering\includegraphics{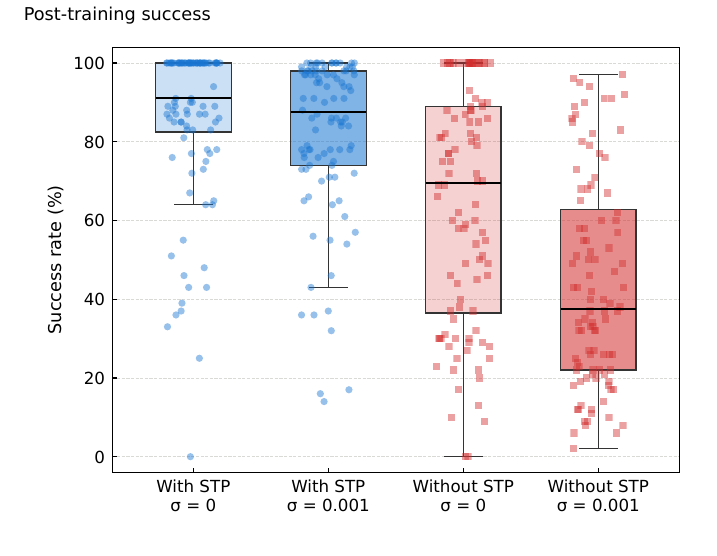}\caption{\textbf{Post-training success under matched training and evaluation noise.} Box plots show success rates over 100 evaluation episodes with learning disabled for With STP and Without STP at $\sigma{=}0$ and $0.001$ ($n{=}100$ paired networks; $\tau{=}1{,}000$\,ms; $\alpha_r{=}3.0$ in both conditions). Networks were trained separately at each noise level and evaluated at the same noise level. Boxes show the median and interquartile range (IQR); whiskers extend to the most extreme values within 1.5 interquartile ranges of the box, and points show individual networks. Lighter boxes denote $\sigma{=}0$; blue circles denote With STP and red squares denote Without STP. Numerical summaries and paired differences are reported in Table~\ref{tab:04}.}\label{fig:03}}\ifdim\dimexpr\ht\OnepageFigureBox+\dp\OnepageFigureBox\relax>\textheight\fi\box\OnepageFigureBox\end{figure}

\paragraph{Robustness under a fixed state-noise level (comparison at the same recurrent scaling)}
\begingroup\setlength{\emergencystretch}{3em}For RQ2 (robustness of task performance), we compared separately trained networks without noise with networks trained and evaluated under Gaussian state noise ($\sigma{=}0.001$). The success-rate reduction was smaller with STP at the same recurrent scaling (for the recurrent scaling selected for each condition, see below and Table~\ref{tab:04}). Without noise, the Without-STP condition still reached $(63.7\pm29.6)\%$, but under noise it decreased by 19.9 percentage points. With state noise, across the 100 seeds, with $\alpha_r{=}3.0$ in both conditions, the final success rate was $(81.9\pm20.6)\%$ in the With-STP condition and $(43.7\pm26.6)\%$ in the Without-STP condition (paired difference 38.2 percentage points, 95\% confidence interval [31.6, 44.2]; paired $t$-test, $p=\ensuremath{1.2\times10^{-19}}$, $d_z{=}1.14$; Table~\ref{tab:04}). In the With-STP condition, the reduction in success rate between the conditions without and with noise was 3.7 percentage points (85.6\% vs 81.9\%), 16.2 percentage points smaller than that in the Without-STP condition (95\% CI [10.5, 21.4]; Fig.~\ref{fig:03}). Even without noise, the With-STP condition outperformed the Without-STP condition, so the advantage of STP was not restricted to the condition with noise. However, the difference became more pronounced under noise. For the predefined primary comparisons between the With-STP and Without-STP conditions, paired sign-flip permutation tests, Wilcoxon signed-rank tests, and bootstrap 95\% confidence intervals of the paired differences yielded conclusions consistent with the paired $t$-tests (all $p < 0.001$; all confidence intervals excluded zero).\par\endgroup

 \paragraph{Comparison with the recurrent scaling selected for each condition} We next compared the two conditions with $\alpha_r$ selected separately for each condition from the same grid (\S\ref{sec:12}; Fig.~\ref{fig:04}). The selected values were $\alpha_r{=}2$ for the With-STP condition and $\alpha_r{=}1.25$ for the Without-STP condition. Success rates were higher in the With-STP condition at the selected values in the condition without noise (Table~\ref{tab:04}). At these values, the success rate at $\sigma{=}0.001$ was $(86.3\pm15.6)\%$ in the With-STP condition and $(71.6\pm21.9)\%$ in the Without-STP condition (paired difference 14.8 percentage points, 95\% CI [9.6, 20.1]; paired $t$-test, $p < 0.001$, $d_z{=}0.56$; Table~\ref{tab:04}). Between $\sigma{=}0$ and $\sigma{=}0.001$ (each used for both training and evaluation), the mean difference in success-rate reductions (Without STP minus With STP) was 1.4 percentage points (95\% BCa CI [\ensuremath{-}3.5, 6.3]). The With-STP success-rate advantage remained when recurrent-scaling selection and evaluation used separate network seeds in both split directions at $\sigma{=}0.001$ (Supplementary Fig.~\ref{fig:37:parta}A).

\begin{table}[htb] \centering \caption{\textbf{Primary comparisons between the With-STP and Without-STP conditions at $\tau{=}1{,}000$\,ms.} Same $\alpha_r$: the comparison at the same recurrent scaling ($\alpha_r{=}3.0$ in both conditions). Selected $\alpha_r$: the comparison with the recurrent scaling selected for each condition from the same grid of 16 values by the mean success rate at $\sigma{=}0.001$ ($\alpha_r{=}2$ With STP, $\alpha_r{=}1.25$ Without STP). These selected scalings were held fixed across noise conditions; networks were trained separately at each noise level. Success rates are mean $\pm$ standard deviation across $n{=}100$ seeds; differences are means of the seed-wise paired differences (With STP minus Without STP, percentage points) with 95\% BCa bootstrap confidence intervals (2{,}000 resamples); $d_z$ is the paired effect size. Paired differences were calculated before rounding and may therefore differ slightly from the difference between the displayed rounded means. In each row, the larger of the two mean success rates is shown in bold.} \label{tab:04} {\small\setlength{\tabcolsep}{3pt}\begin{tabular}{llcccc} \toprule Comparison & \shortstack{$\sigma_{\mathrm{train}}$\\$=\sigma_{\mathrm{eval}}$} & With STP (\%) & Without STP (\%) & Difference [95\% CI] & $d_z$ \\ \midrule Same $\alpha_r$ & 0 & \textbf{85.6}$\,\pm\,$20.7& 63.7$\,\pm\,$29.6& 22.0 [15.7, 27.9] & 0.69\\ Same $\alpha_r$ & 0.001& \textbf{81.9}$\,\pm\,$20.6& 43.7$\,\pm\,$26.6& 38.2 [31.6, 44.2] & 1.14\\ Selected $\alpha_r$ & 0 & \textbf{89.5}$\,\pm\,$13.2& 76.2$\,\pm\,$22.2& 13.3 [8.8, 17.9] & 0.58\\ Selected $\alpha_r$ & 0.001& \textbf{86.3}$\,\pm\,$15.6& 71.6$\,\pm\,$21.9& 14.8 [9.6, 20.1] & 0.56\\ \bottomrule \end{tabular}} \end{table}

\begingroup\begin{figure}[htbp]\centering\includegraphics{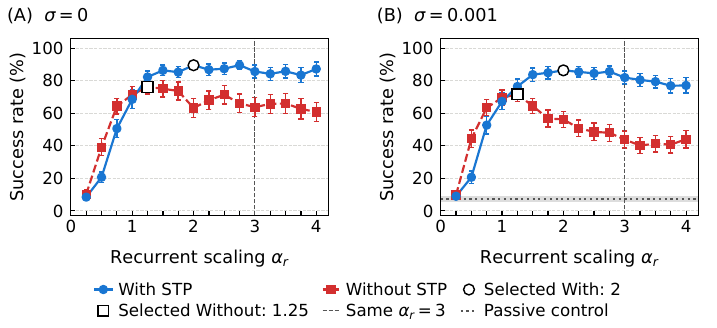}\caption{\textbf{Success rates remained higher in the With-STP condition than in the Without-STP condition when recurrent scaling was selected separately for each condition.} Mean post-training success rate as a function of the recurrent scaling $\alpha_r$ (16 grid values, 0.25--4.0; $n{=}100$ seeds per value and condition; error bars indicate the $t$-based 95\% confidence interval for the mean across seeds) at $\tau{=}1{,}000$\,ms. \textbf{(A)} $\sigma{=}0$ during training and evaluation. \textbf{(B)} $\sigma{=}0.001$. Blue solid lines with circles: With STP; red dashed lines with squares: Without STP. Large open markers with black edges indicate the values selected at $\sigma{=}0.001$ and held fixed in both panels ($\alpha_r{=}2$ With STP, $\alpha_r{=}1.25$ Without STP), and the vertical dashed line indicates the value used in the comparison at the same recurrent scaling ($\alpha_r{=}3.0$). The gray dotted line and shaded band in (B) show the mean and $t$-based 95\% confidence interval of the passive control (Without STP with $\alpha_r{=}0.01$ and without Q-value feedback).}\label{fig:04}\end{figure} \endgroup 

\begin{figure}[p]\ifdefined\OnepageFigureBox\else\newbox\OnepageFigureBox\fi\setbox\OnepageFigureBox=\vbox{\hsize=\textwidth\centering\includegraphics{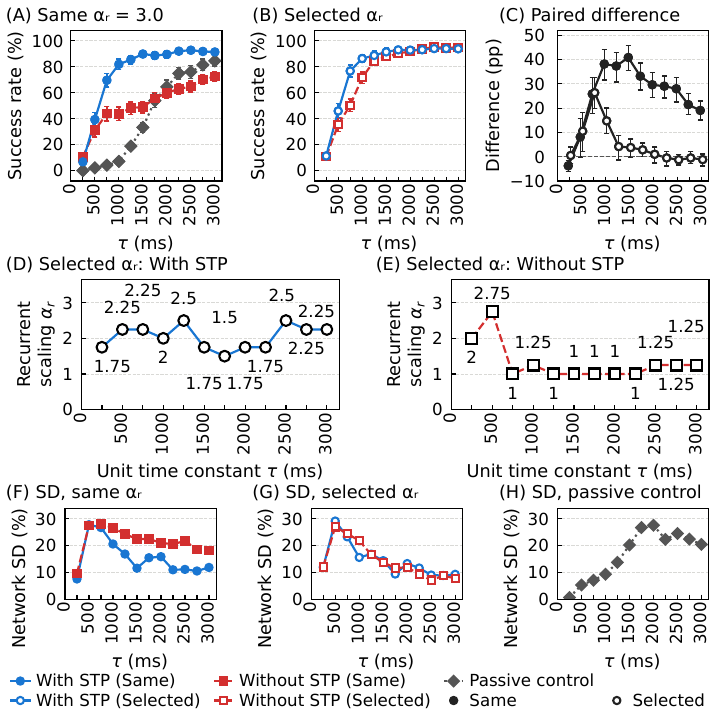}\caption{\textbf{Dependence of the STP effect on the unit time constant.} Post-training success rate at 12 unit time constants ($\tau{=}250$--3{,}000\,ms; $\sigma{=}0.001$; $n{=}100$ seeds per point; mean and $t$-based 95\% confidence interval across seeds). \textbf{(A)} The same recurrent scaling in both conditions ($\alpha_r{=}3.0$): With STP (blue), Without STP (red), and the passive control (gray; Without STP with $\alpha_r{=}0.01$ and without Q-value feedback). \textbf{(B)} $\alpha_r$ selected for each condition at each $\tau$ from the same grid of 16 values (the selected values are shown in panels \textbf{(D)} and \textbf{(E)}). \textbf{(C)} Paired difference (With STP minus Without STP, in percentage points, pp) and its 95\% BCa bootstrap confidence interval (2{,}000 resamples) for the comparison in (A) (filled markers) and in (B) (open markers); the horizontal line indicates zero. The Same and Selected series are displaced by $-25$ and $+25$\,ms for display. The main condition is $\tau{=}1{,}000$\,ms. With the same recurrent scaling, the 95\% CI of the paired difference lay above zero from 750 to 3{,}000\,ms; with the recurrent scaling selected for each condition, from 500 to 1{,}000\,ms. \textbf{(D, E)} Recurrent scaling $\alpha_r$ selected at each $\tau$ for With STP (D) and Without STP (E; separate axis) from the same grid of 16 values by the mean success rate at $\sigma{=}0.001$; the selected value is printed beside each point. \textbf{(F)--(H)} Standard deviation of the success rate across the $n{=}100$ networks, in percentage units: (F) at the same recurrent scaling ($\alpha_r{=}3.0$) and (G) with $\alpha_r$ selected for each condition at each $\tau$, each for With STP (blue circles, solid line) and Without STP (red squares, dashed line); (H) for the passive control (Without STP with $\alpha_r{=}0.01$ and without Q-value feedback; gray diamonds, dotted line).}\label{fig:05}}\ifdim\dimexpr\ht\OnepageFigureBox+\dp\OnepageFigureBox\relax>\textheight\fi\box\OnepageFigureBox\end{figure}

\paragraph{Dependence on the unit time constant}  Both conditions were trained and evaluated at 12 values of $\tau$ from 250 to 3{,}000\,ms, with the same state noise ($\sigma{=}0.001$ per 10-ms integration step) at every $\tau$ (\S\ref{sec:12}; Fig.~\ref{fig:05}; Supplementary Fig.~\ref{fig:20}). We also trained the passive control with recurrent scaling $\alpha_r{=}0.01$ and no Q-value feedback at each unit time constant (\S\ref{sec:12}).\par \par With the same recurrent scaling in both conditions ($\alpha_r{=}3.0$; Fig.~\ref{fig:05}A), success rates were higher in the With-STP condition than in the Without-STP condition at time constants from 500 to 3{,}000\,ms (paired differences between 8.1 and 40.9 percentage points; the 95\% CI of the difference lay above zero at the 10 time constants from 750 to 3{,}000\,ms). At $\tau{=}250$\,ms, success was lower with STP at the same recurrent scaling (Fig.~\ref{fig:05}A, C).\par When $\alpha_r$ was selected separately for each condition at each $\tau$ (Fig.~\ref{fig:05}B), success rates were higher in the With-STP condition at $\tau{=}500$\,ms (45.9\% vs 35.4\%; difference 10.5 percentage points, 95\% CI [2.1, 18.2]). Success rates were also higher with STP at $\tau{=}750$\,ms (76.7\% vs 50.3\%; difference 26.4 percentage points, 95\% CI [19.9, 32.6]). The difference at the main time constant of 1{,}000\,ms was 14.8 percentage points (95\% CI [9.6, 20.1]; see above). At these time constants the passive control solved at most 7.1\% of the trials, so slow unit integration alone did not solve the task. At $\tau\geq 1{,}250$\,ms, both conditions reached success rates of 84.5\% or higher (Fig.~\ref{fig:05}B). At $\tau{=}250$\,ms, success rates were near 10\% in both conditions (11.3\% and 10.8\%).\par \relax

\begin{figure}[t]\ifdefined\OnepageFigureBox\else\newbox\OnepageFigureBox\fi\setbox\OnepageFigureBox=\vbox{\hsize=\textwidth\centering\includegraphics{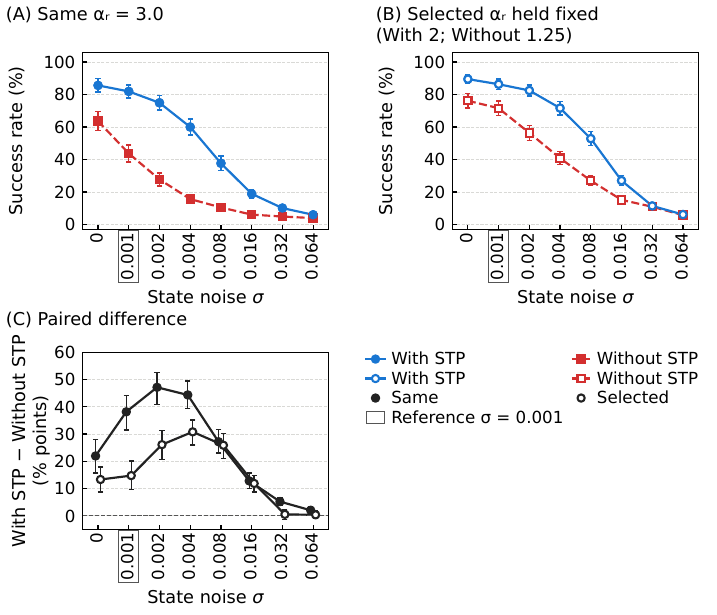}\caption{\textbf{Dependence of task success and the STP effect on state-noise intensity.} Post-training success rate at 8 noise levels ($\sigma{=}0$ and $0.001$--0.064 in factors of two, shown at equal spacing; the same $\sigma$ during training and evaluation; $\tau{=}1{,}000$\,ms; $n{=}100$ seeds per point; mean and $t$-based 95\% confidence interval across seeds). \textbf{(A)} Comparison at the same recurrent scaling ($\alpha_r{=}3.0$ in both conditions): With STP (blue solid line, circles) and Without STP (red dashed line, squares). Success rate declined with increasing state noise in both comparisons, and the paired difference was positive, with its 95\% BCa CI above zero, from $\sigma{=}0$ to $0.016$ in both. The values at $\sigma{=}0$ and $\sigma{=}0.001$ are given in Table~\ref{tab:04}. \textbf{(B)} Comparison with the recurrent scaling selected for each condition, with $\alpha_r$ fixed at the values selected at $\sigma{=}0.001$ ($\alpha_r{=}2$ With STP, $\alpha_r{=}1.25$ Without STP). \textbf{(C)} Paired difference (With STP minus Without STP) and its 95\% BCa bootstrap confidence interval (2{,}000 resamples) for the comparisons in (A) (filled markers) and (B) (open markers, Selected); the horizontal line indicates zero, and the boxed noise label marks the reference level $\sigma{=}0.001$.}\label{fig:06}}\ifdim\dimexpr\ht\OnepageFigureBox+\dp\OnepageFigureBox\relax>\textheight\fi\box\OnepageFigureBox\end{figure}

 \paragraph{Dependence on the noise level} To examine the range of noise over which these differences hold, we trained and evaluated both conditions at 8 state-noise levels from $\sigma{=}0$ to $0.064$, using the same $\sigma$ during training and evaluation (\S\ref{sec:10}; Fig.~\ref{fig:06}). In the comparison at the same recurrent scaling ($\alpha_r{=}3.0$ in both conditions), the paired difference was largest at $\sigma{=}0.002$ (74.9\% vs 27.8\%; 47.1 percentage points, 95\% CI [40.8, 52.5]) and remained large at $\sigma{=}0.004$ (60.0\% vs 15.7\%; 44.3 percentage points) and at $\sigma{=}0.008$ (27.2 percentage points). At the highest level, $\sigma{=}0.064$, the difference was 2.0 percentage points (95\% CI [0.8, 3.3]); the 95\% CI of the difference lay above zero at all 8 levels. With $\alpha_r$ fixed at the values selected at $\sigma{=}0.001$ ($2$ With STP and $1.25$ Without STP), success was higher with STP at each of the six noise levels from $\sigma{=}0$ to $0.016$. The paired differences ranged from 11.9 to 30.8 percentage points, with all corresponding 95\% confidence intervals above zero (Fig.~\ref{fig:06}C). At $\sigma{=}0.032$ and $0.064$, success rates were below 12\% in both conditions (11.5\% and 10.9\% at $\sigma{=}0.032$). The values obtained when $\alpha_r$ was re-selected for each condition at each noise level are shown in Supplementary Fig.~\ref{fig:34}.

We also varied the training and evaluation noise independently (4 training levels $\times$ 7 evaluation levels) and applied two other forms of noise, multiplicative noise and input noise (\S\ref{sec:10}). Across the independently varied combinations of training and evaluation state noise, success rates were higher in the With-STP condition than in the Without-STP condition in the comparison at the same recurrent scaling. The difference (With STP minus Without STP) ranged from 7.5 to 49.0 percentage points, and its 95\% CI lay above zero in all 28 combinations (Supplementary Fig.~\ref{fig:19}). Under multiplicative noise with $s{=}0.05$ and $0.25$ and under input noise with $\sigma_{\mathrm{in}}{=}0.1$, success rates were also higher with STP at the same recurrent scaling (34.9, 23.6, and 15.0 percentage points, respectively; Supplementary Fig.~\ref{fig:25}). At the higher tested input-noise level ($\sigma_{\mathrm{in}}{=}0.5$), success rates were low in both conditions and the confidence interval of the paired difference included zero.

\paragraph{Controls for alternative explanations}
\begingroup\setlength{\emergencystretch}{3em}We next asked which property of the STP-derived modulation carries the With-STP advantage. Separately trained controls compared fixed scaling, prescribed time courses, simpler activity-dependent rules, and steady-state STP. Separate checks varied learning-rate schedules, Q-value feedback and reward values.\par\endgroup \textit{Fixed scaling, learning, and feedback controls.} At $\tau{=}1{,}000$\,ms and $\sigma{=}0.001$, With STP exceeded both gain-matched controls (Supplementary Fig.~\ref{fig:16}); the advantage over the Gain-matched (constant component preserved) control was 22.0 percentage points (95\% BCa CI [15.2, 28.6]). In the primary comparisons, the learning rate was reduced after 633 successful training episodes; this threshold was reached by 90 of 100 With-STP and 32 of 100 Without-STP networks at the same recurrent scaling, and by 93 and 72 networks at the selected scalings. The success-rate advantage also persisted across the fixed learning-rate schedules (paired differences of 27.5--36.9 percentage points at the same recurrent scaling and 13.2--19.4 percentage points at the selected scalings, all with 95\% BCa confidence intervals above zero; $\sigma{=}0.001$; Supplementary Fig.~\ref{fig:39}). At $\tau{=}1{,}000$\,ms, $\alpha_r{=}3.0$ and $\sigma{=}0.001$, With STP retained higher success after evaluation-time feedback removal (Supplementary Fig.~\ref{fig:23:parta}A, B) and when both conditions were trained without Q-value feedback (Supplementary Figs.~\ref{fig:24:parta}B and~\ref{fig:24:partb}C). With feedback removed from trial onset, With STP also retained higher late goal-decoding accuracy and GO3 action-value differences (Supplementary Fig.~\ref{fig:23:partb}D, E). Under state noise at $\alpha_r{=}3.0$, the With-STP success-rate advantage also persisted across the tested feedback strengths (Supplementary section ``Q-value feedback strength''; Supplementary Fig.~\ref{fig:30}). \par \textit{Prescribed recurrent modulation.} At $\alpha_r{=}3.0$, Common prescribed scaling raised Without-STP success from 43.7\% to 71.7\%, and With STP exceeded it by 10.2 percentage points (95\% BCa CI [4.2, 16.1]; Supplementary Fig.~\ref{fig:24:partc}F--H, which also shows Unit-wise prescribed scaling). Thus, under state noise, the With-STP advantage persisted without Q-value feedback and was only partly reproduced by Common prescribed scaling. \par \textit{Activity-dependent modulation rules.} We next compared simpler activity-dependent scaling and STP driven by another unit, with the driver assignment fixed within each network. At the same recurrent scaling ($\alpha_r{=}3.0$) under state noise, the three controls (Population-mean scaling, Own-unit scaling and Other-unit STP) reached 16.3--33.2\% success, compared with 81.9\% with STP and 43.7\% without STP. First-move accuracy and the action-value difference at GO2--GO3 were also higher with STP than with each control (Supplementary Figs.~\ref{fig:29:partb}C, D and~\ref{fig:29:partc}E, F). We selected $\alpha_r$ for each control from the same 16 values, keeping the modulation coefficient and the prescribed time courses obtained at $\alpha_r{=}3.0$ unchanged. The three simpler activity-dependent rules and the two controls with prescribed time-varying scaling reached success rates of 71.3--79.9\%. With STP exceeded each by 6.4--15.0 percentage points (all 95\% BCa confidence intervals above zero; Supplementary Figs.~\ref{fig:33:partb}D, E and~\ref{fig:37:partc}D). At $\alpha_r{=}3.0$, the With-STP success-rate advantage also persisted across changes in the modulation coefficient and relaxation time of the single-variable rules (Supplementary section ``Sensitivity of the single-variable rules''; Supplementary Fig.~\ref{fig:32}), where each variant is compared with its original setting and with STP.

\textit{Acute perturbations of trained networks.} In trained With-STP networks, resetting STP at cue offset or freezing the STP-derived scaling at a goal-independent reference reduced success (Supplementary Fig.~\ref{fig:17}). \par \begingroup\textit{Replay in trained networks.} We tested how assigning the same replayed STP scaling trajectories to different units affected the first action. Replaying mean scaling trajectories reduced the success rate from 82.0\% in a separate intact re-evaluation to 10.1--19.6\% across the replay conditions, so assignments were compared within replay conditions. We examined the first move from the common start position, before actions altered the position input. The cue-matched assignment gave higher first-move accuracy than assignments based on another goal's ranks or randomized permutations (Supplementary Fig.~\ref{fig:26}). Thus, within replay, the unit assignment of STP modulation affected goal-directed action selection.\endgroup \par \begingroup\textit{Steady-state STP control.}  A separately trained steady-state STP control exceeded Without STP at the same recurrent scaling and with the recurrent scaling selected for each condition (Supplementary section ``Steady-state STP modulation''; Supplementary Figs.~\ref{fig:33:parta}A--C, \ref{fig:37:partb}B, C and~\ref{fig:38}). At the same recurrent scaling, steady-state STP reached 83.9\% success, compared with 81.9\% with dynamic STP (the With-STP condition) and 43.7\% without STP (dynamic minus steady-state STP, \ensuremath{-}2.0 percentage points, 95\% BCa CI [\ensuremath{-}6.0, 1.9]; with the recurrent scaling selected for each condition, \ensuremath{-}1.4 percentage points, 95\% BCa CI [\ensuremath{-}4.9, 1.9]). At $\tau{=}500$ and 750\,ms, steady-state STP also gave success similar to that of dynamic STP and higher than without STP, at the same recurrent scaling and with the recurrent scaling selected for each condition (Supplementary Figs.~\ref{fig:33:parta}B, C and~\ref{fig:37:partb}B, C). At $\tau{=}1{,}000$\,ms, $\alpha_r{=}3.0$ and $\sigma{=}0.001$, first-move accuracy (\num[round-mode=places,round-precision=1]{99.37}\% versus \num[round-mode=places,round-precision=1]{99.46}\%) and the GO2--GO3 action-value difference (0.04715 versus 0.04579) were similar for steady-state and dynamic STP (Supplementary Fig.~\ref{fig:38}).\endgroup 

\par To assess sensitivity to reward values, we varied the magnitudes and relative weighting of the progress and nonprogress rewards. We compared all nine reward pairs, which include halving and doubling both rewards, at $\alpha_r{=}3.0$ and at the recurrent scalings selected for each condition under the reference rewards. Success was higher with STP in every comparison, with all 95\% BCa confidence intervals for the paired differences above zero (Supplementary Figs.~\ref{fig:28:parta}B and~\ref{fig:28:partb}E). The action-value difference $D_Q$ averaged over GO2--GO3 (\S\ref{sec:15}) was also higher with STP for every reward pair in both comparisons (all 95\% BCa confidence intervals above zero; Supplementary Figs.~\ref{fig:28:parta}C and~\ref{fig:28:partb}F).

\begingroup\setlength{\emergencystretch}{2em}With recurrent scaling selected for each condition, the steady-state STP control retained the success-rate advantage over Without STP. The controls with prescribed time-varying scaling and the simpler activity-dependent rules remained below With STP (Table~\ref{tab:05}A). For the gain-matched controls, see Supplementary Fig.~\ref{fig:16}.\par\endgroup \begin{table}[t] \centering \caption{\textbf{Control performance and goal use with the recurrent scaling selected for each condition.}} \label{tab:05} \begingroup \small \setlength{\tabcolsep}{3pt} \begin{tabular}{@{}p{.34\linewidth}p{.12\linewidth}p{.14\linewidth}p{.32\linewidth}@{}} \toprule \multicolumn{4}{@{}l}{\textbf{(A) Separately trained conditions}}\\ Condition & Selected $\alpha_r$ & Success (\%) & With STP~$-$ condition (pp) [95\% BCa CI]\\ \midrule With STP & 2.00 & 86.33 & Reference\\ Without STP & 1.25 & 71.58 & 14.75 [9.58, 20.08]\\ Common prescribed scaling & 2.00 & 75.45 & 10.88 [5.80, 16.81]\\ Unit-wise prescribed scaling & 1.75 & 78.51 & 7.82 [3.90, 12.49]\\ Population-mean scaling & 1.25 & 72.38 & 13.95 [9.20, 18.93]\\ Own-unit scaling & 0.75 & 71.29 & 15.04 [9.38, 20.68]\\ Other-unit STP & 0.50 & 79.89 & 6.44 [1.59, 11.23]\\ Steady-state STP & 1.75 & 87.70 & \ensuremath{-}1.37 [\ensuremath{-}4.93, 1.93]\\ \bottomrule \end{tabular} \par\medskip \begin{tabular}{@{}p{.34\linewidth}p{.12\linewidth}p{.14\linewidth}p{.32\linewidth}@{}} \toprule \multicolumn{4}{@{}l}{\textbf{(B) With STP versus Without STP}}\\ Measure & With STP & Without STP & Paired difference [95\% BCa CI]\\ \midrule Late goal decoding, 8--9 s (\%) & 99.57 & 98.74 & 0.83 [0.50, 1.23] pp\\ Action-value difference $D_Q$ at GO2 & 0.0636 & 0.0599 & 0.0037 [0.0003, 0.0074]\\ Action-value difference $D_Q$ at GO3 & 0.0263 & 0.0140 & 0.0123 [0.0079, 0.0167]\\ \bottomrule \end{tabular} \par\smallskip \begin{minipage}{\linewidth} \small Steady-state STP exceeded Without STP by 16.12 percentage points (95\% BCa CI [11.29, 21.01]; Supplementary Fig.~\ref{fig:37:partc}E). All conditions used $\tau{=}1{,}000$\,ms, $\sigma{=}0.001$ during training and evaluation, 1{,}900 training episodes and 100 evaluation episodes per network, with Q-value feedback. STP settings were $U_{\mathrm{SE}}{=}0.1$, $\tau_{\mathrm{R}}{=}100$\,ms and $\tau_{\mathrm{F}}{=}500$\,ms. Results summarize 100 paired networks. Each condition's $\alpha_r$ was selected from the same 16-point grid using mean success on these same 100 networks; the intervals are conditional on the selected values. The prescribed time courses and simpler-rule calibration remained fixed at their $\alpha_r{=}3.0$ reference settings. Intervals are 95\% BCa bootstrap confidence intervals from 2{,}000 paired-seed resamples. In (A), differences are With STP minus each comparison condition; in (B), the selected scalings are $\alpha_r{=}2.00$ with STP and $1.25$ without STP. Decoding is the window average of time-resolved accuracies. $D_Q$ and its differences are in raw Q-value units; success and decoding differences are in percentage points (pp). Values are from Supplementary Figs.~\ref{fig:37:partc}D, E and~\ref{fig:35}A, B. Split-seed validation for With STP, Without STP and steady-state STP is reported separately in Supplementary Figs.~\ref{fig:37:parta}A and~\ref{fig:37:partb}B, C. \end{minipage} \endgroup \end{table}

 \subsection{Experiment 2: behavior and organization of state space}\label{sec:23}

For RQ2, we asked whether goal identity remained decodable from reservoir activity through the later GO windows. We also asked whether the action-value readout favored goal-consistent moves. The analyses in this subsection and in \S\ref{sec:24} use the networks of the comparison at the same recurrent scaling ($\tau{=}1{,}000$\,ms, $\alpha_r{=}3.0$ in both conditions, $\sigma{=}0.001$). Supplementary Figure~\ref{fig:35} compares the internal measures reported below with those obtained when the recurrent scaling was selected separately for each condition. We visualized reservoir trajectories with PCA and measured goal-decoding accuracy. The task was designed so that each goal could be reached from the central start position in three actions. Inspection of evaluation trials showed that successful trials reached the goal by the end of the third GO window, except for 5 trials in the With-STP condition and 39 trials in the Without-STP condition that reached it at the end of the fourth GO window ($t{=}10$--$11$\,s; of the $100{\times}100$ evaluation trials per condition); all other trials extending to the fourth GO window were failed trials (Supplementary Fig.~\ref{fig:36}A). Accordingly, the time-resolved analyses below focus mainly on the interval from $t{=}0$ to $9$\,s.

\begin{figure}[htbp]\centering \setlength{\abovecaptionskip}{3pt}\setlength{\belowcaptionskip}{0pt} \includegraphics{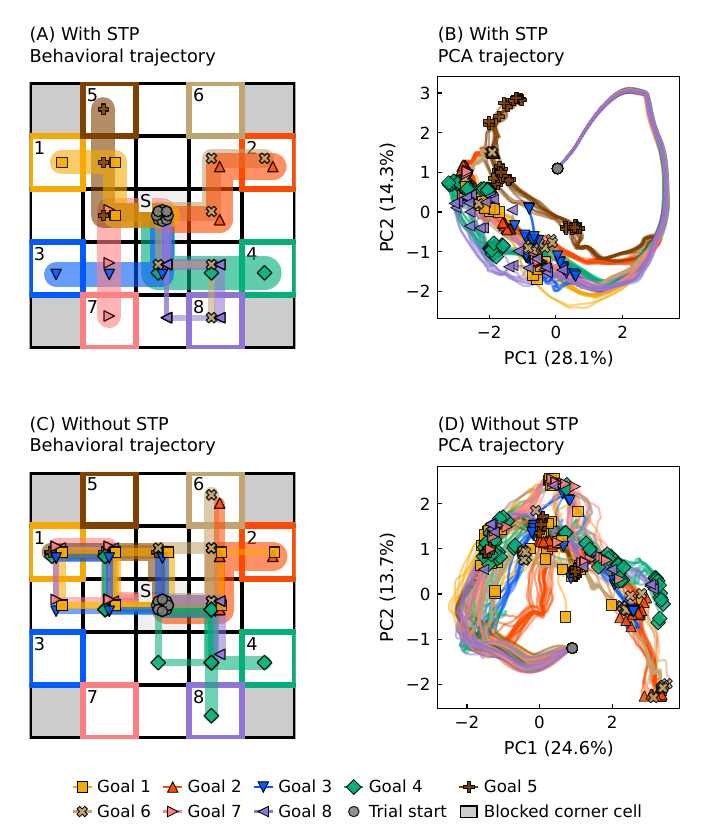} \caption{\textbf{Behavioral and PCA trajectories from a representative seed.} \textbf{(A, B)} With STP. \textbf{(C, D)} Without STP. Behavioral panels (A, C) show reconstructed cursor trajectories from all evaluation trials of the displayed seed on the $5{\times}5$ grid; gray cells indicate blocked corner cells, the start cell is labeled S, and the eight non-corner perimeter goals are numbered 1--8 with cell-color coding. The displayed labels 1--8 correspond to goal indices $g{=}0$--$7$ in \S\ref{sec:07}. Line thickness in the behavioral panels indicates how often each complete displayed path (the start and the first three actions) was observed across evaluation trials. PCA panels (B, D) show normalized reservoir-activity trajectories over $t=0$--$9$\,s projected onto independent 2D PCA axes per condition; gray markers indicate trial start ($t=0$\,s), and goal-colored markers indicate the samples at the action-execution times $t=5$ and $7$\,s and the final sample at $t=8.99$\,s; axis labels give the explained variance of each component. The displayed seed was selected as a representative qualitative example with evaluation performance close to the group medians; statistical inference is based on seed-wise quantitative analyses across $n=100$ networks.} \label{fig:07} \end{figure}

\paragraph{PCA trajectories and time-resolved decoding}
In the PCA visualization, trajectories in the With-STP condition tended to occupy distinct regions for different goals. By contrast, trajectories in the Without-STP condition were diffuse and irregular, and such goal-dependent separation was not clear. Representative examples of both conditions are shown in Fig.~\ref{fig:07}. Note that PCA here is used as a qualitative visualization computed independently for each seed; quantitative evaluation of condition differences is based on time-resolved decoding and scatter ratio. \par 

\begin{figure}[tp]\centering\includegraphics{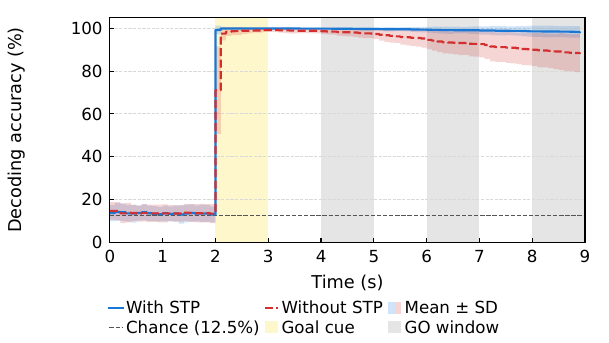}\caption{\textbf{Time course of goal-decoding accuracy.} Eight-class linear decoding of goal identity from normalized reservoir activities was evaluated independently at each sampled time point using five-fold cross-validation (\S\ref{sec:14}). Lines show the mean accuracy across $n{=}100$ networks per condition (blue solid, With STP; red dashed, Without STP); shaded bands show the standard deviation across networks. Accuracy was evaluated every 100\,ms and is drawn as steps. The horizontal dashed line marks chance accuracy (12.5\%). The time axis spans trial onset to $t{=}9$\,s, when the third move is executed and a goal can first be reached. Yellow band: goal cue (2--3\,s); gray bands: GO windows (4--5, 6--7, 8--9\,s). Both conditions rose above chance after the goal cue, and accuracy with STP remained higher during the later GO windows. All networks used $\alpha_r{=}3.0$, $\tau{=}1{,}000$\,ms and $\sigma{=}0.001$.}\label{fig:08}\end{figure}

\paragraph{Time-resolved decoding}
Figure~\ref{fig:08} shows the time course of goal-decoding accuracy (RQ2). Supplementary Figures~\ref{fig:35}A and~\ref{fig:40:parta}B, C report window averages of time-resolved decoding, whereas the delay-period result below decodes activity averaged within the window. Before the goal cue was presented (0--2\,s), both conditions remained near chance level. After cue presentation, decoding accuracy in both conditions rapidly rose to a high level, but the With-STP condition maintained near-ceiling accuracy from the delay period through the later part of the trial. In contrast, accuracy in the Without-STP condition gradually declined from the latter half of the delay into the GO windows, dropping to $89.1 \pm 8.3\%$ in the third GO window ($t{=}8$--$9$\,s), whereas the With-STP condition maintained $98.5 \pm 2.6\%$ (paired $t$-test, $t(99){=}10.8$, $p < 0.001$, $d_z{=}1.08$). For the delay period ($t{=}3$--$4$\,s), the eight-class linear decoding accuracy based on time-averaged normalized activities was $99.9 \pm 0.4\%$ with STP and $99.2 \pm 1.0\%$ without STP. Thus, in both conditions goal identity was highly decodable during the delay period, but later goal-decoding accuracy was higher with STP at the same recurrent scaling. When the recurrent scaling was selected separately for each condition, third-GO-window accuracy was near ceiling in both conditions (99.6\% with STP and 98.7\% without STP), yet success rate and $D_Q$ at GO3 remained higher with STP (Table~\ref{tab:05}B; Supplementary Fig.~\ref{fig:35}A, B; Table~\ref{tab:04}).\par 

\paragraph{Interpretation of the decoding results and their scope}
Between goal-cue offset and the first GO window, the position input was fixed and the goal and GO channels were zero, so goal-dependent activity in this interval reflected internal retention. The summary windows include boundary samples, as specified in \S\ref{sec:20}. In the later part of the trial, by contrast, position input can covary with goal condition, so separation in that interval is more appropriately interpreted as the maintenance of dynamics conditioned jointly on goal and task state.

\begin{figure}[htbp]\centering\includegraphics{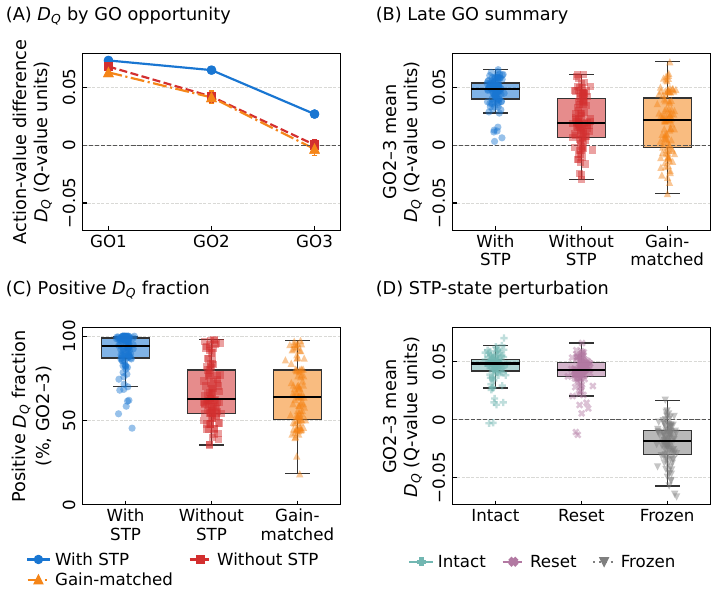}\caption{\textbf{Action-value difference at GO opportunities.} For each evaluable GO decision, $D_Q$ was defined as the maximum Q value among target-consistent valid actions (valid moves that decrease the squared Euclidean distance to the goal, matching the reward shaping in the environment) minus the maximum Q value among all other action indices (Eq.~\eqref{eq:06}); the competitor set therefore includes invalid moves (off-grid or blocked-corner) because the policy uses an all-action argmax. Positive values indicate that the readout assigned a higher value to a target-consistent action than to every other action index. $D_Q$ is used as an operational readout-level index and is not the reinforcement-learning advantage function. All panels use $\tau{=}1{,}000$\,ms and $\sigma{=}0.001$; $\alpha_r{=}3.0$ except for the gain-matched control ($\alpha_r{\approx}4.65$; \S\ref{sec:19}). \textbf{(A)} $D_Q$ at the first three GO action-execution times. Markers and error bars indicate the mean and pointwise normal-approximation 95\% confidence interval across $n{=}100$ paired seeds. With STP is shown in blue (solid line, circles), Without STP in red (dashed line, squares), and the gain-matched Without-STP control in orange (dash-dot line, triangles). \textbf{(B)} Seed-wise summary of $D_Q$ averaged over GO2--GO3. Boxes show the median and interquartile range; jittered points are individual seeds. \textbf{(C)} Fraction of evaluable GO decisions with $D_Q{>}0$, averaged over GO2--GO3. GO2--GO3 summaries give equal weight to the two GO-specific network means or fractions (\S\ref{sec:15}). \textbf{(D)} $D_Q$ under the post-training STP-state perturbation control (intact, STP reset at cue offset, and frozen-STP; teal, purple, and gray). The intact condition is a re-evaluation of the trained networks and is distinct from the original With-STP evaluation in (B). In (B)--(D), boxes show the median and interquartile range, whiskers extend to the most extreme values within 1.5 interquartile ranges of the box, and jittered points show individual seeds ($n{=}100$ per condition); the dashed zero line in (D) marks equal values of the target-consistent and competing maxima.}\label{fig:09}\end{figure}

 \paragraph{Action-value difference at GO opportunities}
To test whether the maintained goal-conditioned structure was reflected in the action-value readout, we computed the action-value difference $D_Q$ defined in \S\ref{sec:15} at GO opportunities (Fig.~\ref{fig:09}). Across the first three GO opportunities, the With-STP condition showed larger positive differences than the standard Without-STP baseline and the gain-matched Without-STP control. Averaged over GO2--GO3, the seed-wise mean $D_Q$ was $0.046$ in the With-STP condition versus $0.021$ in the Without-STP condition (paired $t$-test, $d_z{=}1.07$, $p < 0.001$) and $0.019$ in the gain-matched Without-STP condition (paired $t$-test, $d_z{=}1.05$, $p < 0.001$). The fraction of evaluable GO decisions with $D_Q{>}0$ was correspondingly larger in the With-STP condition ($90.5\%$ over GO2--GO3) than in the Without-STP ($66.5\%$) and gain-matched Without-STP ($65.3\%$) conditions. Post-training perturbations showed a smaller reduction under cue-offset STP reset and a clear reduction under frozen, goal-independent STP scaling (frozen mean $D_Q = \ensuremath{-}0.020$ over GO2--GO3, paired $d_z{=}3.16$, $p < 0.001$). Because the policy selects the action with the largest value among all action indices, $D_Q$ summarizes the readout's preference for progress-making actions over all other action indices, including invalid ones.\par\paragraph{Trial outcomes and errors} To characterize failed trials, we counted the outcome of every evaluation trial and the action selected at each GO opportunity (Supplementary Fig.~\ref{fig:36}A, B). At GO3, a cross-validated linear decoder (\S\ref{sec:14}) identified the goal correctly while the trained action readout selected an erroneous move (a distance-increasing move or an invalid move that left the cursor in place) in 14.6\% of decisions with STP and 38.3\% without STP (means of the seed-wise percentages of all analyzed third-GO decisions across 100 seeds).  We quantified next-move progress toward the goal within the errors generated by each trained model. For each seed, the proportion was the number of erroneous moves at GO1--GO3 followed by a valid move decreasing the squared Euclidean distance to the goal at the next GO opportunity, divided by the number of erroneous moves at GO1--GO3. The next move was directed toward the goal in 61.3\% of cases with STP and 47.6\% without STP (means of the seed-wise proportions across the 88 seeds with at least one erroneous move in both conditions; paired difference 13.7 percentage points, 95\% CI [5.5, 21.5]; Supplementary Fig.~\ref{fig:36}C). Every goal is three moves from the start, so the goal can be reached at the fourth opportunity after an invalid move that leaves the cursor in place. To compare the two conditions on errors made at the same GO opportunity or of the same type, we stratified the measure by GO opportunity and by the type of erroneous move, averaging in each stratum over the seeds with at least one such error in both conditions (Supplementary Fig.~\ref{fig:36}C lists every stratum). Errors at GO3 and distance-increasing moves each made up most errors in both conditions, and within each of these strata the next move was directed toward the goal more often with STP (errors at GO3: paired difference 11.5 percentage points, 95\% CI [2.8, 20.2], 87 seeds; distance-increasing moves: 14.8 percentage points, 95\% CI [6.6, 23.1], 87 seeds). Within the errors generated by each trained condition, next-move progress toward the goal was higher with STP.

\paragraph{Analysis restricted to successful trials}
\begingroup\LayoutKeepFourLines To characterize goal-conditioned activity within successful behavior, we recalculated the scatter ratio and decoding accuracy using only successful trials for each seed in each condition, with decoding based on window-averaged activity (75 paired seeds for scatter ratio and 73 for decoding, following the inclusion criteria in \S\ref{sec:09}). In the later interval ($t{=}8$--$9$\,s), the With-STP condition remained significantly higher in both scatter ratio ($80.7 \pm 181.5$ vs $8.9 \pm 14.3$, $d_z{=}0.40$, $p < 0.001$; median $=26.6$, IQR $=12.3$--$65.9$ vs median $=5.34$, IQR $=3.20$--$8.80$; positive in 69 of 75 pairs, Wilcoxon signed-rank test $p < 0.001$) and decoding accuracy ($99.6 \pm 1.0\%$ vs $96.7 \pm 5.0\%$, $d_z{=}0.57$, $p < 0.001$). For the delay period ($t{=}3$--$4$\,s), both measures were also higher with STP under the same restriction to successful trials (scatter ratio: $d_z{=}0.35$, paired difference 156.2, 95\% BCa CI [95.8, 333.8], paired $t$-test $p{=}0.0032$; Wilcoxon signed-rank test $p < 0.001$, positive in 70 of 75 pairs; decoding accuracy: $d_z{=}0.58$, paired difference 2.1 percentage points, 95\% BCa CI [1.4, 3.3], paired $t$-test $p{=}4.6{\times}10^{-6}$). Within successful trials selected in each condition from the eligible paired networks, the With-STP condition showed greater goal separability and higher decoding accuracy later in the trial (Supplementary Fig.~\ref{fig:14}). The same pattern remained after matching successful-trial counts per goal between conditions (Supplementary section \hyperref[sec:37]{``Successful-trial composition''}). As a control for minor goal-count imbalance in evaluation trials, we repeated the scatter-ratio and late-decoding analyses after balanced-goal resampling ($n_{\min}$ trials per goal, 50 resamples per seed). This analysis yielded the same qualitative conclusions, with paired With-STP versus Without-STP effect sizes closely matching those from the all-trial analysis (Supplementary Fig.~\ref{fig:15}).\par\endgroup

\subsection{Experiment 3: network dynamics and goal-conditioned dynamics}\label{sec:24}

\begin{figure}[tp]\centering\includegraphics{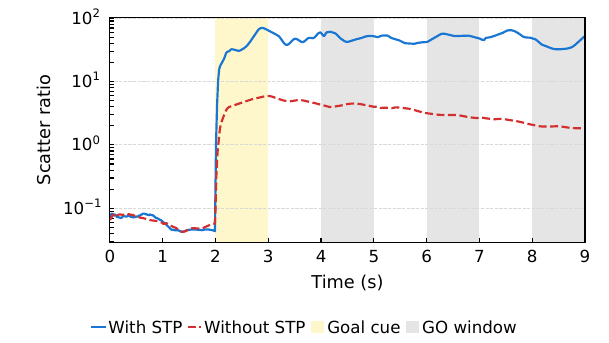}\caption{\textbf{Representative dynamics of goal separability (scatter ratio \(\SB/\SW\)).} Time series of the scatter ratio $\SR(t)$ with and without STP for one representative network, whose delay-period scatter ratio lay within the interquartile range in both conditions. The ratio, computed from all 100 evaluation trials in each condition, is dimensionless; the blue solid line denotes the With-STP condition and the red dashed line the Without-STP condition, and the vertical axis is logarithmic because the two conditions differ greatly in scale. With STP, the ratio increased sharply after goal-cue presentation and remained high during the delay period. Yellow band: goal cue (2--3\,s); gray bands: GO windows (4--5, 6--7, 8--9\,s). The group-level delay-period scatter-ratio difference remained after unit-norm normalization of the state vectors (see main text), indicating that it reflects goal-dependent differences in state-vector direction.}\label{fig:10}\end{figure}

\begin{figure}[htbp]\centering\includegraphics{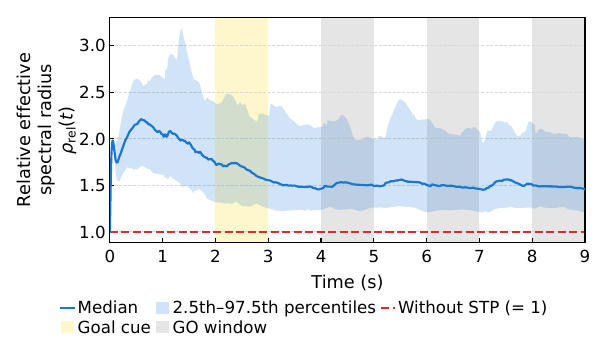}\caption{\textbf{Time series of the relative effective spectral radius.} $\rho_{\mathrm{rel}}(t)=\rho_{\mathrm{eff}}^{\mathrm{STP}}(t)/\rho_{\mathrm{eff}}^{\mathrm{Without\,STP}}$ for $n{=}100$ paired networks at the main parameter setting. The line shows the median across networks; the shaded band spans the 2.5th--97.5th percentiles across networks. The red dashed reference is $\rho_{\mathrm{rel}}{=}1$. The pre-cue increase reflects relaxation from the trial's initial state. Yellow shading indicates the goal cue (2--3~s), and gray shading indicates GO windows (4--5, 6--7, 8--9~s). This operational index describes STP-induced modulation of recurrent weights; goal specificity is assessed separately by $\Delta_{\mathrm{sim}}(t)$ (\S\ref{sec:17}).}\label{fig:11}\end{figure}

\begin{figure}[t]\centering\includegraphics{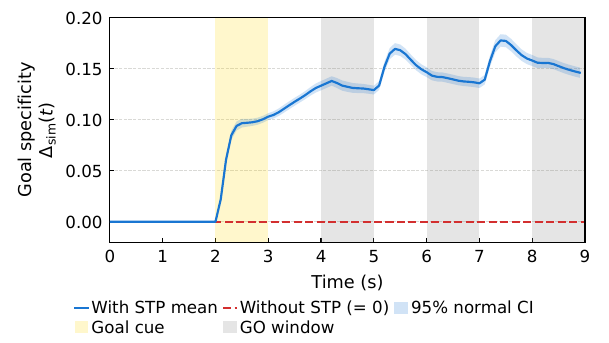}\caption{\textbf{Temporal evolution of goal specificity in effective connectivity.} Time-resolved trajectory of $\Delta_{\mathrm{sim}}(t)$ (cosine similarity for same-goal pairs minus that for different-goal pairs), shown for $n{=}100$ seeds with 95\% confidence intervals. These are pointwise normal-approximation intervals based on the standard error across seeds. Yellow band: goal cue (2--3\,s); gray bands: GO windows (4--5, 6--7, 8--9\,s). In the With-STP condition, $\Delta_{\mathrm{sim}}(t)$ rose rapidly after cue presentation, was maintained during the delay period, and increased further toward the later part of the trial. During the delay period this reflects goal-specific patterning; in the later interval, where goal identity covaries with position and task state, it should be read as goal- and task-state-conditioned patterning. In the Without-STP condition, because $W^{\mathrm{rec}}_{\mathrm{eff}}(t){=}2W^{\mathrm{rec}}$ is time-invariant, $\Delta_{\mathrm{sim}}(t){=}0$ (red dashed line).}\label{fig:12}\end{figure}

We next asked whether the behavioral and readout differences were accompanied by STP-dependent changes in goal-conditioned dynamics (RQ3). Figures~\ref{fig:10} and~\ref{fig:11} show state separation and global recurrent scaling, and Fig.~\ref{fig:12} shows goal-dependent efficacy patterns during the delay. See \S\ref{sec:17} for the operational indices.

\paragraph{Goal-separation dynamics}\mbox{}\\
\begingroup\setlength{\emergencystretch}{2em}In the With-STP condition, the scatter ratio $\SR(t){=}\SB/\SW$ computed within each seed increased after goal-cue presentation ($t{=}2$\,s) and remained high throughout the delay period. The predefined-window mean of $\SR(t)$ during the delay period ($t{=}3$--$4$\,s) was $145.9 \pm 418.2$ in the With-STP condition and $9.04 \pm 26.59$ in the Without-STP condition, a significant difference (paired $t$-test, $p < 0.001$, $d_z{=}0.35$, $n{=}100$ seeds). Because the seed-wise delay-period $\SR$ values were right-skewed, we also report the median and interquartile range for the same delay-period window: With STP, median $=62.8$, IQR $=27.1$--$149.2$; Without STP, median $=4.35$, IQR $=2.28$--$7.12$. The paired difference was positive in 98 of 100 seed pairs (Wilcoxon signed-rank test, $p < 0.001$).  These paired results indicate greater separation of goal-conditioned activity with STP during the delay. Time series with and without STP for one representative network, whose delay-period scatter ratio lay within the interquartile range in both conditions, are shown on a logarithmic axis in Fig.~\ref{fig:10}. After the goal cue, $\SR(t)$ in the With-STP condition remained elevated and varied across subsequent task epochs, whereas it remained lower in the Without-STP condition.  The scatter ratio is invariant to a common rescaling of all state vectors but not to trial-dependent differences in vector magnitude. To assess the contribution of such differences, we recalculated the ratio from the directions of the state vectors alone. After unit-norm normalization of each state vector, the delay-period scatter ratio remained higher in the With-STP condition than in the Without-STP condition ($145.7 \pm 416.6$ vs $9.02 \pm 26.61$, paired $t$-test, $p < 0.001$, $d_z{=}0.35$, $n{=}100$ seeds). The With-STP value remained close to the unnormalized value ($145.9$), and the Without-STP value was likewise nearly unchanged. Goal-dependent separation therefore remained in the directions of the state vectors and was not explained by trial-dependent response amplitudes. The scatter ratio summarizes between-goal separation relative to within-goal variability, whereas linear decoding evaluates whether a classifier can distinguish goals. Accordingly, during the delay the scatter ratio was higher with STP even though goal identity was highly decodable in both conditions.\par\endgroup

 \paragraph{Temporal evolution of the effective spectral radius}\mbox{}\\
In the main $(\tau_{\mathrm{R}}, \tau_{\mathrm{F}}){=}(100, 500)$\,ms setting, the relative effective spectral radius $\rho_{\mathrm{rel}}(t)=\rho_{\mathrm{eff}}^{\mathrm{STP}}(t)/\rho_{\mathrm{eff}}^{\mathrm{Without\,STP}}$ rose rapidly after the initial updates and remained above the fixed Without-STP baseline over the subsequent analyzed interval (Fig.~\ref{fig:11}). STP modulation affected the broader eigenvalue-magnitude distribution as well as its largest value (distributional summaries and normalized spectra in Supplementary Fig.~\ref{fig:41}).\par  In the Without-STP condition, $\rho_{\mathrm{eff}}(t)$ is constant across time, so $\rho_{\mathrm{rel}}(t){=}1$ by construction. Because the goal cue begins at $t{=}2$\,s, the early rise of $\rho_{\mathrm{rel}}(t)$ in the main setting should not be interpreted as goal-specific modulation or as a GO-synchronized preparatory peak. Rather, $\rho_{\mathrm{rel}}(t)$ serves as an operational index of the global STP-derived scaling of recurrent input. The early rise is a transient that follows from the initial conditions of each trial: the units start as described in \S\ref{sec:09}, with activity near the midpoint of the activation function after the first integration step, while the STP variables start from their resting values ($x_j{=}1$, $u_j{=}U_{\mathrm{SE}}$). The activity following the initial update raises $u_j$ and depletes $x_j$; because the depletion lags behind the facilitation, the product $x_j u_j$ first overshoots and then relaxes. With activity held at $0.5$ in all units, the scaling is identical across units and $\rho_{\mathrm{rel}}=x_j u_j/U_{\mathrm{SE}}$. Iterating Eqs.~\eqref{eq:03}--\eqref{eq:04} alone at this activity gives $x_j u_j/U_{\mathrm{SE}}\approx1.98$ at the fifth integration step. The observed $\rho_{\mathrm{rel}}(t)$ agreed with this calculation over the first eight steps (median across networks; maximum deviation 0.0045). $\rho_{\mathrm{rel}}(t)$ then reached a peak of 2.21 at $t{=}0.59$\,s and relaxed thereafter. When the same equations were driven by each unit's own trial-averaged activity, the predicted $\rho_{\mathrm{rel}}(t)$ reproduced this peak (2.21 at $t{=}0.59$\,s; maximum deviation 0.020 over 0--2\,s). By contrast, driving all units with the population-mean activity reproduced only the initial overshoot and gave 1.59 at the time of the later peak. This later peak therefore reflects differences in activity across units while the network relaxes from its initial state. The mean delay-period value of $\rho_{\mathrm{rel}}$ was $1.55 \pm 0.25$, significantly above the baseline value of $1.0$ (one-sample $t$-test, $t(99){=}21.9$, $p < 0.001$, $d{=}2.19$). In the later interval ($t{=}8$--$9$\,s), $\rho_{\mathrm{rel}} = 1.53 \pm 0.20$, comparable to or slightly below the delay-period mean. Goal-specific modulation of the effective recurrent connectivity is assessed separately by $\Delta_{\mathrm{sim}}(t)$ (see below), which rose after goal-cue presentation and increased toward the later part of the trial. Thus, $\rho_{\mathrm{rel}}(t)$ indicated a sustained STP-mediated increase in effective recurrent scaling (\S\ref{sec:17}).

 \paragraph{Temporal evolution of goal specificity in effective connectivity}
In the Without-STP condition, $W^{\mathrm{rec}}_{\mathrm{eff}}(t){=}2W^{\mathrm{rec}}$ is time-invariant, so $\Delta_{\mathrm{sim}}(t){=}0$ is the definitional baseline. Accordingly, inferential statistics focus on whether $\Delta_{\mathrm{sim}}(t)$ is positive in the With-STP condition and whether it increases toward the later part of the trial.  Before goal-cue presentation, $\Delta_{\mathrm{sim}}(t)$ remained near zero; after the cue, it rose rapidly and stayed positive throughout the delay period (Fig.~\ref{fig:12}). It increased further toward the later part of the trial.  These results support the interpretation that STP reorganizes effective connectivity in a goal-specific manner immediately after goal-cue presentation and throughout the delay period; the stronger patterning later in the trial reflects dynamics conditioned jointly on goal and task state. Comparisons within predefined summary windows based on the task schedule also showed that the delay-period mean of $\Delta_{\mathrm{sim}}$ was significantly above the zero baseline ($0.118 \pm 0.018$; one-sample $t$-test, $t(99){=}64.3$, $p < 0.001$, $d{=}6.43$). Moreover, the mean over the later interval was significantly larger than the delay-period mean (later trial: $0.153 \pm 0.024$; paired $t$-test, $t(99){=}14.4$, $p < 0.001$, $d_z{=}1.44$). Because the position input differs across trials after the first move, we use the delay-period window, when the position input is fixed, to assess internally maintained goal specificity (\S\ref{sec:23}).

\subsection{Experiment 4: STP parameter characteristics and the facilitation-dominant band}\label{sec:25}

This subsection addresses RQ4. We examine how success rates depend on the STP time constants $(\tau_{\mathrm{R}}, \tau_{\mathrm{F}})$ and utilization parameter $U_{\mathrm{SE}}$. We then discuss the structure and physiological consistency of the high-success-rate band.

\begin{figure}[t]\centering\includegraphics{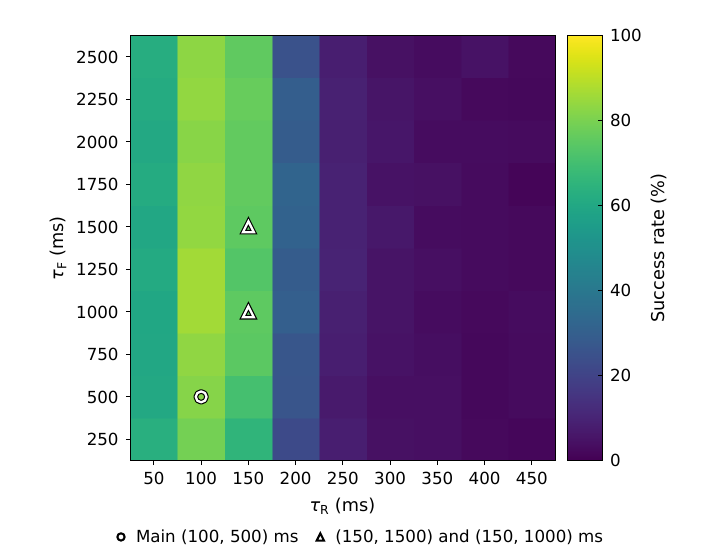}\caption{\textbf{Exploratory STP parameter map.} With STP, $\tau{=}1{,}000$\,ms, $\sigma{=}0.001$, mean over $n{=}100$ seeds per parameter set. Success rate across the grid search over $\tau_{\mathrm{R}}$ (horizontal axis, 50--450\,ms) and $\tau_{\mathrm{F}}$ (vertical axis, 250--2{,}500\,ms) at $\alpha_r{=}3.0$ (90 grid points). High success rates were concentrated in a facilitation-dominant region with small $\tau_{\mathrm{R}}$ (50--150\,ms) over the whole examined range of $\tau_{\mathrm{F}}$. The white circle indicates the parameter pair $(\tau_{\mathrm{R}}, \tau_{\mathrm{F}}){=}(100, 500)$\,ms used for the main comparison; triangles indicate $(150, 1{,}500)$\,ms and $(150, 1{,}000)$\,ms. Color indicates the mean success rate (\%).}\label{fig:13}\end{figure}\paragraph{Facilitation-dominant band}
\begingroup\setlength{\emergencystretch}{2em}Across the 90 grid points (Fig.~\ref{fig:13}), high success rates were concentrated where $\tau_{\mathrm{R}}$ was relatively small (50--150\,ms; highest at $\tau_{\mathrm{R}}{=}100$\,ms, 79.1--86.3\%) over the whole examined range of $\tau_{\mathrm{F}}$ (250--2{,}500\,ms), corresponding to a facilitation-dominant regime. The success rate was 65.3--76.6\% at $\tau_{\mathrm{R}}{=}150$\,ms and 59.5--62.9\% at $\tau_{\mathrm{R}}{=}50$\,ms. Comparisons at $(\tau_{\mathrm{R}}, \tau_{\mathrm{F}}) = (150, 1{,}500)$~ms and $(150, 1{,}000)$~ms, which are grid points of this map, yielded success rates of 75.2\% and 75.4\%, that is, 31.5 and 31.7 percentage points above the Without-STP condition at the same recurrent scaling ($\alpha_r{=}3.0$; 95\% CIs [24.1, 38.8] and [24.9, 38.8]; Supplementary Fig.~\ref{fig:21:partb}F, G, I). Supplementary Figure~\ref{fig:40} summarizes the behavioral and internal measures for the main and alternative STP parameter sets under the same recurrent scaling.\par\endgroup

\paragraph{Utilization parameter} To assess sensitivity to utilization, we varied $U_{\mathrm{SE}}$ over five levels from $0.025$ to $0.4$ (\S\ref{sec:18}), with $(\tau_{\mathrm{R}}, \tau_{\mathrm{F}})$ fixed at $(100, 500)$\,ms. At $\tau{=}1{,}000$\,ms, $\alpha_r{=}3.0$ and $\sigma{=}0.001$, utilization values of $0.05$ and $0.1$ gave higher success than Without STP (Supplementary Figs.~\ref{fig:21:parta}B, D and~\ref{fig:22}A).

\paragraph{Reported STP parameters and the rate scale in the STP update} We also evaluated the parameter sets of \citet{mongillo2008synaptic} and \citet{wang2006heterogeneity} at $\tau{=}1{,}000$\,ms and $\sigma{=}0.001$ (\S\ref{sec:05} and \S\ref{sec:18}). A normalized activity of 1 was mapped to 20\,Hz in the STP update, using the order of magnitude of reported population-mean delay-period PFC firing rates as a physiological reference \citep{funahashi1989mnemonic,compte2003temporally}. With $\alpha_r{=}2$ in both conditions, success rates with both parameter sets were higher than in the Without-STP condition (Supplementary Fig.~\ref{fig:22}D). At the 100-Hz scale used in the other experiments, both parameter sets gave success below 5\% at all four recurrent scalings (Supplementary Fig.~\ref{fig:22}B).  Panels B--D of Supplementary Fig.~\ref{fig:22} report the parameter sets and comparisons at both rate scales and all tested recurrent scalings.

Empirical findings in cortex, including prefrontal cortex, indicate that facilitation typically decays over tens to hundreds of milliseconds, while recovery from depression ranges from hundreds of milliseconds to several seconds (see \S\ref{sec:01} for the relevant references). The relation of the high-success-rate band to these reported values is discussed in \S\ref{sec:30}.

\section{Discussion}\label{sec:26}

With respect to RQ1, the RC--RL model learned the multistep task with sequential GO opportunities (\S\ref{sec:22}). STP increased task success under state noise in this PFC-inspired reservoir. It also strengthened later action-value preferences for goal-consistent moves, at the same recurrent scaling and with the recurrent scaling selected for each condition. Goal identity was highly decodable during the delay in both conditions; at the same recurrent scaling, later decoding was higher with STP, and with the recurrent scaling selected for each condition, it was high in both. The steady-state STP control, with efficacy set by the STP rule from each unit's current activity, retained the success-rate advantage over Without STP, with success similar to that of With STP (\S\ref{sec:22}; \S\ref{sec:31}). These results support STP-derived activity-dependent recurrent modulation as a mechanism that keeps retained goals usable for later action selection in this model. Prior models established synaptic working memory, goal-to-action switching, and improved short-term memory with dynamic synapses \citep{mongillo2008synaptic,katori2011representational,mori2015dynamic}. \citet{kozachkov2022robust} found greater robustness to synaptic ablation with STP but a tendency toward greater process-noise robustness in fixed-synapse networks trained by BPTT. The present results link goal retention to learned action-value preferences and successive goal-directed choices in fixed random recurrent circuits with a TD-trained readout. 

\subsection{Relation to persistent-firing models}\label{sec:27}
In this PFC-inspired model, STP-derived activity-dependent recurrent modulation and recurrent activity support delayed goal use. Classical persistent-firing models maintain information through ongoing activity, while STP modulates synaptic transmission efficacy according to presynaptic activity. In continuous-attractor models, short-term facilitation can reduce noise-driven diffusion and heterogeneity-driven drift of remembered locations \citep{seeholzer2019stability}. In the present model, synaptic efficacy follows each unit's evolving activity and reconfigures effective connectivity in a goal-dependent manner, supporting goal use at later GO opportunities. At the same recurrent scaling, this modulation maintained goal-dependent state structure into the later part of the trial and supported stronger action-value preferences for goal-consistent moves at later GO opportunities (\S\ref{sec:23}; \S\ref{sec:24}). Stable population coding can coexist with heterogeneous single-neuron dynamics in primate PFC \citep{murray2017stable}. Here, STP-derived activity-dependent efficacy provides a functional account of how recurrent activity supports delayed goal use.

\subsection{From dynamic synapses to goal-conditioned dynamics}\label{sec:28}
Because the model is closed-loop through Q-value feedback, the observed state trajectories reflect the interaction between STP-modulated recurrent dynamics, the learned readout, the feedback input, and the action history.  The effective-connectivity analyses isolate the STP-derived recurrent modulation component rather than decomposing the entire closed-loop flow. Without noise, the differences between the conditions depended on feedback strength and recurrent-scaling selection (\S\ref{sec:22}; Supplementary Fig.~\ref{fig:24:partb}C--E; Supplementary section ``Q-value feedback strength''; Supplementary Fig.~\ref{fig:30}). At the same recurrent scaling ($\alpha_r{=}3.0$), the With-STP advantage at $\sigma{=}0.001$ persisted when networks were trained without Q-value feedback and when feedback strength was varied.

In the comparison at the same recurrent scaling ($\alpha_r{=}3.0$), goal identity could be linearly decoded with high accuracy from delay-averaged activity even without STP, but STP supported higher goal-decoding accuracy and stronger action-value preferences for goal-consistent moves during later GO windows. With selected recurrent scalings, later goal decoding was high in both conditions and $D_Q$ at GO3 remained larger with STP (Supplementary Fig.~\ref{fig:35}A, B). These results distinguish goal decodability from the action-value preferences expressed by the TD-trained readout.

\begingroup\setlength{\emergencystretch}{2em}STP scales the fixed recurrent matrix through unit-wise efficacy factors (\S\ref{sec:17}). We assessed goal-dependent effective-connectivity patterns with $\Delta_{\mathrm{sim}}(t)$ and separation of goal-conditioned activity with the scatter ratio. At the main STP setting, $(\tau_{\mathrm{R}}, \tau_{\mathrm{F}}){=}(100, 500)$\,ms, $\rho_{\mathrm{rel}}(t)$ summarized the sustained increase in global recurrent scaling.\par\endgroup

\begingroup\setlength{\emergencystretch}{2em}The time-resolved trajectory of goal specificity in effective connectivity, $\Delta_{\mathrm{sim}}(t)$, rose rapidly after goal-cue presentation, remained positive throughout the delay period, and increased further toward the later part of the trial. The representative scatter-ratio trace illustrates temporal variation, while the group trajectory of $\Delta_{\mathrm{sim}}(t)$ summarizes effective-connectivity patterning across networks.  During the delay period, when the goal cue is off and position is fixed, $\Delta_{\mathrm{sim}}(t)$ reflects goal-specific patterning; in the later interval, where goal identity covaries with position and task state, it should be read as goal- and task-state-conditioned effective-connectivity patterning rather than pure goal specificity. The quantities $\Delta_s(t)$ and $\Delta_{\mathrm{sim}}(t)$ describe the same STP scaling patterns with equal unit weights and squared recurrent-column-norm weights, respectively (\S\ref{sec:17}; Supplementary Fig.~\ref{fig:18}). The two quantities have similar time courses and are strongly correlated across seeds. The fact that $\Delta_{\mathrm{sim}}(t)$ increases in the later part of the trial while $\rho_{\mathrm{rel}}(t)$ remains comparable to its delay-period mean (1.53 at $t{=}8$--$9$\,s versus 1.55; \S\ref{sec:24}) further supports the view that the two indices capture different aspects of the dynamics. \par\endgroup

\subsection{Disambiguating the role of STP from a generic recurrent-scaling increase}\label{sec:29}
Two kinds of controls constrain the interpretation of the STP effect: separately trained comparisons and perturbations of trained With-STP networks (Supplementary Figs.~\ref{fig:16} and~\ref{fig:17}). With STP achieved higher success than both gain-matched controls (\S\ref{sec:22}; Supplementary Fig.~\ref{fig:16}). The advantage over fixed recurrent scaling also held when that scaling was selected for task performance: the With-STP condition exceeded the highest Without-STP success rate obtained across the 16 grid values of $\alpha_r$ (\S\ref{sec:22}; Fig.~\ref{fig:04}).\par In trained With-STP networks, resetting the STP variables at cue offset partially reduced performance. Freezing the STP-derived scaling disrupted task performance.  A separate intact re-evaluation yielded a success rate close to the original With-STP value and served as a procedural check for the post-training perturbations.\par At $\alpha_r{=}3.0$, Common prescribed scaling reproduced part of the success-rate advantage but remained below With STP. Separately trained controls with the two prescribed time-varying scalings and the three simpler activity-dependent rules extend the fixed-scaling comparison (\S\ref{sec:22}; Supplementary Figs.~\ref{fig:24:partc}F and~\ref{fig:29}). At the same recurrent scaling ($\alpha_r{=}3.0$) under state noise, STP gave higher success, first-move accuracy, and later action-value differences than each of the three simpler activity-dependent rules; with $\alpha_r$ selected for each condition, success was also higher with STP than with each control (Supplementary Fig.~\ref{fig:37:partc}D). A separately trained steady-state STP control also exceeded Without STP and reached success similar to that of With STP (Table~\ref{tab:05}A; Supplementary Figs.~\ref{fig:33}, \ref{fig:37} and~\ref{fig:38}).\par The delay-period $\Delta_s(t)$ and $\Delta_{\mathrm{sim}}(t)$ analyses identify goal-dependent differences in the STP scaling patterns, and the replay intervention showed greater first-move accuracy with the cue-matched unit assignment than after rank-based or randomized reassignment (\S\ref{sec:22}; Supplementary Fig.~\ref{fig:26}). This comparison shows sensitivity of the first action to the unit assignment of replayed STP-derived scaling. In the intact model, this assignment is generated by STP from the network's own activity following the cue.  The action-value-difference analysis (\S\ref{sec:15}, Fig.~\ref{fig:09}) provides a readout-level link between the maintained goal-conditioned state structure and action selection.

\subsection{Physiological consistency of time scales}\label{sec:30}
  The high-success-rate band of the parameter search (\S\ref{sec:25}) is broadly consistent with the cortical STP time scales summarized in \S\ref{sec:01} \citep{markram1998differential,varela1997differential,wang2006heterogeneity,jackman2017mechanisms,hempel2000multiple}. The representative pair $(\tau_{\mathrm{R}}, \tau_{\mathrm{F}})=(100,500)$\,ms used in the main 100-seed comparison is close to the facilitating-synapse parameters used by \citet{tsodyks1998neural} ($\tau_{\mathrm{R}}\approx 130$\,ms, $\tau_{\mathrm{F}}\approx 530$\,ms), providing a cellular-scale reference for this choice. The facilitation time constant of the main setting, $\tau_{\mathrm{F}}{=}500$\,ms, is close to the value measured for facilitating connections between layer-5 pyramidal cells in ferret medial prefrontal cortex (507\,ms; \citealp{wang2006heterogeneity}). $U_{\mathrm{SE}}{=}0.1$ and $\tau_{\mathrm{R}}{=}100$\,ms lie below the group means of the same connections ($0.28$ and $194$\,ms). Slower facilitation time constants, $\tau_{\mathrm{F}}\approx 1{,}500$\,ms with $\tau_{\mathrm{R}}\approx 200$\,ms, were used in a prior PFC working-memory model \citep{mongillo2008synaptic}. With the parameter set of \citet{mongillo2008synaptic} and the measured group means of \citet{wang2006heterogeneity}, evaluated with a normalized activity of 1 mapped to 20\,Hz, the With-STP condition also exceeded the Without-STP condition when both conditions used $\alpha_r{=}2$ (\S\ref{sec:25}). The rate scale and STP parameters jointly determine the efficacy applied to recurrent activity; Supplementary Fig.~\ref{fig:22}B compares both rate scales for each parameter set. The main STP parameter pair and the two alternative pairs specified in \S\ref{sec:18} combined high success rates with mean delay-period relative effective spectral radii above one and larger $D_Q$ at GO2 and GO3 than in the Without-STP condition (Supplementary Figs.~\ref{fig:40:parta}A, D, E and \ref{fig:40:partb}G). These results support a \emph{functionally facilitation-dominant} interpretation of the high-success-rate band: in these settings, facilitation-induced modulation strengthens effective recurrent connectivity and supports goal use at later GO opportunities. In the STP equations, $\tau_{\mathrm{F}}$ sets the relaxation of utilization toward $U_{\mathrm{SE}}$ in the absence of presynaptic activity; within the recurrent network, $\tau_{\mathrm{F}}$ also shapes how the steady-state efficacy depends on activity (Supplementary section ``Steady-state STP modulation'').

\subsection{Candidate mechanism: STP-induced effective-connectivity modulation and goal-conditioned dynamics}\label{sec:31}
The controls support STP-derived activity-dependent recurrent modulation as a mechanism that keeps retained goals usable for later action selection in this model. In the separately trained steady-state STP control, efficacy has no state of its own, and recurrent activity carries history. At $\sigma{=}0.001$, this control preserved the success-rate advantage over Without STP at $\tau{=}500$, 750 and 1{,}000\,ms, with success rates similar to those with STP (\S\ref{sec:22}; Supplementary Figs.~\ref{fig:37:partb}B, C and~\ref{fig:38}). In this control, efficacy still changed with recurrent activity, so effective connectivity remained activity-dependent. At $\tau{=}1{,}000$\,ms, the other separately trained controls all remained below With STP (\S\ref{sec:22}; Supplementary Figs.~\ref{fig:16}, \ref{fig:24:partc}F--H, \ref{fig:29} and~\ref{fig:37:partc}D). Goal-specific efficacy patterns and acute perturbations describe how the trained With-STP network uses its modulation. The replay comparison further shows that the unit assignment of replayed scaling trajectories affects first-move accuracy. Together, these observations link STP-derived activity-dependent recurrent modulation to retained goals and subsequent goal-directed action selection.

\subsection{A dynamical-systems account of robustness under noise}\label{sec:32}
At the same recurrent scaling, STP reduced the loss in success rate between separately trained conditions without and with noise. With selected recurrent scalings, STP retained higher success at both noise levels, and the difference in the reduction in success rate (Without STP minus With STP) was small (1.4 percentage points; 95\% BCa CI [\ensuremath{-}3.5, 6.3]; Table~\ref{tab:04}) (\S\ref{sec:22}). When the selected scalings were held fixed across noise levels, the With-STP advantage was largest at $\sigma{=}0.004$ (30.8 percentage points; Fig.~\ref{fig:06}C). At the same recurrent scaling, elevated $\rho_{\mathrm{rel}}$ and goal-specific efficacy patterns during the delay accompanied higher success under state noise (\S\ref{sec:28}).

\subsection{\texorpdfstring{Relation to reservoir computing and fading memory}{Relation to reservoir computing and fading memory}}\label{sec:33}
Standard reservoir computing relies on fading memory and rich transient responses. In a reservoir with fixed recurrent weights, how long a past input remains usable is a property of the reservoir's design rather than a fixed limit of the approach. The present study asks whether, in this task and parameter regime, the same network keeps the goal representation usable for action selection better with STP than without it. With the unit time constant fixed at $\tau{=}1{,}000$\,ms, STP added activity-dependent, goal-specific multiplicative modulation of the fixed recurrent weights. Across the unit-time-constant sweep, paired-difference intervals at the same recurrent scaling ($\alpha_r{=}3.0$) were above zero for the tested values from 750 to 3{,}000\,ms. With the recurrent scaling selected for each condition, the advantage of STP was confirmed at unit time constants of 500, 750 and 1{,}000\,ms (\S\ref{sec:22}; Fig.~\ref{fig:05}). With these selected scalings, both conditions reached high success rates at the tested time constants $\tau\geq 1{,}250$\,ms (Fig.~\ref{fig:05}; Supplementary Fig.~\ref{fig:20}). 

Closed-loop forecasting repeatedly feeds the predicted signal back as input, as demonstrated for chaotic laser dynamics \citep{li2023prediction,kai2023forecasting}. In the present task, by contrast, the goal cue is presented once and removed after $t=3$\,s; Q-value feedback returns action values, not the cue, and the retained goal guides actions several seconds later.

Compression and dimensionality reduction can reduce computational cost or support prediction in reservoir computing. Examples include compressed input acquisition \citep{yamaguchi2023timedomain}, eigenvector-based compression and decompression \citep{zhou2026compression}, reduced masking in deep time-delay reservoirs \citep{zhang2024accelerated}, and dimensionality reduction for parallel reservoir prediction \citep{fleddermann2025improving}. Here, PCA is used to visualize the geometry of trained reservoir trajectories (\S\ref{sec:13}).

\subsection{Limitations}\label{sec:34} \paragraph{Task complexity} In the grid without interior obstacles, a greedy shortest-path policy can select a correct move from the retained goal and current position; the task assesses delayed goal use during sequential action selection. Larger grids with longer routes would test how the success-rate advantage depends on the number of decisions and the duration of goal retention. Requirements for extending the learning framework to mazes with interior obstacles and dead ends are discussed in \S\ref{sec:35}.

\paragraph{Task scope}
The conclusions of this study apply to a $5{\times}5$ grid task with four blocked corner cells, eight non-corner perimeter targets, and sequential GO opportunities. In this task, the cue specifies one of eight goals for action selection at each of the first three GO opportunities, 2, 4, and 6\,s after cue offset. During the delay ($3.0<t<4.0$\,s), the goal and GO inputs are zero and position input is fixed, so decoding in this interval assesses internal goal retention. After the first move, position input can covary with the cued goal, so later state separation reflects goal- and task-state-dependent activity. The GO schedule was fixed across trials with four GO opportunities, and the time-resolved analyses focused mainly on the first three GO windows, using all evaluation trials.

\paragraph{Biological abstraction}
The model is an abstraction inspired by the PFC and is not intended as a one-to-one anatomical reconstruction. It uses a minimal Tsodyks--Markram STP model with one facilitation time constant and one recovery time constant, and does not include the heterogeneity and multiple processes observed in cortical synapses, such as augmentation and post-tetanic components. Recurrent connectivity is random and unstructured, and inputs are one-hot vectors. Each reservoir unit represents collective neural activity at a phenomenological rate level, with representative resource and utilization states associated with its outgoing recurrent connections. The signed recurrent weights describe effective interactions among these activity units. The STP variables are defined for each presynaptic unit rather than for each synapse. Because $x_j$ and $u_j$ are driven only by the activity of presynaptic unit $j$ and all connections share the same STP parameters, synapse-wise variables would take identical values for all synapses made by the same unit. This reduction omits STP whose properties depend on the postsynaptic target \citep{markram1998differential,blackman2013target}. As a consequence, the effective recurrent connectivity changes through column-wise scaling factors, which provide $N_m{=}100$ degrees of freedom for about 1{,}000 recurrent connections. The readout dimension is unchanged: the action-value readout in each condition receives $N_m$ reservoir-activity inputs, so the unit-wise constraint concerns how the synaptic state reshapes recurrent interactions. The biological interpretation concerns the computational contribution of STP at an algorithmic level.
\paragraph{Operational indices and limits of dynamical interpretation}
The quantity $\rho_{\mathrm{eff}}$ is not the spectral radius of the full Jacobian including the derivative of $\tanh$ and the leak term, but an operational index of STP-mediated scaling of recurrent coupling. The scatter ratio is an index of separability among goal-conditioned states and depends on the geometry of each network's internal state space. Accordingly, representative time-series plots are positioned as qualitative visualizations, whereas group comparisons are based on summary measures within predefined time windows and on indices that do not depend on dimensionality reduction. Because group averaging can smooth out the sharp temporal structure seen in individual networks, the group-level claims concern maintenance of effective-connectivity patterning during the delay period and its increase toward the later part of the trial, rather than the existence of steep GO-synchronized peaks. ``Goal-conditioned dynamics'' is an operational construct evaluated through aligned behavioral, state-representation, action-value, and effective-connectivity analyses, and does not include direct demonstration of fixed points or local contraction under perturbation. The successful-trial comparison uses trials selected separately in each condition and describes goal- and task-state-dependent activity during successful behavior (\S\ref{sec:23}). The operational indices and summary windows are defined in \S\ref{sec:17} and \S\ref{sec:20}, respectively. In addition, the time-resolved decoding analysis in this study evaluates the decodability of goal identity at each time point and does not directly show that the same representational format is maintained invariantly across time.

\paragraph{Design limitations of the experiment}
The reward function used here consists only of a small distance-based shaping term ($+0.1$ for moves that decrease the squared Euclidean distance to the goal and $-0.3$ otherwise), so the present results characterize robust goal-conditioned action selection under shaped rewards rather than under fully sparse-reward general planning. Because explicit epsilon-greedy exploration was not introduced, exploration early in learning arose from small random readout initialization, reward-dependent updating, and state noise; the viability of this scheme depends on the action space and reward design of the present task, and additional testing will be needed for tasks with larger action spaces or sparser rewards. Comparisons with persistent-firing models and models with learned gates will extend the range of memory mechanisms considered.  Because the model includes feedback connectivity $W^{\mathrm{back}}$ from the output Q values back to the reservoir, differences in state trajectories across conditions may reflect not only STP-induced modulation of recurrent connectivity but also closed-loop feedback through differences in the learned output weights. At the same recurrent scaling under state noise, With STP retained higher success when the Q-value feedback was removed during evaluation and when both conditions were trained without the feedback pathway (\S\ref{sec:22}); the effective-connectivity analysis in this study focuses on STP-induced modulation of recurrent connectivity.

\subsection{Future directions}\label{sec:35}
\begingroup\setlength{\emergencystretch}{2em}The mechanistic interpretation proposed here should be tested further through the following extensions. First, the dependence of goal use on the STP efficacy curve and ongoing recurrent activity should be examined across delay lengths and synaptic parameter settings. Second, heterogeneous distributions of cortical synaptic time constants, augmentation components, and synapse-specific or target-specific STP should be introduced to determine how far the conclusions from the model with homogeneous STP parameters and unit-wise synaptic states generalize. Third, the robustness of STP-induced goal-conditioned dynamics should be tested under task settings with variable delay lengths and distractor stimuli, and the task should be extended to larger grids with more GO opportunities while retaining the present unit and synaptic time constants. Extensions to mazes with interior obstacles and dead ends should combine rewards permitting detours, information about obstacle layouts, and a readout and exploration mechanism supporting state-dependent replanning. Such task settings would also test whether the time course of STP, beyond the activity-dependent efficacy captured by the steady-state STP control, is required for action selection, for example when an action toward an obstacle is not executed and an alternative action must be selected without a change in the sensory input. It will also be important to test how the operational spectral modulation of $W^{\mathrm{rec}}_{\mathrm{eff}}(t)$ is expressed in the true local flow field, using local Jacobian analysis on the augmented state $[\mathbf{m},\mathbf{x},\mathbf{u}]$. Fourth, the present readout-level evidence based on the action-value difference $D_Q$ motivates more complete closed-loop decompositions, for example by analyzing the joint contribution of the trained output weights $W^{\mathrm{out}}$ and the feedback pathway $W^{\mathrm{back}}$, and by comparing STP with persistent-firing models or recurrent models with learned gates.\par\endgroup

\section{Conclusion}\label{sec:36}
\begingroup\setlength{\emergencystretch}{2em}This study tested whether STP stabilizes goal representations as goal-conditioned dynamics in a PFC-inspired RC--RL model for multistep goal-directed action planning. The model learned the task with sequential GO opportunities, and, in the comparison at the same recurrent scaling, the With-STP condition showed higher robustness than the standard reservoir baseline with fixed recurrent scaling (the Without-STP condition), especially under state noise. When the recurrent scaling was selected for each condition from the same grid, the With-STP condition also reached a higher success rate than the Without-STP condition, both without and with noise. At the default state noise, the advantage persisted when selection and evaluation used separate network seeds (Supplementary Fig.~\ref{fig:37:parta}A). Although goal identity was already decodable during the delay period in both conditions, STP maintained goal-dependent structure into the later part of the trial at the same recurrent scaling. With separately selected recurrent scalings, later goal decoding was near ceiling in both conditions, while STP retained larger action-value differences at the third GO opportunity.\par\endgroup

Goal specificity in the effective connectivity was already present during the delay period, with stronger goal- and task-state-conditioned patterning later in the trial, and the relative effective spectral radius indicated that STP-derived recurrent scaling remained elevated. The exploratory parameter search further indicated that high success rates were concentrated in a functionally facilitation-dominant region, qualitatively compatible with cortical STP time scales.

The steady-state STP control preserved the success-rate advantage at unit time constants of 500, 750 and 1{,}000\,ms. In this control, recurrent activity retained the influence of past inputs, and efficacy was set by the STP rule from each unit's current activity. At $\tau{=}1{,}000$\,ms, success remained higher in the With-STP condition than with the gain-matched controls, the two prescribed time-varying scalings and the three simpler activity-dependent rules. Together, these results support STP-derived activity-dependent recurrent modulation as a mechanism that stabilizes goal-conditioned dynamics, keeping retained goals usable for successive actions under state noise in this model.

\section*{Declaration of competing interest}
The authors declare that they have no known competing financial interests or personal relationships that could have appeared to influence the work reported in this paper.

\section*{Author contributions (CRediT)}
\textbf{Jin Nakamura:} Conceptualization, Methodology, Software, Validation, Formal analysis, Investigation, Data curation, Writing---original draft, Visualization.

\textbf{Yuichi Katori:} Conceptualization, Methodology, Validation, Investigation, Writing---review and editing, Supervision, Project administration, Funding acquisition.

\section*{Funding}
This work was supported by JSPS KAKENHI Grant Numbers 24H02330, 25H00447, and 23H03468, and JST ALCA-Next Grant Number JPMJAN23F3. The funders had no role in study design, data collection, analysis or interpretation, manuscript preparation, or the decision to submit the article for publication.

\section*{Data availability}
The data used in this study were generated by simulation. The parameters and evaluation conditions required to reproduce the main analyses are reported in the main text. Network structures were generated for 100 independent seeds, and paired With-STP and Without-STP conditions were evaluated for each seed. Results are reported with means, standard deviations, and statistical tests. The random seeds, processed summary data, and additional output files are available from the corresponding author upon reasonable request.

\section*{Code availability}
The custom simulation and analysis code used to generate the main results can be made available to editors and reviewers upon request during peer review. After publication, the code will be available from the corresponding author upon reasonable request.

\enlargethispage{\baselineskip}\section*{Declaration of generative AI and AI-assisted\texorpdfstring{\\}{ }technologies in the manuscript preparation process}
During the preparation of this work, the authors used OpenAI's ChatGPT and Anthropic's Claude to improve readability and language, translate and revise text, and check the manuscript for journal-formatting requirements. After using these tools, the authors reviewed and edited the content as needed and take full responsibility for the content of the published article.

\setlength{\bibsep}{5pt plus 2pt minus 1pt}\bibliography{ms}

\clearpage \begin{center}{\large\bfseries Supplementary Material:\\[0.4em] Short-Term Synaptic Plasticity Stabilizes Goal-Conditioned Dynamics in a Prefrontal Cortex--Inspired Reservoir Model for Multistep Goal-Directed Action Planning\par}\vspace{0.5em}{\normalsize Jin Nakamura \quad Yuichi Katori\par}\end{center} %
\ComposeAppendixSetup\renewcommand{\floatpagefraction}{0.72}\setlength{\textfloatsep}{22pt plus 2pt minus 4pt}

\begingroup\begingroup\begin{figure}[tp]\begin{composepage}[fig:14|1|pg_fig_14_1]\begin{composerow}\begin{composecell}[121.254000000]\begin{composegraphic}\includegraphics{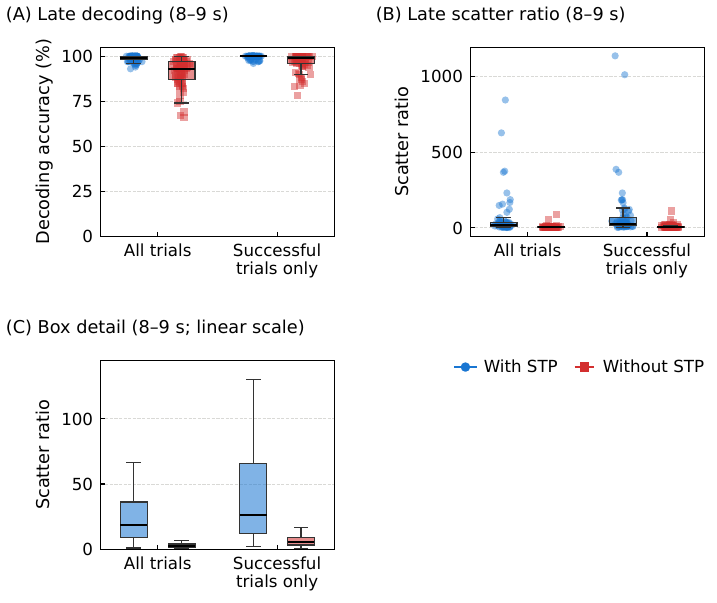}\end{composegraphic}\caption{\textbf{The later-trial STP advantage remains even when the analysis is restricted to successful trials.} \textbf{(A)} Goal-decoding accuracy from window-averaged activity in the later interval ($t{=}8$--$9$\,s). For each panel, the all-trial and successful-trials-only analyses are shown side by side for the With-STP and Without-STP conditions. The successful-trials-only analysis used the 75 paired seeds (73 for decoding) that met the inclusion criteria given in \S\ref{sec:09}. The all-trial analysis was also recalculated on the same seed subsets: all-trial decoding used all evaluation trials from the same 73 paired seeds, whereas decoding restricted to successful trials used only the goals that met the inclusion criteria in both conditions. The With-STP condition remained significantly higher for both decoding accuracy ($d_z{=}0.57$, $p < 0.001$) and scatter ratio ($d_z{=}0.40$, $p < 0.001$).  Each point represents one seed, and the box plots show the median and interquartile range. Whiskers: 1.5 IQR. Both conditions used $\alpha_r{=}3.0$ and $\sigma{=}0.001$. \textbf{(B)} Scatter ratio $\SR$ in the same interval, shown with all points. \textbf{(C)} The same boxes as in (B), enlarged on a linear axis.}\label{fig:14}\end{composecell}\end{composerow}\end{composepage}\end{figure}\endgroup \begingroup\begin{figure}[tp]\begin{composepage}[fig:15|1|pg_fig_15_1]\begin{composerow}\begin{composecell}[121.254000000]\begin{composegraphic}\includegraphics{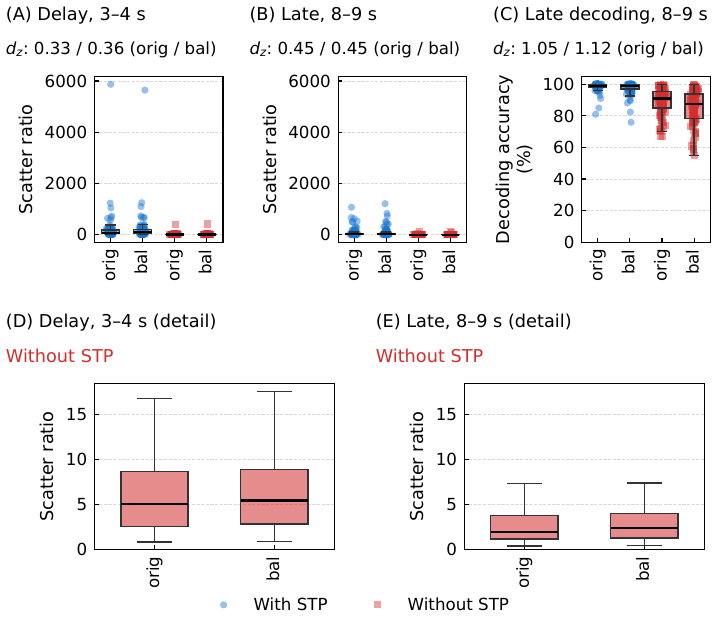}\end{composegraphic}\caption{\textbf{Balanced-goal resampling confirms that the main condition differences are not explained by minor goal-count imbalance.} Within each seed and condition, evaluation trials were balanced across goals by sampling $n_{\min}$ trials per goal without replacement within each resample (50 resamples per seed, then averaged), where $n_{\min}$ is the smallest per-goal count. \textbf{(A--C)} Scatter ratio in the delay ($t{=}3$--$4$\,s) and late ($t{=}8$--$9$\,s) windows, and late goal-decoding accuracy (\%, $t{=}8$--$9$\,s, stratified cross-validation with $K=\min(5,n_{\min})$ shuffled folds). Paired With-STP versus Without-STP effect sizes are shown separately for the all-trial (orig) and balanced (bal) analyses above each panel. All-trial $d_z$ values use the same pipeline as the balanced analysis and may differ slightly from main-text summaries. Here, we first averaged activity within each window and then computed the scatter ratio and decoding accuracy. In the corresponding time-resolved summaries in the main text, each measure was computed at each time point and then averaged within the same window. Both analyses used the same networks and evaluation trials. Boxes show the median and IQR across the 100 paired seeds, and points show the individual networks. \textbf{(D, E)} Below panels (A)--(C), the detail panels for (A) and (B) show the Without-STP boxes on separate linear axes.}\label{fig:15}\end{composecell}\end{composerow}\end{composepage}\end{figure}\endgroup \begingroup\begin{figure}[tp]\begin{composepage}[fig:16|1|pg_fig_16_1]\begin{composerow}\begin{composecell}[121.254000000]\begin{composegraphic}\includegraphics{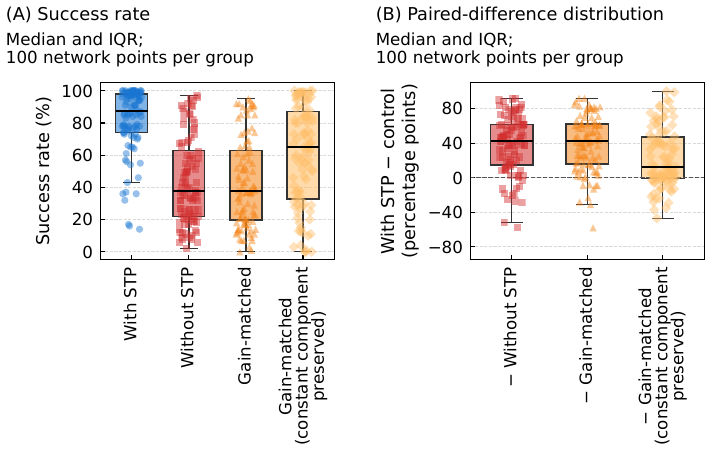}\end{composegraphic}\par\nointerlineskip\vspace{0.5mm}\begin{composetable}\begingroup\fontsize{8}{9}\selectfont\setlength{\tabcolsep}{2pt}%
\setbox\FinalTabularBox=\hbox{%
\begin{tabular}{@{}>{\raggedright\arraybackslash}p{36mm}rrrr@{}}\toprule%
\multicolumn{5}{l}{\textbf{(C, D) Paired success-rate statistics}}\\%
Setting & \shortstack{(C) Difference\\(pp)} & \shortstack{(C) 95\%\\BCa CI} & (D) $d_z$ & (D) $p$\\\midrule%
\rule{0pt}{4mm}With STP − Without STP & $38.16$ & $[31.55, 44.20]$ & $1.14$ & $1.2\times10^{-19}$\\ %
\rule{0pt}{4mm}With STP − Gain-matched & $39.39$ & $[32.92, 45.31]$ & $1.24$ & $6.9\times10^{-22}$\\ %
\rule{0pt}{4mm}With STP − Gain-matched (constant component preserved) & $22.02$ & $[15.18, 28.62]$ & $0.64$ & $4.4\times10^{-9}$\\ %
\bottomrule\end{tabular}}%
\vbox{\hsize=121mm\parindent=0pt\box\FinalTabularBox\par%
\noindent\parbox{121mm}{Paired n = $100$.}\par%
}\endgroup%
\end{composetable}\caption{\textbf{With STP exceeded both gain-matched controls under state noise.} \textbf{(A)} Post-training success rates for the original With-STP condition, the original Without-STP condition, and a gain-matched Without-STP condition. Both panels also include Gain-matched (constant component preserved), which uses $W^{\mathrm{rec}}(2c\bar{\mathbf r}-\mathbf 1)$ with the same calibrated $c=1.5506$ and a readout trained from scratch. \textbf{(B)} Paired success-rate differences relative to the original With-STP condition (positive values indicate that With STP exceeded the control). Each point represents one of $n{=}100$ paired seeds. Boxes show the median and interquartile range. \textbf{(C)} Means of the three paired differences with 95\% BCa bootstrap confidence intervals (2{,}000 resamples; \S\ref{sec:20}). \textbf{(D)} The differences, the interval endpoints, paired Cohen's $d_z$, the two-sided paired $t$-test $p$ values, and $n$ for the same comparisons. All comparisons use the same $n{=}100$ paired seeds at $\tau{=}1{,}000$\,ms and $\sigma{=}0.001$, with $\alpha_r{=}3.0$ in the original With-STP and Without-STP conditions; the gain-matched Without-STP control was trained with $\alpha_r{=}4.6518$ (\S\ref{sec:19}). Success rates are in percent and differences in percentage points. The means, SDs, paired difference, BCa CI and $d_z$ of the main With-STP versus Without-STP comparison are given in Table~\ref{tab:04}; its $p$ value is given in (D).}\label{fig:16}\end{composecell}\end{composerow}\end{composepage}\end{figure}\endgroup \begingroup\begin{figure}[tp]\begin{composepage}[fig:17|1|pg_fig_17_1]\begin{composerow}\begin{composecell}[121.254000000]\begin{composegraphic}\includegraphics{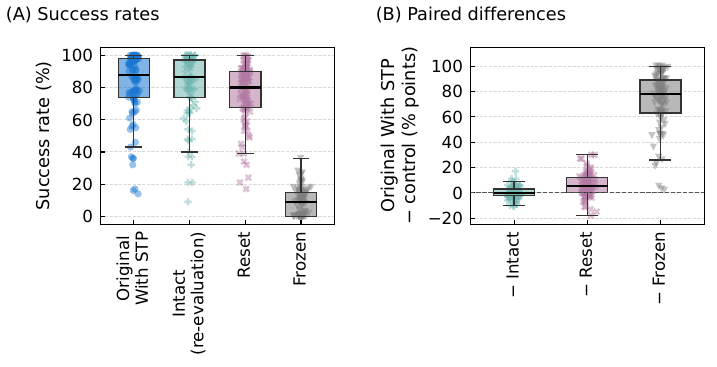}\end{composegraphic}\end{composecell}\end{composerow}\vspace{2mm}\begin{composerow}\begin{composecell}[121.254000000]\begin{composetable}\begingroup\fontsize{8}{9}\selectfont\setlength{\tabcolsep}{2pt}%
\setbox\FinalTabularBox=\hbox{%
\begin{tabular}{@{}>{\raggedright\arraybackslash}p{24mm}rrrrrr@{}}\toprule%
\multicolumn{7}{l}{\textbf{(C) Paired success-rate statistics}}\\%
Setting & \shortstack{Mean\\(\%)} & \shortstack{SD\\(\%)} & \shortstack{Original With STP\\minus condition\\(pp)} & \shortstack{95\%\\BCa CI} & $d_z$ & $p$\\\midrule%
\rule{0pt}{4mm}Intact (re-evaluation) & $81.94$ & $20.33$ & $-0.04$ & $[-0.93, 0.88]$ & $-0.01$ & $0.93$\\ %
\rule{0pt}{4mm}Reset & $75.71$ & $18.64$ & $6.19$ & $[4.46, 8.01]$ & $0.67$ & $1.1\times10^{-9}$\\ %
\rule{0pt}{4mm}Frozen & $8.92$ & $8.33$ & $72.98$ & $[67.79, 76.57]$ & $3.27$ & $6.4\times10^{-55}$\\ %
\bottomrule\end{tabular}}%
\vbox{\hsize=121mm\parindent=0pt\box\FinalTabularBox\par%
\noindent\parbox{121mm}{Paired n = $100$.}\par%
}\endgroup%
\end{composetable}\caption{\textbf{Resetting or freezing STP during evaluation reduced success in the trained With-STP networks.} These evaluation-time perturbations test how the trained With-STP networks use their modulation with all learned weights held fixed. \textbf{(A)} Post-training success rates for the original With-STP condition, an intact re-evaluation control, an STP-reset perturbation, and a frozen-STP perturbation. Whiskers: 1.5 IQR. \textbf{(B)} Paired success-rate differences relative to the original With-STP condition (positive values indicate that With STP exceeded the control). Resetting the STP variables at cue offset partially reduced the success rate of the trained With-STP networks. Freezing the STP-derived scaling reduced the success rate by 72.98 percentage points (C). Each point represents one of $n{=}100$ paired seeds. Boxes show the median and interquartile range. \textbf{(C)} For each condition, the mean success rate $\pm$ SD, the mean paired difference with its 95\% BCa bootstrap confidence interval (2{,}000 resamples; \S\ref{sec:20}), paired Cohen's $d_z$, the two-sided paired $t$-test $p$ value, and $n$. All comparisons use the same $n{=}100$ paired seeds at $\tau{=}1{,}000$\,ms and $\sigma{=}0.001$ with $\alpha_r{=}3.0$; the original With-STP values are given in Table~\ref{tab:04}. Success rates are in percent and differences in percentage points.}\label{fig:17}\end{composecell}\end{composerow}\end{composepage}\end{figure}\endgroup \begingroup\begin{figure}[tp]\begin{composepage}[fig:18|1|pg_fig_18_1]\begin{composerow}\begin{composecell}[121.254000000]\begin{composegraphic}\includegraphics{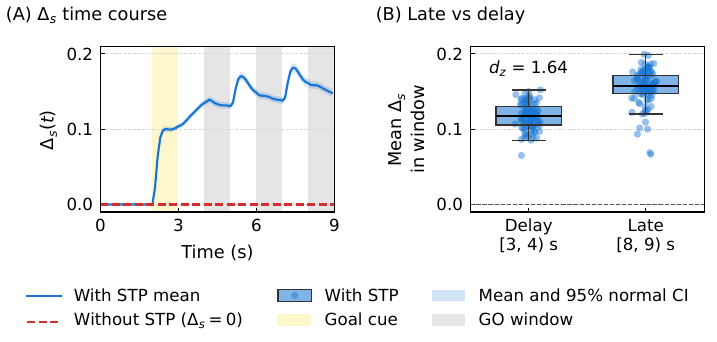}\end{composegraphic}\end{composecell}\end{composerow}\vspace{2mm}\begin{composerow}\begin{composecell}[121.254000000]\begin{composegraphic}\includegraphics{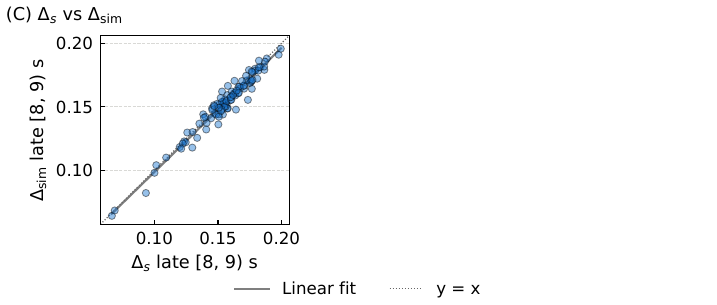}\end{composegraphic}\caption{\textbf{Late goal-separation indices with equal and squared-column-norm weighting were strongly correlated across networks.} This figure compares unweighted and recurrent-column-norm-weighted descriptions of the same STP scaling patterns. The main-text $\Delta_{\mathrm{sim}}(t)$ (Fig.~\ref{fig:12}) uses squared recurrent-column norms as cosine weights, while $\Delta_s$ is computed directly on $\mathbf{s}(t){=}2\mathbf{x}(t)\odot\mathbf{u}(t)/U_{\mathrm{SE}}$ with equal unit weights (\S\ref{sec:17}). \textbf{(A)} Time course of $\Delta_s(t)$ (mean across $n{=}100$ seeds, shaded band 95\% CI). The shaded band is a pointwise normal-approximation interval based on the standard error across seeds. Yellow band: goal cue; gray bands: GO windows. The red dashed line marks the Without-STP condition ($\Delta_s{=}0$). \textbf{(B)} Late versus delay comparison of $\Delta_s$ in the delay window ($t{=}3$--$4$\,s) and the late window ($t{=}8$--$9$\,s). Boxes show median and IQR; jittered points are individual seeds. Late-window values were significantly larger than delay-window values (paired $t$-test; $d_z$ shown above the panel as late minus delay). $\Delta_s$ and $\Delta_{\mathrm{sim}}$ are both defined as the same-goal minus different-goal cosine similarity of evaluation episodes within the same seed; in the Without-STP condition $\mathbf{s}(t)$ is constant, so $\Delta_s{=}0$ by construction. \textbf{(C)} Per-seed scatter of $\Delta_s$ versus $\Delta_{\mathrm{sim}}$ in the late window ($t{=}8$--$9$\,s), with the identity line ($y{=}x$, dotted) and a linear fit (solid) drawn for reference. The two quantities were strongly correlated across seeds (Pearson $r{=}0.98$, Spearman $\rho{=}0.96$, both $p<10^{-40}$).}\label{fig:18}\end{composecell}\end{composerow}\end{composepage}\end{figure}\endgroup \begingroup\begin{figure}[tp]\begin{composepage}[fig:19|1|pg_fig_19_1]\ComposePart{a}\begin{composerow}\begin{composecell}[121.254000000]\begin{composegraphic}\includegraphics{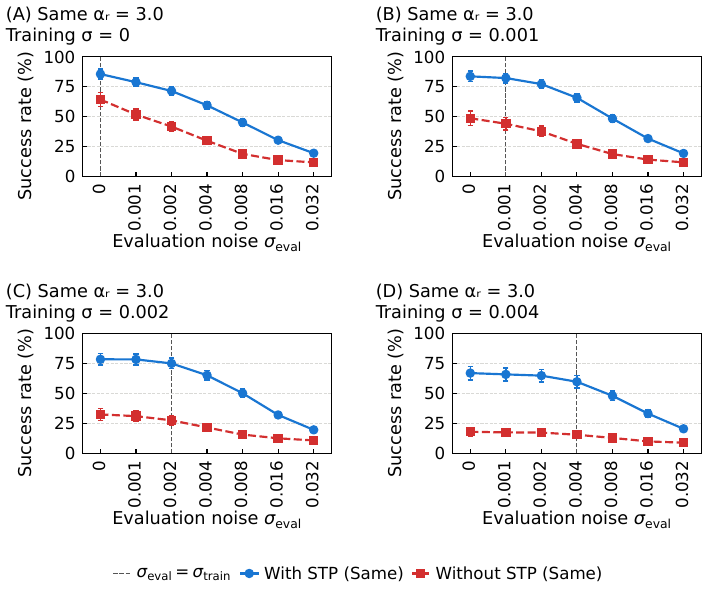}\end{composegraphic}\caption{\textbf{At the same recurrent scaling, With STP retained its success advantage when training and evaluation noise differed.} Post-training success rate of the With-STP and Without-STP conditions for networks evaluated, with learning disabled, at each of 7 state-noise levels ($n{=}100$ paired seeds; $\tau{=}1{,}000$\,ms). \textbf{(A--D)} Comparison at the same recurrent scaling ($\alpha_r{=}3.0$ in both conditions) for networks trained at $\sigma_{\mathrm{train}}{=}0$, $0.001$, $0.002$, and $0.004$; markers and error bars show the mean and $t$-based 95\% confidence interval across seeds. The evaluation noise levels are shown at equal spacing, and the vertical dashed line marks $\sigma_{\mathrm{eval}}{=}\sigma_{\mathrm{train}}$. All values come from a separate set of 100 evaluation episodes of the trained networks (\S\ref{sec:10}); for $\sigma_{\mathrm{eval}}{=}\sigma_{\mathrm{train}}$, the original evaluations are shown in Fig.~\ref{fig:06} and Supplementary Fig.~\ref{fig:34}.}\ComposePartLabel{a}\label{fig:19}\end{composecell}\end{composerow}\end{composepage}\end{figure}\endgroup \begingroup\begin{figure}[tp]\begin{composepage}[fig:19|2|pg_fig_19_2]\ComposePart{b}\begin{composerow}\begin{composecell}[121.254000000]\begin{composegraphic}\includegraphics{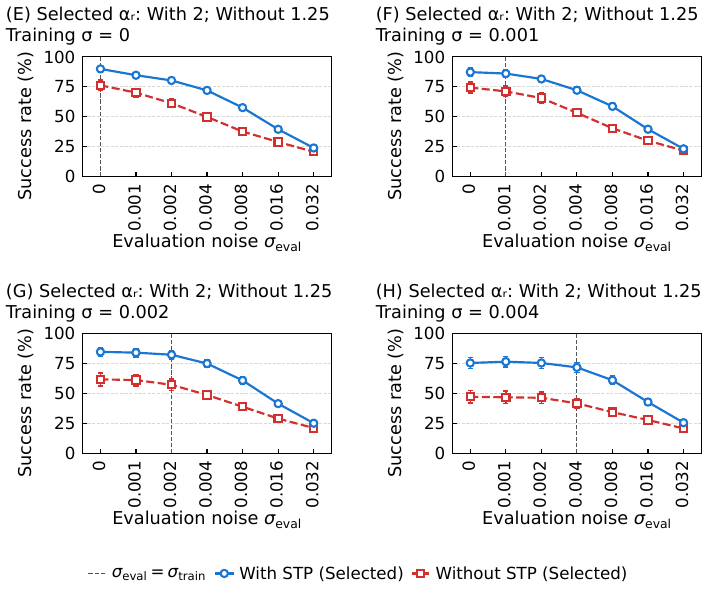}\end{composegraphic}\caption{\textbf{(E--H)} Comparison with the recurrent scaling selected for each condition, with $\alpha_r$ fixed at the values selected at $\sigma{=}0.001$ ($\alpha_r{=}2$ With STP, $\alpha_r{=}1.25$ Without STP), for networks trained at $\sigma_{\mathrm{train}}{=}0$, $0.001$, $0.002$, and $0.004$; markers and error bars show the mean and $t$-based 95\% confidence interval across seeds ($n{=}100$ paired seeds; $\tau{=}1{,}000$\,ms).}\ComposePartLabel{b}\end{composecell}\end{composerow}\end{composepage}\end{figure}\endgroup \begingroup\begin{figure}[tp]\begin{composepage}[fig:19|3|pg_fig_19_3]\ComposePart{c}\begin{composerow}\begin{composecell}[121.254000000]\begin{composegraphic}\includegraphics{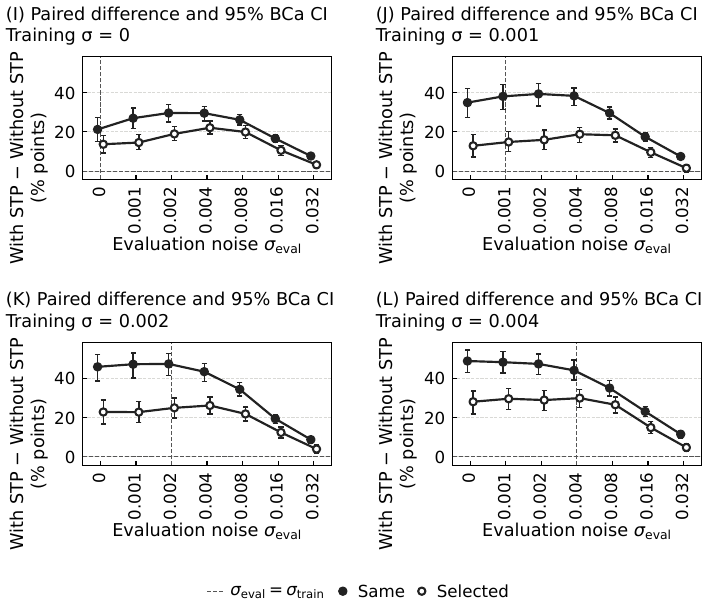}\end{composegraphic}\caption{\textbf{(I--L)} Paired difference (With STP minus Without STP) and its 95\% BCa bootstrap confidence interval (2{,}000 resamples; $n{=}100$ paired seeds; $\tau{=}1{,}000$\,ms) for the comparisons at the same recurrent scaling ($\alpha_r{=}3.0$ in both conditions; filled markers) and with the recurrent scaling fixed at the values selected at $\sigma{=}0.001$ ($\alpha_r{=}2$ With STP, $\alpha_r{=}1.25$ Without STP; open markers); the horizontal line indicates zero. Panels (I)--(L) show $\sigma_{\mathrm{train}}{=}0$, $0.001$, $0.002$, and $0.004$, respectively.}\ComposePartLabel{c}\end{composecell}\end{composerow}\end{composepage}\end{figure}\endgroup \endgroup 
 \par\par \paragraph{Successful-trial composition}\label{sec:37} To assess sensitivity to successful-trial composition, we matched the common goals and successful-episode counts per goal between conditions over 20 draws per network within the originally eligible subset. Across 73 paired networks, the With-STP minus Without-STP differences remained positive in both windows (scatter ratio: delay, 162.0 [98.9, 324.2]; late, 105.2 [65.9, 180.0]; decoding: delay, 2.1 [1.4, 3.2] percentage points; late, 2.9 [1.9, 4.2] percentage points; all intervals are 95\% BCa CIs). The scatter-ratio difference was positive in 69 and 67 of 73 seed pairs in the delay and late windows, respectively. Both Wilcoxon signed-rank tests gave $p < 0.001$. These matched comparisons show greater goal separability and higher goal-decoding accuracy with STP among successful trials from the eligible paired networks. When at least five successful trials per shared goal were required in both conditions (63 paired networks), the decoding differences also remained positive (delay, 0.84 [0.28, 1.36] percentage points; late, 1.53 [0.86, 2.49] percentage points; 95\% BCa CIs).

\par \begingroup\par \paragraph{Gain-matched controls} The two separately trained gain-matched controls (Fig.~\ref{fig:16}) apply the same calibrated multiplier to the full recurrent input or to its activity term. Here, fixed recurrent scaling refers to the scalar multiplier $\alpha_r$ applied to the fixed recurrent matrix $W^{\mathrm{rec}}$ in the Without-STP model. The gain-matched condition used $\alpha_r{\approx}4.65$, computed as $\alpha_r^{\mathrm{matched}}=\rho_{\mathrm{rel}}^{\mathrm{delay}}\,\alpha_r^{\mathrm{default}}$ using the ratio of the group-mean delay-period effective spectral radii of the With-STP and Without-STP conditions (see Methods). Both gain-matched Without-STP conditions remained below the With-STP condition, arguing against a simple fixed recurrent-scaling explanation of the STP advantage. The BCa CI for the constant-component-preserving control is conditional on the same fixed calibration $c=1.5506$.\par \par\endgroup \begingroup\par \paragraph{Evaluation-time STP perturbations} The intact re-evaluation condition re-ran the original With-STP model through the same post-training re-evaluation procedure without modifying the STP variables. In the reset condition, $x_j$ and $u_j$ were reset to their baseline values, $x_j{=}1$ and $u_j{=}U_{\mathrm{SE}}$, at cue offset ($t{=}3$\,s), and then evolved according to the standard STP equations. In the frozen-STP condition, the dynamic STP-derived scaling $s_j(t){=}2x_j(t)u_j(t)/U_{\mathrm{SE}}$ was replaced throughout each trial by a static, time-invariant, goal-independent per-unit reference vector $\bar s_j$ computed within each seed (see Methods).\par

\endgroup   

\par %

\par \paragraph{Training and evaluation noise} We varied training and evaluation noise independently to distinguish noise effects on learning from those on later goal use (Fig.~\ref{fig:19}).\par \relax  \relax \begingroup\begingroup\normalsize\normalbaselines \begingroup\begin{figure}[tp]\begin{composepage}[fig:20|1|pg_fig_20_1]\begin{composerow}\begin{composecell}[121.254000000]\begin{composegraphic}\includegraphics{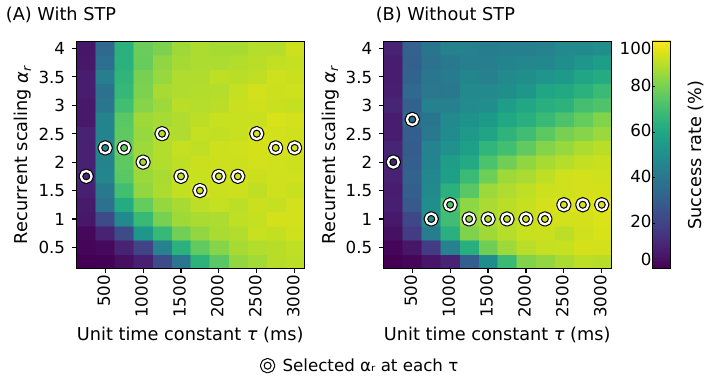}\end{composegraphic}\end{composecell}\end{composerow}\vspace{2mm}\begin{composerow}\begin{composecell}[121.254000000]\begin{composegraphic}\includegraphics{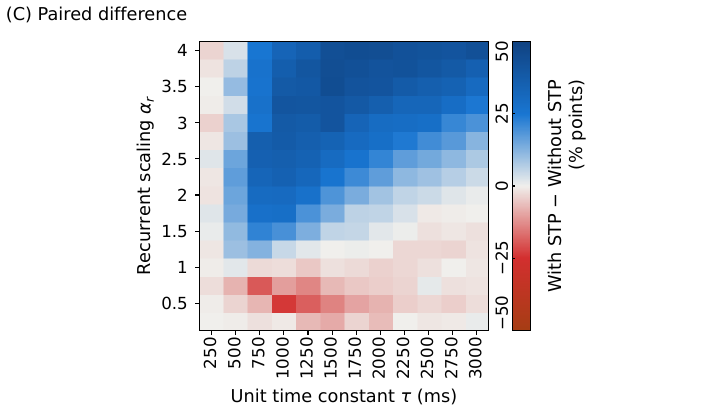}\end{composegraphic}\caption{\textbf{With the recurrent scaling selected for each condition, STP increased success at the tested unit time constants from 500 to 1{,}000\,ms.} Mean post-training success rate ($n{=}100$ networks per grid point; $\sigma{=}0.001$) as a function of the unit time constant $\tau$ (horizontal axis, 12 values) and the recurrent scaling $\alpha_r$ (vertical axis, 16 values) for the With-STP condition, the Without-STP condition, and their paired difference (With STP minus Without STP). \textbf{(A)} With STP, (B) Without STP, and (C) their paired difference. In (A) and (B) the color gives the success rate (\%, 0--100); in (C) it gives the difference in percentage points (negative values red or orange, zero light gray, positive values blue). Rings in (A) and (B) mark, for each $\tau$, the value of $\alpha_r$ selected for each condition (highest mean success rate). The main setting is $\tau{=}1{,}000$\,ms and $\alpha_r{=}3.0$. The comparisons at the same recurrent scaling and with the selected scalings, and the passive control, are shown for each $\tau$ in Fig.~\ref{fig:05}.}\label{fig:20}\end{composecell}\end{composerow}\end{composepage}\end{figure}\endgroup \section*{Unit time constant and recurrent scaling} \label{sec:38} The complete success-rate grid in Fig.~\ref{fig:20} complements the comparisons at the same and selected recurrent scalings in Fig.~\ref{fig:05}; the grid and shared settings are defined in \S\ref{sec:12}.\par\endgroup\endgroup

\par \begingroup{}  \par  \par

\section*{Noise level} \label{sec:39} The state-noise level, applied identically during training and evaluation, was varied over 8 values from $\sigma{=}0$ to $\sigma{=}0.064$ at the main unit time constant. Figure~\ref{fig:06} shows the comparison at the same recurrent scaling and the comparison with the selected recurrent scalings held fixed, and Supplementary Fig.~\ref{fig:34} shows the comparison with the recurrent scaling re-selected at each noise level. \par

\begingroup\begin{figure}[tp]\begin{composepage}[fig:21|1|pg_fig_21_1]\ComposePart{a}\begin{composerow}\begin{composecell}[121.254000000]\begin{composegraphic}\includegraphics{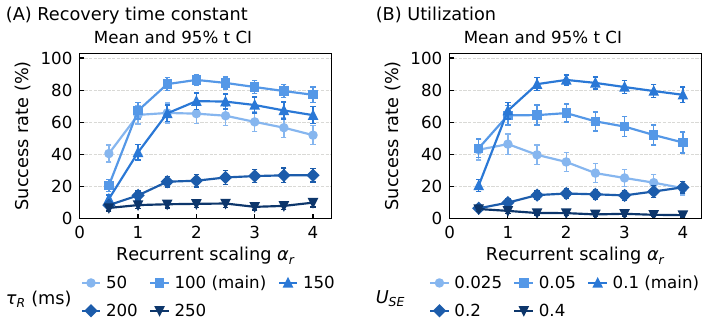}\end{composegraphic}\end{composecell}\end{composerow}\vspace{2mm}\begin{composerow}\begin{composecell}[121.254000000]\begin{composetable}\begingroup\fontsize{8}{9}\selectfont\setlength{\tabcolsep}{2pt}%
\setbox\FinalTabularBox=\hbox{%
\begin{tabular}{@{}>{\raggedright\arraybackslash}p{32mm}rr@{}}\toprule%
\multicolumn{3}{l}{\textbf{(C) Recovery: With STP minus Without STP (pp)}}\\%
\shortstack{$\tau_R$\\(ms)} & \shortstack{Difference\\(pp)} & \shortstack{95\%\\BCa CI}\\\midrule%
\rule{0pt}{4mm}50 & $16.44$ & $[7.77, 24.94]$\\ %
\rule{0pt}{4mm}100 (main) & $38.16$ & $[31.55, 44.20]$\\ %
\rule{0pt}{4mm}150 & $26.95$ & $[19.16, 34.23]$\\ %
\rule{0pt}{4mm}200 & $-17.31$ & $[-23.99, -11.08]$\\ %
\rule{0pt}{4mm}250 & $-36.48$ & $[-42.00, -31.12]$\\ %
\bottomrule\end{tabular}}%
\vbox{\hsize=121mm\parindent=0pt\box\FinalTabularBox\par%
\noindent\parbox{121mm}{Paired n = $100$.}\par%
}\endgroup%
\end{composetable}\end{composecell}\end{composerow}\vspace{2mm}\begin{composerow}\begin{composecell}[121.254000000]\begin{composetable}\begingroup\fontsize{8}{9}\selectfont\setlength{\tabcolsep}{2pt}%
\setbox\FinalTabularBox=\hbox{%
\begin{tabular}{@{}>{\raggedright\arraybackslash}p{32mm}rr@{}}\toprule%
\multicolumn{3}{l}{\textbf{(D) Utilization: With STP minus Without STP (pp)}}\\%
$U_{SE}$ & \shortstack{Difference\\(pp)} & \shortstack{95\%\\BCa CI}\\\midrule%
\rule{0pt}{4mm}0.025 & $-18.49$ & $[-25.37, -10.47]$\\ %
\rule{0pt}{4mm}0.05 & $13.52$ & $[5.13, 22.38]$\\ %
\rule{0pt}{4mm}0.1 (main) & $38.16$ & $[31.55, 44.20]$\\ %
\rule{0pt}{4mm}0.2 & $-29.36$ & $[-35.66, -22.73]$\\ %
\rule{0pt}{4mm}0.4 & $-40.81$ & $[-46.28, -35.57]$\\ %
\bottomrule\end{tabular}}%
\vbox{\hsize=121mm\parindent=0pt\box\FinalTabularBox\par%
\noindent\parbox{121mm}{Paired n = $100$.}\par%
}\endgroup%
\end{composetable}\caption{\textbf{At the same recurrent scaling, the STP success advantage depended on recovery and utilization settings.} \textbf{(A)} Mean success rate of the With-STP condition as a function of the recurrent scaling $\alpha_r$ for recovery time constants $\tau_{\mathrm{R}}{=}50$--$250$\,ms ($\tau_{\mathrm{F}}{=}500$\,ms, $U_{\mathrm{SE}}{=}0.1$). \textbf{(B)} The same for five values of the utilization parameter $U_{\mathrm{SE}}$ ($0.025$--$0.4$; main time constants). All panels: $\tau{=}1{,}000$\,ms, $\sigma{=}0.001$, $n{=}100$ networks per parameter set; error bars in (A) and (B) show the $t$-based 95\% confidence interval across networks. The rate scale in the STP update is 100\,Hz. \textbf{(C, D)} Paired differences (With STP minus Without STP at the same $\alpha_r$, in percentage points) for the recovery time constants and the utilization values at $\alpha_r{=}3.0$. Each row lists the mean paired difference with its 95\% BCa bootstrap confidence interval (2{,}000 resamples; \S\ref{sec:20}).}\ComposePartLabel{a}\label{fig:21}\end{composecell}\end{composerow}\end{composepage}\end{figure}\endgroup \begingroup\begin{figure}[tp]\begin{composepage}[fig:21|2|pg_fig_21_2]\ComposePart{b}\begin{composerow}\begin{composecell}[121.254000000]\begin{composegraphic}\includegraphics{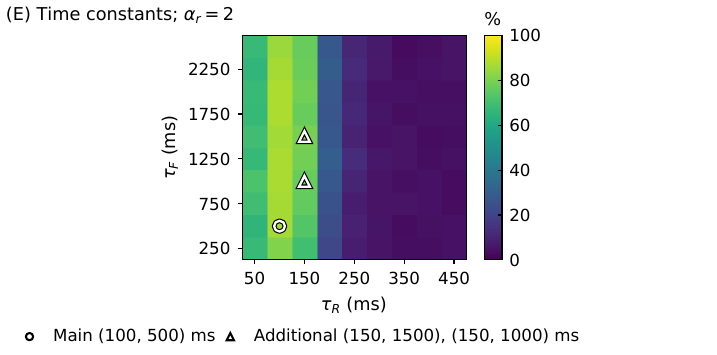}\end{composegraphic}\end{composecell}\end{composerow}\vspace{2mm}\begin{composerow}\begin{composecell}[121.254000000]\begin{composetable}\begingroup\fontsize{8}{9}\selectfont\setlength{\tabcolsep}{2pt}%
\setbox\FinalTabularBox=\hbox{%
\begin{tabular}{@{}>{\raggedright\arraybackslash}p{32mm}rr@{}}\toprule%
\multicolumn{3}{l}{\textbf{(F, G) Success rate (\%)}}\\%
Setting & $\alpha_r=2$ & $\alpha_r=3$\\\midrule%
\rule{0pt}{4mm}(F) (150, 1500) & $79.39$ & $75.19$\\ %
\rule{0pt}{4mm}(G) (150, 1000) & $76.62$ & $75.39$\\ %
\bottomrule\end{tabular}}%
\vbox{\hsize=121mm\parindent=0pt\box\FinalTabularBox\par%
\noindent\parbox{121mm}{Paired n = $100$.}\par%
}\endgroup%
\end{composetable}\end{composecell}\end{composerow}\vspace{2mm}\begin{composerow}\begin{composecell}[121.254000000]\begin{composetable}\begingroup\fontsize{8}{9}\selectfont\setlength{\tabcolsep}{2pt}%
\setbox\FinalTabularBox=\hbox{%
\begin{tabular}{@{}>{\raggedright\arraybackslash}p{32mm}rr@{}}\toprule%
\multicolumn{3}{l}{\textbf{(H) $\alpha_r{=}2$: With STP $-$ Without STP (pp)}}\\%
\shortstack{$(\tau_R,\tau_F)$\\(ms)} & \shortstack{Difference\\(pp)} & \shortstack{95\%\\BCa CI}\\\midrule%
\rule{0pt}{4mm}(150, 1500) & $23.22$ & $[17.50, 28.91]$\\ %
\rule{0pt}{4mm}(150, 1000) & $20.45$ & $[13.98, 26.33]$\\ %
\rule{0pt}{4mm}(100, 500) main & $30.16$ & $[24.55, 35.85]$\\ %
\bottomrule\end{tabular}}%
\vbox{\hsize=121mm\parindent=0pt\box\FinalTabularBox\par%
\noindent\parbox{121mm}{Paired n = $100$.}\par%
}\endgroup%
\end{composetable}\end{composecell}\end{composerow}\vspace{2mm}\begin{composerow}\begin{composecell}[121.254000000]\begin{composetable}\begingroup\fontsize{8}{9}\selectfont\setlength{\tabcolsep}{2pt}%
\setbox\FinalTabularBox=\hbox{%
\begin{tabular}{@{}>{\raggedright\arraybackslash}p{32mm}rr@{}}\toprule%
\multicolumn{3}{l}{\textbf{(I) $\alpha_r{=}3$: With STP $-$ Without STP (pp)}}\\%
\shortstack{$(\tau_R,\tau_F)$\\(ms)} & \shortstack{Difference\\(pp)} & \shortstack{95\%\\BCa CI}\\\midrule%
\rule{0pt}{4mm}(150, 1500) & $31.45$ & $[24.10, 38.83]$\\ %
\rule{0pt}{4mm}(150, 1000) & $31.65$ & $[24.91, 38.77]$\\ %
\rule{0pt}{4mm}(100, 500) main & $38.16$ & $[31.55, 44.20]$\\ %
\bottomrule\end{tabular}}%
\vbox{\hsize=121mm\parindent=0pt\box\FinalTabularBox\par%
\noindent\parbox{121mm}{Paired n = $100$.}\par%
}\endgroup%
\end{composetable}\caption{\textbf{(E)} Map of the mean success rate over $(\tau_{\mathrm{R}},\tau_{\mathrm{F}})$ for the value of $\alpha_r$ that is not shown in Fig.~\ref{fig:13} ($\alpha_r{=}2$); a circle marks the main pair $(100,500)$\,ms and triangles mark the additional pairs $(150,1500)$ and $(150,1000)$\,ms. \textbf{(F)} Mean success rates for the additional time-constant pair $(\tau_{\mathrm{R}},\tau_{\mathrm{F}}){=}(150,1500)$\,ms at $\alpha_r{=}2$ and $3$, and  \textbf{(G)} for $(150,1000)$\,ms, both with $U_{\mathrm{SE}}{=}0.1$; the values are group means. Panels (H) and (I) list the differences with their intervals; the main comparison at $\alpha_r{=}3.0$ is given in Table~\ref{tab:04} and Fig.~\ref{fig:03}. \textbf{(H)} Paired differences for the additional pairs and the main pair at $\alpha_r{=}2$. \textbf{(I)} Paired differences for the additional pairs and the main pair at $\alpha_r{=}3.0$.}\ComposePartLabel{b}\end{composecell}\end{composerow}\end{composepage}\end{figure}\endgroup \begingroup\begin{figure}[tp]\begin{composepage}[fig:22|1|pg_fig_22_1]\begin{composerow}\begin{composecell}[121.254000000]\begin{composegraphic}\includegraphics{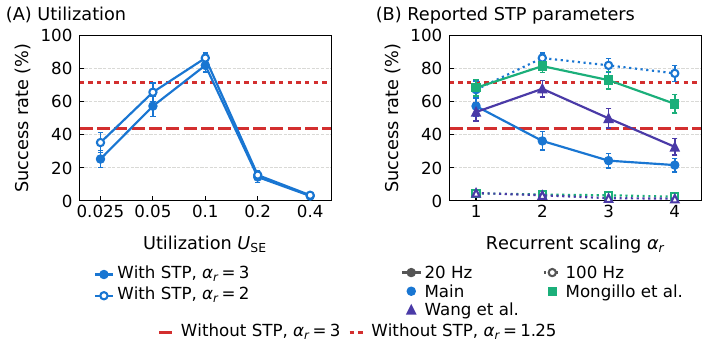}\end{composegraphic}\end{composecell}\end{composerow}\vspace{2mm}\begin{composerow}\begin{composecell}[121.254000000]\begin{composetable}\begingroup\fontsize{8}{9}\selectfont\setlength{\tabcolsep}{1pt}
\begin{tabular}{@{}l@{\hspace{2pt}}r@{\hspace{5pt}}r@{\hspace{2pt}}r@{\hspace{2.5pt}}r@{\hspace{4pt}}r@{\hspace{5pt}}r@{}}\toprule
\shortstack[l]{STP\\parameters} & \shortstack{With STP\\Mean $\pm$ SD (\%)} & \shortstack{Without STP\\Mean $\pm$ SD (\%)} & \shortstack{Difference\\(pp)} & 95\% BCa CI & $d_z$ & $p_t$\\\midrule
\multicolumn{7}{l}{\rule{0pt}{4.5mm}(C) Success rate (\%), same $\alpha_r=3$}\\ %
\rule{0pt}{4.5mm}Mongillo et al. & $\underline{\mathbf{72.96}}\,\pm\,26.33$ & $43.74\,\pm\,26.63$ & $29.22$ & $[22.00, 36.08]$ & $0.80$ & $2.0\times10^{-12}$\\ %
\rule{0pt}{4.5mm}Wang et al. & $\underline{\mathbf{49.86}}\,\pm\,29.15$ & $43.74\,\pm\,26.63$ & $6.12$ & $[-1.42, 13.50]$ & $0.16$ & $0.10$\\ %
\rule{0pt}{4.5mm}Main & $\underline{\mathbf{81.90}}\,\pm\,20.61$ & $43.74\,\pm\,26.63$ & $38.16$ & $[31.55, 44.20]$ & $1.14$ & $1.2\times10^{-19}$\\ %
\midrule
\multicolumn{7}{l}{\rule{0pt}{4.5mm}(D) Success rate (\%), same $\alpha_r=2$}\\ %
\rule{0pt}{4.5mm}Mongillo et al. & $\underline{\mathbf{81.46}}\,\pm\,19.24$ & $56.17\,\pm\,25.12$ & $25.29$ & $[19.31, 31.15]$ & $0.84$ & $3.8\times10^{-13}$\\ %
\rule{0pt}{4.5mm}Wang et al. & $\underline{\mathbf{67.72}}\,\pm\,24.75$ & $56.17\,\pm\,25.12$ & $11.55$ & $[5.35, 18.09]$ & $0.35$ & $7.2\times10^{-4}$\\ %
\rule{0pt}{4.5mm}Main & $\underline{\mathbf{86.33}}\,\pm\,15.59$ & $56.17\,\pm\,25.12$ & $30.16$ & $[24.55, 35.85]$ & $1.04$ & $1.9\times10^{-17}$\\ %
\bottomrule\end{tabular}\endgroup
\par\vspace{2mm}
\begingroup\fontsize{8}{9}\selectfont\noindent\parbox[t]{345pt}{\(n=100\) paired networks; 2{,}000 BCa resamples.\\
Difference: With STP minus Without STP; \(p_t\): paired \(t\)-test.}\endgroup
\end{composetable}\caption{\textbf{At the same $\alpha_r=2$ and 20\,Hz, both reported STP settings exceeded Without STP.} All panels: $\tau{=}1{,}000$\,ms, $\sigma{=}0.001$, mean over $n{=}100$ seeds per parameter set. \textbf{(A)} Success rate as a function of the utilization parameter $U_{\mathrm{SE}}$ (0.025--0.4, logarithmic axis) with $(\tau_{\mathrm{R}}, \tau_{\mathrm{F}}){=}(100, 500)$\,ms, at $\alpha_r{=}3.0$ and at $\alpha_r{=}2$. Red dashed lines show the Without-STP condition at $\alpha_r{=}3.0$ (43.7\%) and at its selected value $\alpha_r{=}1.25$ (71.6\%). \textbf{(B)} STP parameters taken from \citet{mongillo2008synaptic} and \citet{wang2006heterogeneity} and the main parameter set, evaluated at $\alpha_r{=}1$--$4$ with a normalized activity of 1 mapped to 100\,Hz or to 20\,Hz in the STP update; red horizontal lines as in (A). Error bars in (A) and (B) indicate the $t$-based 95\% confidence interval across seeds. \textbf{(C, D)} Paired success-rate differences, With STP minus Without STP, at the same recurrent scaling, $\alpha_r{=}3.0$ in (C) and $\alpha_r{=}2$ in (D), for the reported parameter sets at 20\,Hz and the main parameter set at 100\,Hz. Each row lists the mean paired difference in percentage points with its 95\% BCa bootstrap confidence interval over 100 paired networks (2{,}000 resamples; \S\ref{sec:20}). Panels (C) and (D) list the difference [95\% BCa CI] in the same row order. Bold underlined values mark the larger mean in each row.}\label{fig:22}\end{composecell}\end{composerow}\end{composepage}\end{figure}\endgroup \section*{Sensitivity to the STP parameters} \label{sec:40} We varied STP recovery and utilization across recurrent scalings to locate settings supporting goal use under state noise (\S\ref{sec:18}; Fig.~\ref{fig:21}).   \begingroup  \leavevmode{}  \endgroup       To relate parameter sensitivity to reported synaptic kinetics, we compared the reported parameter sets at two rate scales (Fig.~\ref{fig:22}).

\begingroup\begin{figure}[tp]\begin{composepage}[fig:23|1|pg_fig_23_1]\ComposePart{a}\begin{composerow}\begin{composecell}[121.254000000]\begin{composegraphic}\includegraphics{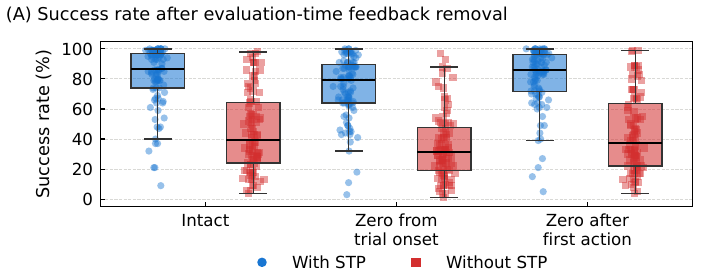}\end{composegraphic}\end{composecell}\end{composerow}\vspace{2mm}\begin{composerow}\begin{composecell}[121.254000000]\begin{composetable}\begingroup\fontsize{8}{9}\selectfont\setlength{\tabcolsep}{2pt}%
\setbox\FinalTabularBox=\hbox{%
\begin{tabular}{@{}>{\raggedright\arraybackslash}p{24mm}rrr@{}}\toprule%
\multicolumn{4}{l}{\textbf{(B) Paired success-rate statistics}}\\%
Setting & \shortstack{With STP\\mean $\pm$ SD (\%)} & \shortstack{Without STP\\mean $\pm$ SD (\%)} & \shortstack{Difference (pp)\\{[95\% BCa CI]}}\\\midrule%
\rule{0pt}{4mm}Intact re-evaluation & $81.94\pm20.33$ & $44.81\pm26.18$ & \begin{tabular}[t]{@{}c@{}}$37.13$\\$[30.06, 43.29]$\end{tabular}\\ %
\rule{0pt}{4mm}Zero from trial onset & $74.03\pm20.67$ & $35.10\pm22.09$ & \begin{tabular}[t]{@{}c@{}}$38.93$\\$[32.73, 44.72]$\end{tabular}\\ %
\rule{0pt}{4mm}Zero after first action & $80.73\pm19.89$ & $43.61\pm25.71$ & \begin{tabular}[t]{@{}c@{}}$37.12$\\$[30.44, 43.08]$\end{tabular}\\ %
\bottomrule\end{tabular}}%
\vbox{\hsize=121mm\parindent=0pt\box\FinalTabularBox\par%
\noindent\parbox{121mm}{Paired n = $100$; Resamples = $2{,}000$.}\par%
}\endgroup%
\end{composetable}\caption{\textbf{With STP retained higher success when Q-value feedback was removed from trial onset.} Post-training success rate of the With-STP and Without-STP networks when the Q-value feedback pathway $W^{\mathrm{back}}$ was left intact, set to zero from trial onset, or set to zero after the first action execution during evaluation, together with the scatter ratio, goal-decoding accuracy, and action-value difference $D_Q$ at the second and third GO opportunities ($n{=}100$ paired seeds). \textbf{(A)} Success rate; boxes show the median and interquartile range, and points are individual networks. Blue circles denote With STP and red squares denote Without STP; filled markers denote the comparison at the same recurrent scaling ($\alpha_r{=}3.0$), which all panels show. All values come from a separate set of 100 evaluation episodes of the trained networks, with the feedback kept in the Intact condition; the original evaluation of the same networks is given in Table~\ref{tab:04} and shown in Fig.~\ref{fig:03}. Whiskers: 1.5 IQR. \textbf{(B)} For the Intact, zero-from-trial-onset and zero-after-first-action conditions, the table lists the success rate of each condition (mean $\pm$ SD across networks, \%) and the paired difference, With STP minus Without STP, in percentage points with its 95\% BCa bootstrap confidence interval (2{,}000 resamples; \S\ref{sec:20}).}\ComposePartLabel{a}\label{fig:23}\end{composecell}\end{composerow}\end{composepage}\end{figure}\endgroup \begingroup\begin{figure}[tp]\begin{composepage}[fig:23|2|pg_fig_23_2]\ComposePart{b}\begin{composerow}\begin{composecell}[121.254000000]\begin{composetable}\begingroup\fontsize{8}{9}\selectfont\setlength{\tabcolsep}{2pt}%
\setbox\FinalTabularBox=\hbox{%
\begin{tabular}{@{}>{\raggedright\arraybackslash}p{24mm}rrrr@{}}\toprule%
\multicolumn{5}{l}{\textbf{(C) Scatter ratio: median [Q1, Q3]}}\\%
Setting & \shortstack{GO2\\With STP} & \shortstack{GO2\\Without STP} & \shortstack{GO3\\With STP} & \shortstack{GO3\\Without STP}\\\midrule%
\rule{0pt}{4mm}Intact & \begin{tabular}[t]{@{}c@{}}$23.72$\\$[10.87, 57.74]$\end{tabular} & \begin{tabular}[t]{@{}c@{}}$2.62$\\$[1.33, 4.34]$\end{tabular} & \begin{tabular}[t]{@{}c@{}}$19.08$\\$[8.94, 38.32]$\end{tabular} & \begin{tabular}[t]{@{}c@{}}$1.81$\\$[1.15, 3.48]$\end{tabular}\\ %
\rule{0pt}{4mm}Zero from trial onset & \begin{tabular}[t]{@{}c@{}}$20.13$\\$[9.29, 60.56]$\end{tabular} & \begin{tabular}[t]{@{}c@{}}$2.10$\\$[1.12, 3.54]$\end{tabular} & \begin{tabular}[t]{@{}c@{}}$13.22$\\$[6.39, 28.90]$\end{tabular} & \begin{tabular}[t]{@{}c@{}}$1.51$\\$[0.83, 2.45]$\end{tabular}\\ %
\rule{0pt}{4mm}Zero after first action & \begin{tabular}[t]{@{}c@{}}$24.78$\\$[11.56, 74.01]$\end{tabular} & \begin{tabular}[t]{@{}c@{}}$2.73$\\$[1.31, 4.34]$\end{tabular} & \begin{tabular}[t]{@{}c@{}}$18.59$\\$[8.54, 46.15]$\end{tabular} & \begin{tabular}[t]{@{}c@{}}$2.00$\\$[1.06, 3.38]$\end{tabular}\\ %
\bottomrule\end{tabular}}%
\vbox{\hsize=121mm\parindent=0pt\box\FinalTabularBox\par%
\noindent\parbox{121mm}{Paired n = $100$.}\par%
}\endgroup%
\end{composetable}\end{composecell}\end{composerow}\vspace{2mm}\begin{composerow}\begin{composecell}[121.254000000]\begin{composetable}\begingroup\fontsize{8}{9}\selectfont\setlength{\tabcolsep}{2pt}%
\setbox\FinalTabularBox=\hbox{%
\begin{tabular}{@{}>{\raggedright\arraybackslash}p{24mm}rrrr@{}}\toprule%
\multicolumn{5}{l}{\textbf{(D) Goal-decoding accuracy (\%): mean [$t$-based 95\% CI]}}\\%
Setting & \shortstack{GO2\\With STP} & \shortstack{GO2\\Without STP} & \shortstack{GO3\\With STP} & \shortstack{GO3\\Without STP}\\\midrule%
\rule{0pt}{4mm}Intact & \begin{tabular}[t]{@{}c@{}}$99.39$\\$[99.17, 99.61]$\end{tabular} & \begin{tabular}[t]{@{}c@{}}$93.50$\\$[92.49, 94.50]$\end{tabular} & \begin{tabular}[t]{@{}c@{}}$98.61$\\$[98.19, 99.03]$\end{tabular} & \begin{tabular}[t]{@{}c@{}}$88.81$\\$[87.22, 90.40]$\end{tabular}\\ %
\rule{0pt}{4mm}Zero from trial onset & \begin{tabular}[t]{@{}c@{}}$99.27$\\$[99.04, 99.51]$\end{tabular} & \begin{tabular}[t]{@{}c@{}}$92.69$\\$[91.42, 93.95]$\end{tabular} & \begin{tabular}[t]{@{}c@{}}$98.34$\\$[97.91, 98.77]$\end{tabular} & \begin{tabular}[t]{@{}c@{}}$87.10$\\$[85.15, 89.04]$\end{tabular}\\ %
\rule{0pt}{4mm}Zero after first action & \begin{tabular}[t]{@{}c@{}}$99.36$\\$[99.12, 99.60]$\end{tabular} & \begin{tabular}[t]{@{}c@{}}$93.65$\\$[92.61, 94.68]$\end{tabular} & \begin{tabular}[t]{@{}c@{}}$98.68$\\$[98.18, 99.17]$\end{tabular} & \begin{tabular}[t]{@{}c@{}}$89.03$\\$[87.45, 90.60]$\end{tabular}\\ %
\bottomrule\end{tabular}}%
\vbox{\hsize=121mm\parindent=0pt\box\FinalTabularBox\par%
\noindent\parbox{121mm}{Paired n = $100$.}\par%
}\endgroup%
\end{composetable}\end{composecell}\end{composerow}\vspace{2mm}\begin{composerow}\begin{composecell}[121.254000000]\begin{composetable}\begingroup\fontsize{8}{9}\selectfont\setlength{\tabcolsep}{2pt}%
\setbox\FinalTabularBox=\hbox{%
\begin{tabular}{@{}>{\raggedright\arraybackslash}p{24mm}rrrr@{}}\toprule%
\multicolumn{5}{l}{\textbf{(E) Action-value difference: mean [$t$-based 95\% CI]}}\\%
Setting & \shortstack{GO2\\With STP} & \shortstack{GO2\\Without STP} & \shortstack{GO3\\With STP} & \shortstack{GO3\\Without STP}\\\midrule%
\rule{0pt}{4mm}Intact & \begin{tabular}[t]{@{}c@{}}$0.0652$\\$[0.0623, 0.0682]$\end{tabular} & \begin{tabular}[t]{@{}c@{}}$0.0424$\\$[0.0378, 0.0470]$\end{tabular} & \begin{tabular}[t]{@{}c@{}}$0.0263$\\$[0.0230, 0.0296]$\end{tabular} & \begin{tabular}[t]{@{}c@{}}$0.0008$\\$[-0.0038, 0.0054]$\end{tabular}\\ %
\rule{0pt}{4mm}Zero from trial onset & \begin{tabular}[t]{@{}c@{}}$0.0607$\\$[0.0575, 0.0639]$\end{tabular} & \begin{tabular}[t]{@{}c@{}}$0.0319$\\$[0.0271, 0.0367]$\end{tabular} & \begin{tabular}[t]{@{}c@{}}$0.0222$\\$[0.0187, 0.0257]$\end{tabular} & \begin{tabular}[t]{@{}c@{}}$-0.0047$\\$[-0.0089, -0.0004]$\end{tabular}\\ %
\rule{0pt}{4mm}Zero after first action & \begin{tabular}[t]{@{}c@{}}$0.0659$\\$[0.0628, 0.0689]$\end{tabular} & \begin{tabular}[t]{@{}c@{}}$0.0429$\\$[0.0381, 0.0476]$\end{tabular} & \begin{tabular}[t]{@{}c@{}}$0.0250$\\$[0.0216, 0.0285]$\end{tabular} & \begin{tabular}[t]{@{}c@{}}$-0.0002$\\$[-0.0048, 0.0044]$\end{tabular}\\ %
\bottomrule\end{tabular}}%
\vbox{\hsize=121mm\parindent=0pt\box\FinalTabularBox\par%
\noindent\parbox{121mm}{Paired n = $100$.}\par%
}\endgroup%
\end{composetable}\caption{\textbf{(C)} Scatter ratio (median and interquartile range across networks). Within each GO window, scatter ratio and decoding accuracy were averaged over $t{=}6$--$7$\,s (GO2) or $t{=}8$--$9$\,s (GO3); $D_Q$ was evaluated at the corresponding action time (\S\ref{sec:15}). Panels (C)--(E) show GO2 and GO3 in separate columns. \textbf{(D)} Goal-decoding accuracy and (E) action-value difference $D_Q$ (mean and $t$-based 95\% confidence interval across networks).}\ComposePartLabel{b}\end{composecell}\end{composerow}\end{composepage}\end{figure}\endgroup \begingroup\begin{figure}[tp]\begin{composepage}[fig:24|1|pg_fig_24_1]\ComposePart{a}\begin{composerow}\begin{composecell}[121.254000000]\centerline{\includegraphics{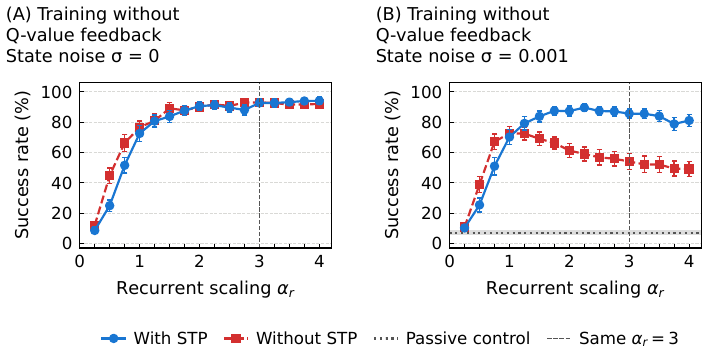}}\caption{\textbf{At the same recurrent scaling under state noise, With STP retained its success advantage in learning without Q-value feedback.} All panels: $\tau{=}1{,}000$\,ms, $n{=}100$ seeds per condition. \textbf{(A)} Networks trained without the Q-value feedback pathway ($\alpha_b{=}0$): mean post-training success rate as a function of the recurrent scaling $\alpha_r$ (16 grid values, 0.25--4.0; error bars indicate the $t$-based 95\% confidence interval across seeds) at $\sigma{=}0$. Blue: With STP; red: Without STP. The vertical dashed line indicates the value used in the comparison at the same recurrent scaling ($\alpha_r{=}3.0$). \textbf{(B)} The same at $\sigma{=}0.001$. The gray horizontal line in (B) shows the passive control (Without STP with $\alpha_r{=}0.01$ and without Q-value feedback).}\ComposePartLabel{a}\label{fig:24}\end{composecell}\end{composerow}\end{composepage}\end{figure}\endgroup \begingroup\begin{figure}[tp]\begin{composepage}[fig:24|2|pg_fig_24_2]\ComposePart{b}\begin{composerow}\begin{composecell}[121.254000000]\begin{composetable}\begingroup\fontsize{8}{9}\selectfont\setlength{\tabcolsep}{2pt}%
\setbox\FinalTabularBox=\hbox{%
\begin{tabular}{@{}llllll@{}}\toprule%
\multicolumn{6}{l}{\textbf{(C, D, E) Training without Q-value feedback}}\\%
State noise $\sigma$ & \shortstack[l]{$\alpha_r$\\With} & \shortstack[l]{$\alpha_r$\\Without} & \shortstack[l]{With STP\\mean $\pm$ SD (\%)} & \shortstack[l]{Without STP\\mean $\pm$ SD (\%)} & \shortstack[l]{Difference [95\%\\BCa CI] (pp)}\\\midrule%
\multicolumn{6}{l}{\textbf{(C) Same recurrent scaling}}\\%
0 & 3.00 & 3.00 & $92.66\pm12.66$ & $93.06\pm10.87$ & $-0.40\ [-3.73, 2.86]$\\%
0.001 & 3.00 & 3.00 & $85.58\pm14.94$ & $54.04\pm25.84$ & $31.54\ [26.08, 36.75]$\\%
0.002 & 3.00 & 3.00 & $78.07\pm19.05$ & $31.44\pm20.69$ & $46.63\ [41.17, 51.45]$\\%
0.004 & 3.00 & 3.00 & $60.82\pm24.31$ & $15.80\pm12.31$ & $45.02\ [40.03, 49.98]$\\\midrule%
\multicolumn{6}{l}{\textbf{(D) Recurrent scaling selected at $\sigma=0.001$ and held fixed}}\\%
0 & 2.25 & 1.00 & $91.20\pm13.18$ & $76.58\pm21.43$ & $14.62\ [9.72, 19.89]$\\%
0.001 & 2.25 & 1.00 & $89.56\pm12.37$ & $72.58\pm22.47$ & $16.98\ [12.35, 22.18]$\\%
0.002 & 2.25 & 1.00 & $81.27\pm17.62$ & $68.23\pm21.33$ & $13.04\ [8.38, 17.76]$\\%
0.004 & 2.25 & 1.00 & $69.34\pm22.02$ & $51.72\pm22.25$ & $17.62\ [12.29, 22.71]$\\\midrule%
\multicolumn{6}{l}{\textbf{(E) Recurrent scaling re-selected at $\sigma=0$}}\\%
0 & 4.00 & 3.00 & $94.04\pm12.04$ & $93.06\pm10.87$ & $0.98\ [-2.20, 3.92]$\\%
\bottomrule\end{tabular}}%
\vbox{\hsize=121mm\parindent=0pt\box\FinalTabularBox\par%
\noindent\parbox{121mm}{Paired n = $100$.}\par%
}\endgroup%
\end{composetable}\caption{\textbf{(C--E)} Success rates (mean $\pm$ SD across networks) and paired differences (With STP minus Without STP, with 95\% BCa bootstrap confidence intervals) for the networks trained without the Q-value feedback pathway, at $\sigma{=}0$, $0.001$, $0.002$ and $0.004$ during training and evaluation. In (C), $\alpha_r{=}3.0$ in both conditions. In (D), the recurrent scalings selected among these networks at $\sigma{=}0.001$ ($\alpha_r{=}2.25$ with STP and $\alpha_r{=}1$ without STP) are held fixed across the noise levels. In (E), the scalings are re-selected among these networks at $\sigma{=}0$ ($\alpha_r{=}4$ with STP and $\alpha_r{=}3$ without STP).}\ComposePartLabel{b}\end{composecell}\end{composerow}\end{composepage}\end{figure}\endgroup \begingroup\begin{figure}[tp]\begin{composepage}[fig:24|3|pg_fig_24_3]\ComposePart{c}\begin{composerow}\begin{composecell}[121.254000000]\centerline{\includegraphics{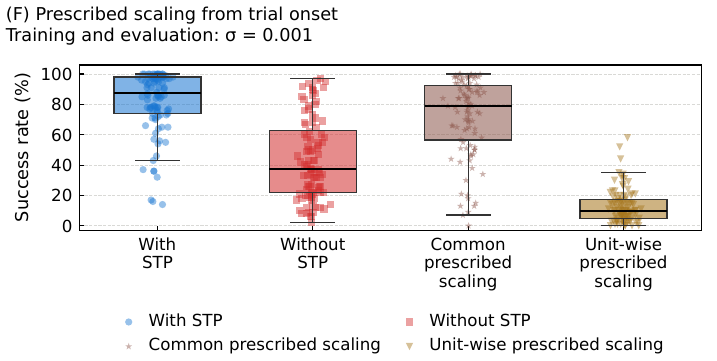}}\end{composecell}\end{composerow}\vspace{2mm}\begin{composerow}\begin{composecell}[121.254000000]\begin{composetable}\begingroup\fontsize{8}{8.6}\selectfont\setlength{\tabcolsep}{2pt}%
\setbox\FinalTabularBox=\hbox{%
\begin{tabular}{@{}>{\raggedright\arraybackslash}p{56mm}@{\hspace{3mm}}r@{\hspace{3mm}}r@{}}\toprule%
\multicolumn{3}{l}{\textbf{(G) Prescribed scaling from trial onset}}\\%
Setting & \shortstack{Control mean\\$\pm$ SD (\%)} & \shortstack{With STP minus\\control (pp)\\{[95\% BCa CI]}}\\\midrule%
No prescribed scaling & $43.74\pm26.63$ & $38.16\;[31.55, 44.20]$\\ %
Common prescribed scaling & $71.67\pm25.86$ & $10.23\;[4.21, 16.12]$\\ %
Unit-wise prescribed scaling & $12.73\pm11.10$ & $69.17\;[63.96, 73.14]$\\ %
\midrule%
\multicolumn{3}{l}{\rule{0pt}{3mm}\textbf{(H) Prescribed scaling minus Without STP}}\\%
Setting & Difference (pp) & 95\% BCa CI\\\midrule%
Common prescribed minus Without STP & $27.93$ & $[20.60, 35.37]$\\ %
Unit-wise prescribed minus Without STP & $-31.01$ & $[-36.36, -25.54]$\\ %
\bottomrule\end{tabular}}%
\vbox{\hsize=121mm\parindent=0pt\box\FinalTabularBox\par%
\noindent\parbox{121mm}{Paired n = $100$.}\par%
}\endgroup%
\end{composetable}\caption{\textbf{(F)} Success rate at $\sigma{=}0.001$ and $\alpha_r{=}3.0$ for the With-STP and Without-STP conditions and for Without-STP networks trained with a goal-independent prescribed time-varying scaling taken from the With-STP networks (\S\ref{sec:12}), as Common prescribed scaling or Unit-wise prescribed scaling. Boxes show the median and interquartile range; each point represents one of the $n{=}100$ seeds. \textbf{(G, H)} Comparisons with prescribed time-varying scaling at $\alpha_r{=}3.0$ and $\sigma{=}0.001$: (G) With STP minus each control with prescribed time-varying scaling, and (H) each control with prescribed time-varying scaling minus Without STP. In (G, H), each entry lists the mean paired difference with its 95\% BCa bootstrap confidence interval (2{,}000 resamples; \S\ref{sec:20}). The original Without-STP reference is the original evaluation of the trained networks. The With-STP and No-prescribed-scaling references (mean $\pm$ SD) are given in Table~\ref{tab:04}. Comparisons with the recurrent scaling selected for each of these controls are shown in Supplementary Figs.~\ref{fig:33:partb}E and~\ref{fig:37:partc}D, E.}\ComposePartLabel{c}\end{composecell}\end{composerow}\end{composepage}\end{figure}\endgroup \begingroup\begin{figure}[tp]\begin{composepage}[fig:25|1|pg_fig_25_1]\begin{composerow}\begin{composecell}[121.254000000]\begin{composetable}\begingroup\fontsize{8}{8.6}\selectfont\setlength{\tabcolsep}{2pt}\setlength{\aboverulesep}{0.15ex}\setlength{\belowrulesep}{0.25ex}\renewcommand{\arraystretch}{0.95}%
\setbox\FinalTabularBox=\hbox{%
\begin{tabular}{@{}>{\raggedright\arraybackslash}p{15mm}rrr@{}}\toprule%
\multicolumn{4}{l}{\textbf{(A, C) Multiplicative noise}}\\%
Level & \shortstack{(A) With STP (\%)\\mean $\pm$ SD\\{[$t$-based 95\% CI]}} & \shortstack{(A) Without STP (\%)\\mean $\pm$ SD\\{[$t$-based 95\% CI]}} & \shortstack{(C) Difference (pp)\\{[95\% BCa CI]}}\\\midrule%
0 & \begin{tabular}[t]{@{}c@{}}$82.19\pm20.06$\\$[78.21, 86.17]$\end{tabular} & \begin{tabular}[t]{@{}c@{}}$44.05\pm26.38$\\$[38.82, 49.28]$\end{tabular} & $38.14\;[31.50, 44.26]$\\ %
0.05 & \begin{tabular}[t]{@{}c@{}}$75.18\pm18.47$\\$[71.52, 78.84]$\end{tabular} & \begin{tabular}[t]{@{}c@{}}$40.31\pm23.59$\\$[35.63, 44.99]$\end{tabular} & $34.87\;[28.70, 40.34]$\\ %
0.25 & \begin{tabular}[t]{@{}c@{}}$44.28\pm13.20$\\$[41.66, 46.90]$\end{tabular} & \begin{tabular}[t]{@{}c@{}}$20.64\pm10.21$\\$[18.61, 22.67]$\end{tabular} & $23.64\;[20.73, 26.45]$\\ %
\midrule%
\multicolumn{4}{l}{\textbf{(B, D) Input noise}}\\%
Level & \shortstack{(B) With STP (\%)\\mean $\pm$ SD\\{[$t$-based 95\% CI]}} & \shortstack{(B) Without STP (\%)\\mean $\pm$ SD\\{[$t$-based 95\% CI]}} & \shortstack{(D) Difference (pp)\\{[95\% BCa CI]}}\\\midrule%
0 & \begin{tabular}[t]{@{}c@{}}$82.19\pm20.06$\\$[78.21, 86.17]$\end{tabular} & \begin{tabular}[t]{@{}c@{}}$44.05\pm26.38$\\$[38.82, 49.28]$\end{tabular} & $38.14\;[31.50, 44.26]$\\ %
0.1 & \begin{tabular}[t]{@{}c@{}}$28.73\pm10.06$\\$[26.73, 30.73]$\end{tabular} & \begin{tabular}[t]{@{}c@{}}$13.70\pm5.43$\\$[12.62, 14.78]$\end{tabular} & $15.03\;[12.92, 17.19]$\\ %
0.5 & \begin{tabular}[t]{@{}c@{}}$10.31\pm3.95$\\$[9.53, 11.09]$\end{tabular} & \begin{tabular}[t]{@{}c@{}}$9.55\pm4.22$\\$[8.71, 10.39]$\end{tabular} & $0.76\;[-0.26, 1.78]$\\ %
\bottomrule\end{tabular}}%
\vbox{\hsize=121mm\parindent=0pt\box\FinalTabularBox\par%
\noindent\parbox{121mm}{Paired n = $100$; Resamples = $2{,}000$.}\par%
}\endgroup%
\end{composetable}\caption{\textbf{The With-STP success advantage extended to multiplicative noise and to the lower tested input-noise level.} Post-training success rate of the With-STP and Without-STP conditions and their paired difference under multiplicative noise and under input noise (\S\ref{sec:10}), as a function of the noise level ($n{=}100$ paired seeds; $\tau{=}1{,}000$\,ms). \textbf{(A)} Multiplicative noise and (B) input noise: success rate in the comparison at the same recurrent scaling ($\alpha_r{=}3.0$ in both conditions; mean and $t$-based 95\% confidence interval across networks). Level 0 is the evaluation with the additive state noise alone ($\sigma{=}0.001$). All values come from a separate set of 100 evaluation episodes of the trained networks (\S\ref{sec:10}). Panels (A) and (B) list, for each noise level, the success rate (mean $\pm$ SD across networks, \%) of the two conditions and the paired difference [95\% BCa CI]; the level-0 values coincide with the combination $\sigma_{\mathrm{train}}{=}\sigma_{\mathrm{eval}}{=}0.001$ of Supplementary Fig.~\ref{fig:19:partc}J and are listed here. \textbf{(C, D)} Paired difference (With STP minus Without STP) and its 95\% BCa bootstrap confidence interval (2{,}000 resamples).}\label{fig:25}\end{composecell}\end{composerow}\end{composepage}\end{figure}\endgroup \section*{Additional controls} \label{sec:41} We tested whether goal-independent prescribed time courses reproduce the With-STP advantage, using the separately trained controls defined in \S\ref{sec:12}. At $\alpha_r{=}3.0$ and $\sigma{=}0.001$, Common prescribed scaling exceeded the original Without-STP condition by 27.9 percentage points (95\% BCa CI [20.6, 35.4]), and With STP exceeded Common prescribed scaling by 10.2 percentage points (95\% BCa CI [4.2, 16.1]; Fig.~\ref{fig:24:partc}F--H). With $\alpha_r$ selected for each control from the same 16 values ($\alpha_r{=}2.0$ and 1.75), Common and Unit-wise prescribed scaling reached 75.5\% and 78.5\% success, and With STP exceeded them by 10.9 (95\% BCa CI [5.8, 16.8]) and 7.8 percentage points (95\% BCa CI [3.9, 12.5]; Figs.~\ref{fig:33:partb}E and~\ref{fig:37:partc}D, E).

\begingroup\setlength{\emergencystretch}{1em}We removed Q-value feedback from trial onset during evaluation to test whether the later differences between the trained With-STP and Without-STP networks require that pathway (Fig.~\ref{fig:23}). \endgroup \begingroup\begingroup\normalsize\normalbaselines \ComposeKeep{\parindent=15pt \tolerance=200 \emergencystretch=10pt \clubpenalty=150 \parskip=0pt plus 1pt \hsize=\textwidth} We trained networks without Q-value feedback to test whether the success advantage under state noise requires that pathway (Figs.~\ref{fig:24:parta}A, B and~\ref{fig:24:partb}C--E).\par\endgroup\endgroup  \par \begingroup\par\endgroup \begingroup\begingroup\normalsize\normalbaselines \ComposeKeep{\parindent=15pt \tolerance=200 \emergencystretch=10pt \clubpenalty=150 \parskip=0pt plus 1pt \hsize=\textwidth} We evaluated the trained networks under multiplicative and input noise while keeping additive state noise at its training level, to test whether the success-rate advantage extended to these noise sources (Fig.~\ref{fig:25}).\par\endgroup\endgroup  

\par  \par

\par  \par \par  \par \par  \par

  \par  \begingroup\begingroup\normalsize\normalbaselines \ComposeKeep{\parindent=15pt \tolerance=200 \emergencystretch=10pt \clubpenalty=150 \parskip=0pt plus 1pt \hsize=\textwidth} \begingroup\begin{figure}[tp]\begin{composepage}[fig:26|1|pg_fig_26_1]\ComposePart{a}\begin{composerow}\begin{composecell}[121.254000000]\begin{composegraphic}\includegraphics{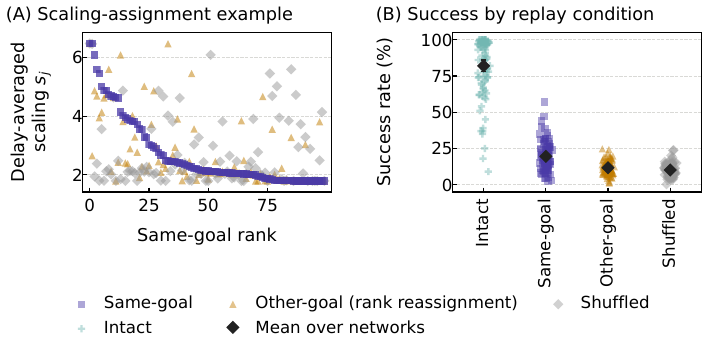}\end{composegraphic}\end{composecell}\end{composerow}\vspace{2mm}\begin{composerow}\begin{composecell}[121.254000000]\begin{composetable}\begingroup\fontsize{8}{8.6}\selectfont\setlength{\tabcolsep}{2pt}\setlength{\aboverulesep}{0.15ex}\setlength{\belowrulesep}{0.25ex}\renewcommand{\arraystretch}{0.95}%
\begin{tabular}{>{\raggedright\arraybackslash}p{17mm}>{\raggedright\arraybackslash}p{30mm}>{\raggedright\arraybackslash}p{35mm}>{\raggedright\arraybackslash}p{30mm}}\toprule
\shortstack[l]{Condition} & \shortstack[l]{Scaling values} & \shortstack[l]{Unit assignment} & \shortstack[l]{Comparison}\\\midrule
\multicolumn{4}{p{116.21751764217518mm}}{\textbf{(C) Scaling definitions}}\\
Intact & Network's own & Network's own activity history & Reference for replay\\ %
Same-goal & Cued-goal mean $M_{g,j}(t)$ & Assignment matched to the cued goal & Intact: effect of mean replay\\ %
Other-goal & Same multiset as Same-goal & Fixed reassignment using another goal's delay-period ranks & Same-goal: effect of rank reassignment\\ %
Shuffled & Same multiset as Same-goal & Randomly relabeled permutation with matched cycle structure & Same-goal: effect of randomized reassignment\\ %
\bottomrule\end{tabular}
\endgroup
\end{composetable}\end{composecell}\end{composerow}\vspace{2mm}\begin{composerow}\begin{composecell}[121.254000000]\begin{composetable}\begingroup\fontsize{8}{8.6}\selectfont\setlength{\tabcolsep}{2pt}\setlength{\aboverulesep}{0.15ex}\setlength{\belowrulesep}{0.25ex}\renewcommand{\arraystretch}{0.95}%
\begin{tabular}{p{70.8mm}p{13mm}p{23mm}}\toprule
\shortstack[l]{Comparison} & \shortstack[l]{Difference} & \shortstack[l]{95\% BCa CI}\\\midrule
\multicolumn{3}{p{109.62335685623357mm}}{\textbf{(D) Success-rate differences (percentage points)}}\\
Intact \ensuremath{-} Same-goal & $62.47$ & $[58.42, 65.78]$\\ %
Same-goal \ensuremath{-} Other-goal & $7.88$ & $[6.07, 10.06]$\\ %
Same-goal \ensuremath{-} Shuffled & $9.45$ & $[7.53, 11.54]$\\ %
\bottomrule\end{tabular}
\par\vspace{1mm}\noindent\begin{minipage}{111.02919607029196mm}\fontsize{8}{9}\selectfont
n = $100$\par
resamples = $2{,}000$\par
\end{minipage}
\endgroup
\end{composetable}\caption{\textbf{Cue-matched replay supported higher success rates than rank-based or randomized reassignment of the same scaling trajectories.} \textbf{(A)} A scaling-assignment example for one network, cued goal 0 (zero-based index), and Other-goal donor offset $d{=}1$. The 100 units are ordered by decreasing Same-goal mean scaling over integration steps 301--399; points compare Same-goal, Other-goal, and Shuffled assignments of the same values. Other-goal uses the cued-goal trajectories assigned according to another goal's ranks; Shuffled uses the corresponding cycle-matched randomized permutation. The four conditions are defined in (C). The vertical axis shows delay-averaged scaling $s_j$. \textbf{(B)} Success rate for Intact, Same-goal, Other-goal, and Shuffled, with means of 82.0\%, 19.6\%, 11.7\%, and 10.1\%. \textbf{(C)} Definition of the replay conditions.  \textbf{(D)} Paired success-rate differences in the comparison order of Fig.~\ref{fig:26:partb}F: Intact minus Same-goal, Same-goal minus Other-goal, and Same-goal minus Shuffled, in percentage points. Rows in (D) and Fig.~\ref{fig:26:partc}(G--I) list paired mean differences and 95\% BCa bootstrap confidence intervals (2{,}000 resamples across network pairs). Rates are percentages and use all 100 recipient trials; differences are in percentage points.}\ComposePartLabel{a}\label{fig:26}\end{composecell}\end{composerow}\end{composepage}\end{figure}\endgroup \begingroup\begin{figure}[p]\begin{composepage}[fig:26|2|pg_fig_26_2]\ComposePart{b}\begin{composerow}\begin{composecell}[121.254000000]\begin{composegraphic}\includegraphics{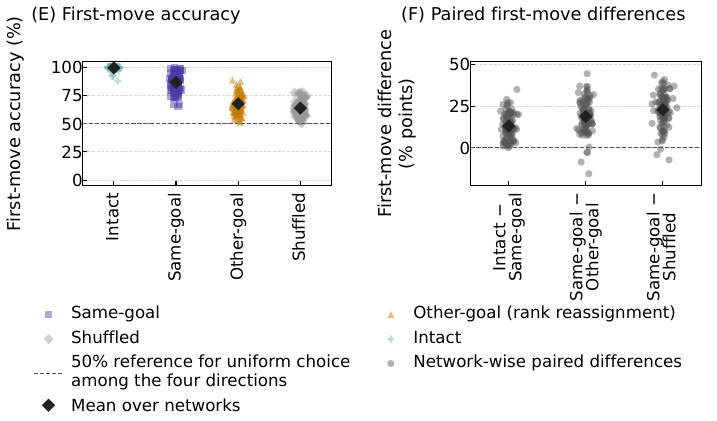}\end{composegraphic}\caption{\textbf{(E)} First-move accuracy for the same conditions, using all 100 recipient trials as the denominator; a move is correct when it reduces the Manhattan distance to the cued goal by one grid step. Means were 99.5\%, 86.5\%, 67.7\%, and 63.7\%; the horizontal line shows the 50\% reference for uniform choice among the four directions. Panels~\ref{fig:26:parta}B and (E) show individual network values and their means. \textbf{(F)} Network-wise paired differences in first-move accuracy (Intact minus Same-goal, Same-goal minus Other-goal, and Same-goal minus Shuffled), with means and 95\% BCa bootstrap confidence intervals (2{,}000 resamples across networks); the horizontal line marks zero. Mean differences were 13.0 [11.6, 14.4], 18.8 [16.8, 20.6], and 22.8 [20.8, 24.7] percentage points, in the order shown in (F).}\ComposePartLabel{b}\label{fig:27}\end{composecell}\end{composerow}\end{composepage}\end{figure}\endgroup \begingroup\begin{figure}[tp]\begin{composepage}[fig:26|3|pg_fig_26_3]\ComposePart{c}\begin{composerow}\begin{composecell}[121.254000000]\begin{composetable}\begingroup\fontsize{8}{8.6}\selectfont\setlength{\tabcolsep}{2pt}\setlength{\aboverulesep}{0.15ex}\setlength{\belowrulesep}{0.25ex}\renewcommand{\arraystretch}{0.95}%
\begin{tabular}{p{40mm}p{14mm}p{14mm}p{13mm}p{23mm}}\toprule
\shortstack[l]{Comparison} & \shortstack[l]{Other-goal\\{}mean (\%)} & \shortstack[l]{Shuffled\\{}mean (\%)} & \shortstack[l]{Difference} & \shortstack[l]{95\% BCa CI}\\\midrule
\multicolumn{5}{p{109.62335685623357mm}}{\textbf{(G) Reached h by GO3 (percent; difference in percentage points)}}\\
Other-goal $-$ Shuffled & $4.61$ & $6.24$ & $-1.63$ & $[-2.37, -0.89]$\\ %
\multicolumn{5}{p{109.62335685623357mm}}{\textbf{(H) Reach selectivity (percentage points)}}\\
Other-goal $-$ Shuffled &  &  & $-1.73$ & $[-2.56, -0.86]$\\ %
\multicolumn{5}{p{109.62335685623357mm}}{\textbf{(I) First move toward h (percent; difference in percentage points)}}\\
Other-goal $-$ Shuffled & $41.81$ & $48.55$ & $-6.74$ & $[-8.35, -5.03]$\\ %
\bottomrule\end{tabular}
\par\vspace{1mm}\noindent\begin{minipage}{111.02919607029196mm}\fontsize{8}{9}\selectfont
n = $100$\par
resamples = $2{,}000$\par
\end{minipage}
\endgroup
\end{composetable}\caption{\textbf{(G)} Probability of reaching the other goal $h$ by GO3, with the condition means and the Other-goal minus Shuffled difference. The goal $h$ is the other goal that defines the corresponding rank reassignment; the matched Shuffled variant uses the same $h$. \textbf{(H)} Reach selectivity, reported as the Other-goal minus Shuffled contrast: the probability of reaching $h$ by GO3 minus the mean probability of reaching each of the remaining 6 goals apart from the cued goal and $h$. \textbf{(I)} Probability that the first move is toward $h$, with the condition means and the Other-goal minus Shuffled difference.}\ComposePartLabel{c}\end{composecell}\end{composerow}\end{composepage}\end{figure}\endgroup \begin{figure}\begin{composepage}[fig:28|1|pg_fig_28_1]\ComposePart{a}\begin{composerow}\begin{composecell}[121.254000000]\begin{composetable}\begingroup\fontsize{8}{9}\selectfont\setlength{\tabcolsep}{2pt}\renewcommand{\arraystretch}{0.93}%
\begin{tabular}{p{6mm}p{8mm}p{8mm}p{14mm}p{14mm}p{25mm}p{32mm}}\toprule
\shortstack[l]{$p$} & \shortstack[l]{$n$} & \shortstack[l]{$\lvert n\rvert/p$} & \shortstack[l]{With STP\\{}(\%)} & \shortstack[l]{Without\\{}STP (\%)} & \shortstack[l]{(B) Success\\{}difference [95\%\\{}BCa CI] (pp)} & \shortstack[l]{(C) $D_Q$ difference\\{}[95\% BCa CI] (Q\\{}units)}\\\midrule
\multicolumn{7}{p{115.43503528435035mm}}{\textbf{(A, B, C) Same recurrent scaling}}\\
$0.05$ & $-0.15$ & $3.00$ & $87.20$ & $48.81$ & \begin{tabular}[t]{@{}l@{}}$38.39$\\$[32.50, 43.26]$\end{tabular} & \begin{tabular}[t]{@{}l@{}}$0.0115$\\$[0.0095, 0.0134]$\end{tabular}\\ %
$0.05$ & $-0.30$ & $6.00$ & $81.46$ & $43.23$ & \begin{tabular}[t]{@{}l@{}}$38.23$\\$[30.40, 45.01]$\end{tabular} & \begin{tabular}[t]{@{}l@{}}$0.0211$\\$[0.0153, 0.0263]$\end{tabular}\\ %
$0.05$ & $-0.60$ & $12.00$ & $72.35$ & $30.95$ & \begin{tabular}[t]{@{}l@{}}$41.40$\\$[34.44, 47.30]$\end{tabular} & \begin{tabular}[t]{@{}l@{}}$0.0487$\\$[0.0385, 0.0573]$\end{tabular}\\ %
$0.10$ & $-0.15$ & $1.50$ & $79.25$ & $48.74$ & \begin{tabular}[t]{@{}l@{}}$30.51$\\$[25.33, 35.16]$\end{tabular} & \begin{tabular}[t]{@{}l@{}}$0.0123$\\$[0.0096, 0.0149]$\end{tabular}\\ %
$0.10$ & $-0.30$ & $3.00$ & $81.90$ & $43.74$ & \begin{tabular}[t]{@{}l@{}}$38.16$\\$[31.55, 44.20]$\end{tabular} & \begin{tabular}[t]{@{}l@{}}$0.0247$\\$[0.0203, 0.0293]$\end{tabular}\\ %
$0.10$ & $-0.60$ & $6.00$ & $74.06$ & $34.15$ & \begin{tabular}[t]{@{}l@{}}$39.91$\\$[32.81, 46.28]$\end{tabular} & \begin{tabular}[t]{@{}l@{}}$0.0518$\\$[0.0414, 0.0618]$\end{tabular}\\ %
$0.20$ & $-0.15$ & $0.75$ & $63.90$ & $39.47$ & \begin{tabular}[t]{@{}l@{}}$24.43$\\$[19.01, 29.44]$\end{tabular} & \begin{tabular}[t]{@{}l@{}}$0.0233$\\$[0.0180, 0.0286]$\end{tabular}\\ %
$0.20$ & $-0.30$ & $1.50$ & $77.90$ & $42.27$ & \begin{tabular}[t]{@{}l@{}}$35.63$\\$[28.91, 41.06]$\end{tabular} & \begin{tabular}[t]{@{}l@{}}$0.0319$\\$[0.0254, 0.0388]$\end{tabular}\\ %
$0.20$ & $-0.60$ & $3.00$ & $71.92$ & $34.81$ & \begin{tabular}[t]{@{}l@{}}$37.11$\\$[28.99, 43.65]$\end{tabular} & \begin{tabular}[t]{@{}l@{}}$0.0526$\\$[0.0410, 0.0638]$\end{tabular}\\ %
\bottomrule\end{tabular}
\par\vspace{1mm}\noindent\begin{minipage}{116.84087449840874mm}\fontsize{8}{9}\selectfont
Network pairs = $100$\par
resamples = $2{,}000$\par
\end{minipage}
\endgroup
\end{composetable}\caption{\textbf{Success-rate differences across reward pairs.} Rows list the rewards $p$ and $n$ and the mean paired differences (With STP minus Without STP) for 100 network pairs, in percentage points for success and raw Q-value units for $D_Q$. At action execution, when the trial continues, $p$ applies to moves reducing squared Euclidean distance to the goal and $n$ to other moves. \textbf{(A)} $\alpha_r=3.0$ in both conditions. The reference rewards are $(0.1,\ensuremath{-}0.3)$; $(0.05,\ensuremath{-}0.15)$ and $(0.2,\ensuremath{-}0.6)$ halve and double both, respectively. Networks used $\tau=1{,}000$\,ms, $\sigma=0.001$, 1{,}900 training and 100 evaluation episodes. \textbf{(B)} Mean success-rate differences with 95\% BCa bootstrap confidence intervals (2{,}000 resamples for each interval; \S\ref{sec:20}) at the same recurrent scaling, and \textbf{(C)} the corresponding mean differences in $D_Q$, in raw Q-value units, averaged over GO2--GO3 across evaluation trials within each network (\S\ref{sec:15}).}\ComposePartLabel{a}\label{fig:28}\end{composecell}\end{composerow}\end{composepage}\end{figure}\begin{figure}[tp]\begin{composepage}[fig:28|2|pg_fig_28_2]\ComposePart{b}\begin{composerow}\begin{composecell}[121.254000000]\begin{composetable}\begingroup\fontsize{8}{9}\selectfont\setlength{\tabcolsep}{2pt}\renewcommand{\arraystretch}{0.93}%
\begin{tabular}{p{6mm}p{8mm}p{8mm}p{14mm}p{14mm}p{25mm}p{32mm}}\toprule
\shortstack[l]{$p$} & \shortstack[l]{$n$} & \shortstack[l]{$\lvert n\rvert/p$} & \shortstack[l]{With STP\\{}(\%)} & \shortstack[l]{Without\\{}STP (\%)} & \shortstack[l]{(E) Success\\{}difference [95\%\\{}BCa CI] (pp)} & \shortstack[l]{(F) $D_Q$ difference\\{}[95\% BCa CI] (Q\\{}units)}\\\midrule
\multicolumn{7}{p{115.43503528435035mm}}{\textbf{(D, E, F) Selected recurrent scalings}}\\
$0.05$ & $-0.15$ & $3.00$ & $86.64$ & $72.11$ & \begin{tabular}[t]{@{}l@{}}$14.53$\\$[10.06, 19.33]$\end{tabular} & \begin{tabular}[t]{@{}l@{}}$0.0040$\\$[0.0027, 0.0052]$\end{tabular}\\ %
$0.05$ & $-0.30$ & $6.00$ & $88.50$ & $67.69$ & \begin{tabular}[t]{@{}l@{}}$20.81$\\$[15.43, 25.78]$\end{tabular} & \begin{tabular}[t]{@{}l@{}}$0.0117$\\$[0.0087, 0.0149]$\end{tabular}\\ %
$0.05$ & $-0.60$ & $12.00$ & $79.47$ & $61.15$ & \begin{tabular}[t]{@{}l@{}}$18.32$\\$[12.53, 24.26]$\end{tabular} & \begin{tabular}[t]{@{}l@{}}$0.0178$\\$[0.0115, 0.0251]$\end{tabular}\\ %
$0.10$ & $-0.15$ & $1.50$ & $77.83$ & $65.17$ & \begin{tabular}[t]{@{}l@{}}$12.66$\\$[7.56, 17.55]$\end{tabular} & \begin{tabular}[t]{@{}l@{}}$0.0058$\\$[0.0037, 0.0078]$\end{tabular}\\ %
$0.10$ & $-0.30$ & $3.00$ & $86.33$ & $71.58$ & \begin{tabular}[t]{@{}l@{}}$14.75$\\$[9.58, 20.08]$\end{tabular} & \begin{tabular}[t]{@{}l@{}}$0.0080$\\$[0.0046, 0.0114]$\end{tabular}\\ %
$0.10$ & $-0.60$ & $6.00$ & $82.21$ & $58.86$ & \begin{tabular}[t]{@{}l@{}}$23.35$\\$[17.36, 29.51]$\end{tabular} & \begin{tabular}[t]{@{}l@{}}$0.0275$\\$[0.0195, 0.0365]$\end{tabular}\\ %
$0.20$ & $-0.15$ & $0.75$ & $61.45$ & $50.78$ & \begin{tabular}[t]{@{}l@{}}$10.67$\\$[5.07, 16.27]$\end{tabular} & \begin{tabular}[t]{@{}l@{}}$0.0117$\\$[0.0072, 0.0160]$\end{tabular}\\ %
$0.20$ & $-0.30$ & $1.50$ & $74.73$ & $65.04$ & \begin{tabular}[t]{@{}l@{}}$9.69$\\$[4.34, 14.53]$\end{tabular} & \begin{tabular}[t]{@{}l@{}}$0.0109$\\$[0.0068, 0.0149]$\end{tabular}\\ %
$0.20$ & $-0.60$ & $3.00$ & $81.28$ & $63.82$ & \begin{tabular}[t]{@{}l@{}}$17.46$\\$[10.94, 24.06]$\end{tabular} & \begin{tabular}[t]{@{}l@{}}$0.0247$\\$[0.0164, 0.0334]$\end{tabular}\\ %
\bottomrule\end{tabular}
\par\vspace{1mm}\noindent\begin{minipage}{116.84087449840874mm}\fontsize{8}{9}\selectfont
Network pairs = $100$\par
resamples = $2{,}000$\par
\end{minipage}
\endgroup
\end{composetable}\caption{\textbf{(D)} $\alpha_r=\num[round-mode=places,round-precision=1]{2}$ With STP and 1.25 Without STP, selected under reference rewards and held fixed. The nine pairs combine $p\in\{0.05,0.1,0.2\}$ with $n\in\{\ensuremath{-}0.15,\ensuremath{-}0.3,\ensuremath{-}0.6\}$ and cover penalty-to-progress ratios $|n|/p$ from 0.75 to 12. \textbf{(E)} Mean success-rate differences at the selected recurrent scaling, and \textbf{(F)} the corresponding mean differences in $D_Q$. Panels (E) and (F) list, for every pair with its ratio $|n|/p$, the success rates (mean across networks, \%), the success-rate difference [95\% BCa CI] and the $D_Q$ difference [95\% BCa CI]; the intervals describe the seed-wise paired differences.}\ComposePartLabel{b}\end{composecell}\end{composerow}\end{composepage}\end{figure}\begin{figure}[tp]\begin{composepage}[fig:29|1|pg_fig_29_1]\ComposePart{a}\begin{composerow}\begin{composecell}[121.254000000]\centerline{\includegraphics{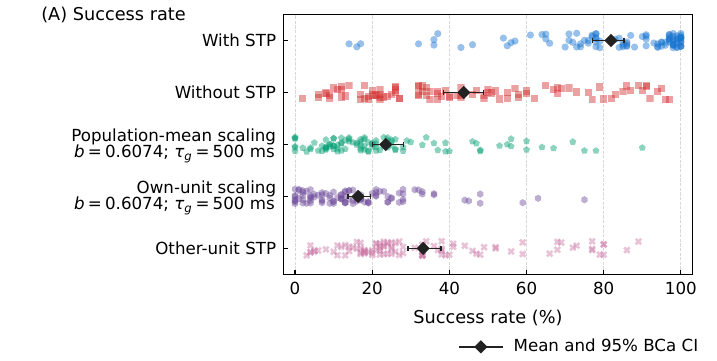}}\end{composecell}\end{composerow}\vspace{2mm}\begin{composerow}\begin{composecell}[121.254000000]\begin{composetable}\begingroup\fontsize{8}{9}\selectfont\setlength{\tabcolsep}{2pt}
\begin{tabular}{>{\raggedright\arraybackslash}p{40mm}p{15mm}p{32mm}}\toprule
\shortstack[l]{Comparison} & \shortstack[l]{Difference} & \shortstack[l]{95\% BCa CI}\\\midrule
\multicolumn{3}{p{89.81167842811679mm}}{\textbf{(B) Success-rate differences (percentage points)}}\\
With STP minus Population-mean scaling & $58.40$ & $[51.90, 63.00]$\\ %
With STP minus Own-unit scaling & $65.57$ & $[59.42, 69.79]$\\ %
With STP minus Other-unit STP & $48.68$ & $[41.97, 54.06]$\\ %
Population-mean scaling minus Without STP & $-20.24$ & $[-26.64, -13.53]$\\ %
Own-unit scaling minus Without STP & $-27.41$ & $[-33.50, -21.58]$\\ %
Other-unit STP minus Without STP & $-10.52$ & $[-17.13, -3.64]$\\ %
Own-unit scaling minus Population-mean scaling & $-7.17$ & $[-10.60, -4.14]$\\ %
\bottomrule\end{tabular}
\par\vspace{1mm}\begin{minipage}{91.21751764217518mm}\fontsize{8}{9}\selectfont
n = $100$\par
resamples = $2{,}000$\par
\end{minipage}
\endgroup
\end{composetable}\caption{\textbf{STP coupled to each unit's own activity supported more accurate goal-directed choices than the three simpler activity-dependent rules.} \textbf{(A)} Success rate for the five conditions (With STP, Without STP, Population-mean scaling, Own-unit scaling, and Other-unit STP). Each point represents one of 100 networks; black markers and bars show means and 95\% BCa bootstrap confidence intervals across network seeds (2{,}000 resamples). The same seeds were used in all conditions. All networks used $\alpha_r=3.0$, $\tau=1{,}000$\,ms, $\sigma=0.001$, 1{,}900 training episodes and 100 evaluation episodes. \textbf{(B)} Paired differences in success rate, for seven comparisons; each row subtracts the second named condition from the first and lists the paired mean difference and its 95\% BCa CI. Across Supplementary Figs.~\ref{fig:29:parta}--\ref{fig:29:partc}, differences in success rate and first-move accuracy are in percentage points, and $D_Q$ and its differences are in raw Q-value units.}\ComposePartLabel{a}\label{fig:29}\end{composecell}\end{composerow}\end{composepage}\end{figure}\begin{figure}[tp]\begin{composepage}[fig:29|2|pg_fig_29_2]\ComposePart{b}\begin{composerow}\begin{composecell}[121.254000000]\centerline{\includegraphics{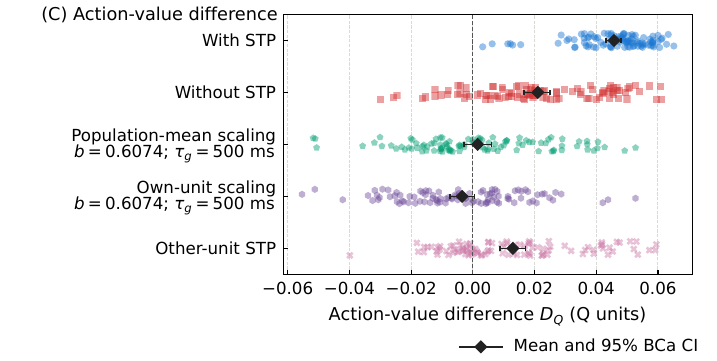}}\end{composecell}\end{composerow}\vspace{2mm}\begin{composerow}\begin{composecell}[121.254000000]\begin{composetable}\begingroup\fontsize{8}{9}\selectfont\setlength{\tabcolsep}{2pt}
\begin{tabular}{>{\raggedright\arraybackslash}p{68mm}p{15mm}p{32mm}}\toprule
\shortstack[l]{Comparison} & \shortstack[l]{Difference} & \shortstack[l]{95\% BCa CI}\\\midrule
\multicolumn{3}{p{89.81167842811679mm}}{\textbf{(D) Action-value differences (Q units)}}\\
With STP minus Population-mean scaling & $0.0442$ & $[0.0390, 0.0489]$\\ %
With STP minus Own-unit scaling & $0.0493$ & $[0.0442, 0.0535]$\\ %
With STP minus Other-unit STP & $0.0328$ & $[0.0281, 0.0375]$\\ %
Population-mean scaling minus Without STP & $-0.0195$ & $[-0.0257, -0.0133]$\\ %
Own-unit scaling minus Without STP & $-0.0246$ & $[-0.0299, -0.0188]$\\ %
Other-unit STP minus Without STP & $-0.0081$ & $[-0.0140, -0.0025]$\\ %
Own-unit scaling minus Population-mean scaling & $-0.0050$ & $[-0.0095, -0.0010]$\\ %
\bottomrule\end{tabular}
\par\vspace{1mm}\begin{minipage}{91.21751764217518mm}\fontsize{8}{9}\selectfont
n = $100$\par
resamples = $2{,}000$\par
\end{minipage}
\endgroup
\end{composetable}\caption{\textbf{(C)} Action-value difference $D_Q$ averaged over GO2--GO3; \textbf{(D)} its paired differences.}\ComposePartLabel{b}\end{composecell}\end{composerow}\end{composepage}\end{figure}\begin{figure}[tp]\begin{composepage}[fig:29|3|pg_fig_29_3]\ComposePart{c}\begin{composerow}\begin{composecell}[121.254000000]\begin{composegraphic}\includegraphics{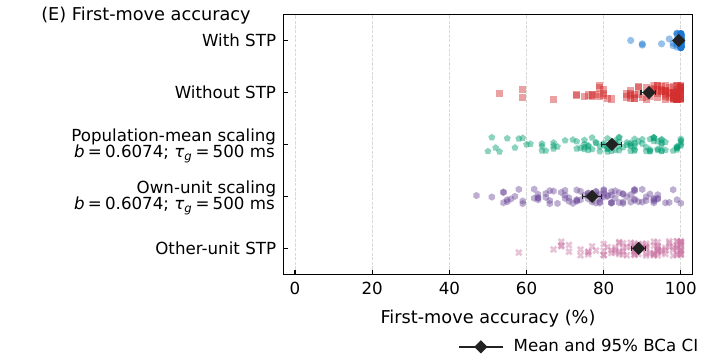}\end{composegraphic}\end{composecell}\end{composerow}\vspace{2mm}\begin{composerow}\begin{composecell}[121.254000000]\begin{composetable}\begingroup\fontsize{8}{9}\selectfont\setlength{\tabcolsep}{2pt}
\begin{tabular}{>{\raggedright\arraybackslash}p{68mm}p{15mm}p{32mm}}\toprule
\shortstack[l]{Comparison} & \shortstack[l]{Difference} & \shortstack[l]{95\% BCa CI}\\\midrule
\multicolumn{3}{p{89.81167842811679mm}}{\textbf{(F) First-move differences (percentage points)}}\\
With STP minus Population-mean scaling & $17.35$ & $[14.86, 19.84]$\\ %
With STP minus Own-unit scaling & $22.47$ & $[20.07, 24.96]$\\ %
With STP minus Other-unit STP & $10.39$ & $[8.62, 12.32]$\\ %
Population-mean scaling minus Without STP & $-9.62$ & $[-12.95, -6.52]$\\ %
Own-unit scaling minus Without STP & $-14.74$ & $[-17.61, -12.08]$\\ %
Other-unit STP minus Without STP & $-2.66$ & $[-4.71, -0.29]$\\ %
Own-unit scaling minus Population-mean scaling & $-5.12$ & $[-7.90, -2.42]$\\ %
\bottomrule\end{tabular}
\par\vspace{1mm}\begin{minipage}{91.21751764217518mm}\fontsize{8}{9}\selectfont
n = $100$\par
resamples = $2{,}000$\par
\end{minipage}
\endgroup
\end{composetable}\end{composecell}\end{composerow}\vspace{2mm}\begin{composerow}\begin{composecell}[121.254000000]\begin{composetable}\begingroup\fontsize{8}{9}\selectfont\setlength{\tabcolsep}{2pt}%
\begin{tabular}{p{40mm}p{20mm}p{22mm}p{24mm}}\toprule
\shortstack[l]{Condition} & \shortstack[l]{Mean relative\\{}scaling} & \shortstack[l]{$b$} & \shortstack[l]{$\tau_g$ (ms)}\\\midrule
\multicolumn{4}{p{106.21751764217518mm}}{\textbf{(G) Realized scaling}}\\
With STP & $1.3030$ & not applicable & not applicable\\ %
Without STP & $1.0000$ & not applicable & not applicable\\ %
Population-mean scaling & $1.3010$ & $0.6074$ & $500$\\ %
Own-unit scaling & $1.3020$ & $0.6074$ & $500$\\ %
Other-unit STP & $1.2810$ & not applicable & not applicable\\ %
\bottomrule\end{tabular}
\endgroup
\end{composetable}\caption{\textbf{(E)} First-move accuracy; \textbf{(F)} its paired differences. \textbf{(G)} Mean relative scaling $\langle s_j/2\rangle$ of the five conditions, averaged over $3\leq t<4$\,s with equal weighting of represented goals within networks; the Without-STP ratio is 1, which holds by definition. Population-mean scaling and Own-unit scaling use $b{=}0.6074$ and $\tau_g{=}500$\,ms, and parameters that do not apply to a condition are marked as such.}\ComposePartLabel{c}\end{composecell}\end{composerow}\end{composepage}\end{figure}\section*{Replay of STP-derived scaling} To test whether assigning STP scaling to particular units affects the first action, we replayed scaling trajectories in trained networks. The cue-matched assignment supported more accurate first moves from the common start position (Fig.~\ref{fig:26}).\par\endgroup \par The acute replay intervention used $n{=}100$ networks at $\tau{=}1{,}000$\,ms, $\alpha_r{=}3.0$, and $\sigma{=}0.001$, with the same 100 recipient trials and noise realizations per network across conditions. A separate donor bank used 12 trials per goal. Other-goal and Shuffled results were averaged over 7 and 7 variants within each network, respectively. All weights and the input and feedback rules were held fixed; $x$ and $u$ continued to evolve with recipient activity, and the network's own scaling was restored from integration step 900. Intact uses the network's own scaling values, set by its own STP variables, and is the reference for replay. Same-goal replays the cued-goal mean scaling $M_{g,j}(t)$ with the assignment matched to the cued goal (comparison with Intact: effect of mean replay). Other-goal keeps the same collection of scaling values and reassigns it to the units by another goal's delay-period ranks (comparison with Same-goal: effect of rank reassignment). Shuffled keeps the same collection and uses a randomly relabeled permutation that matches the cycle structure of the corresponding Other-goal permutation (comparison with Same-goal: effect of randomized reassignment). Replay was applied over integration steps 301--899. Means are across the trained With-STP networks from a separate re-evaluation.\par \begingroup\normalsize\normalbaselines \ComposeKeep{\parindent=15pt \tolerance=200 \emergencystretch=10pt \clubpenalty=150 \parskip=0pt plus 1pt \hsize=\textwidth} To distinguish cue-consistent selection from redirection toward another goal, we measured progress toward the other goal whose unit ranks defined the reassignment. Relative to cycle-matched shuffling, Other-goal reassignment reduced the probability of reaching that goal by GO3, reach selectivity and first-move progress toward that goal (Fig.~\ref{fig:26:partc}G--I); it therefore did not redirect behavior toward the other goal.\par\endgroup\endgroup   \par           \begingroup\begingroup\normalsize\normalbaselines \ComposeKeep{\parindent=15pt \tolerance=200 \emergencystretch=10pt \clubpenalty=150 \parskip=0pt plus 1pt \hsize=\textwidth}   \section*{Reward values} We varied the positive and negative learning rewards across nine pairs, including joint halving and doubling of the reference rewards. The With-STP advantage in success and the later action-value difference was retained across these pairs at both the same and selected recurrent scalings (Figs.~\ref{fig:28:parta}B, C and~\ref{fig:28:partb}E, F). \leavevmode\par\endgroup \endgroup   \par    \section*{Comparisons with simpler activity-dependent rules}\label{sec:42} These controls examine whether the modulation rule and the identity of its driving activity affect goal-directed action selection. We trained controls with single-variable rules driven by population-mean activity or each unit's own activity. In the Other-unit STP control, each unit's modulation was driven by another unit's activity, with the assignment fixed within each network. At the same recurrent scaling ($\alpha_r{=}3.0$), success, first-move accuracy and the GO2--GO3 action-value difference were higher with STP than with all three controls (Fig.~\ref{fig:29}). With $\alpha_r$ selected for each rule ($\alpha_r{=}1.25$, 0.75 and 0.5 for Population-mean scaling, Own-unit scaling and Other-unit STP), With STP exceeded each rule by 6.4--15.0 percentage points (all 95\% BCa confidence intervals above zero; Fig.~\ref{fig:37:partc}D). Relative to Without STP at its selected scaling, Other-unit STP was higher by 8.3 percentage points (95\% BCa CI [2.6, 14.1]), whereas the intervals for the two single-variable rules included zero (Figs.~\ref{fig:33:partb}D and~\ref{fig:37:partc}E). \par \begingroup\par {First-move accuracy is the percentage of evaluation trials whose first move reduced the Manhattan distance to the cued goal, and $D_Q$ is averaged over GO2--GO3 within each network. \leavevmode}\par \par\endgroup \begingroup\par\endgroup  The single-variable rules were \[ s_j^k=2(1+bz_j^k),\qquad z_j^{k+1}=z_j^k+\frac{\Delta t}{\tau_g}(d_j^k-z_j^k),\qquad z_j^0=0, \] with $d_j^k=N_m^{-1}\sum_\ell\bar r_\ell^k$ for population activity and $d_j^k=\bar r_j^k$ for own activity. Here $k$ indexes integration steps, $\bar r^k$ denotes the activity entering the recurrent update, $\Delta t=10$\,ms and $\tau_g=500$\,ms. The shared coefficient was chosen so that the mean single-variable scaling computed on the reference With-STP activity matched its mean STP scaling over $3\leq t<4$\,s: \[ b=\frac{\mu-1}{H}\simeq0.607,\qquad \mu=\left\langle xu/U_{\mathrm{SE}}\right\rangle\simeq1.303,\qquad H=\langle z_{\mathrm{own}}\rangle\simeq0.500. \] Here $z_{\mathrm{own}}$ denotes $z_j$ driven by own-unit activity. The calibration averages were taken over this interval and across units, with equal weight given to represented goals within each network. These network-level averages were then averaged across the 100 networks. Figure~\ref{fig:29:partc}G reports the scaling realized after each control was trained. For Other-unit STP, we retained Eqs.~\eqref{eq:03}--\eqref{eq:04} and replaced only the driving rate by $r_{\pi(j)}^{\mathrm{STP}}$, where the permutation $\pi$ was fixed within each network and $\pi(j)\ne j$. The scaling $s_j=2x_ju_j/U_{\mathrm{SE}}$ still multiplied the activity of unit $j$. All conditions shared the network seeds and base weight matrices and were trained for 1{,}900 episodes, followed by 100 evaluation episodes, with $\tau=1{,}000$\,ms and $\sigma=0.001$, at each of the same 16 recurrent scalings; each rule's $\alpha_r$ was selected by mean success across the same 100 networks used for the comparisons, with ties resolved by the smaller value, as for the other separately trained controls. The coefficient $b$ was kept at its value calibrated at $\alpha_r=3.0$. \par  \leavevmode{}        \begingroup\begin{figure}[tp]\begin{composepage}[fig:30|1|pg_fig_30_1]\begin{composerow}\begin{composecell}[121.254000000]\begin{composegraphic}\includegraphics{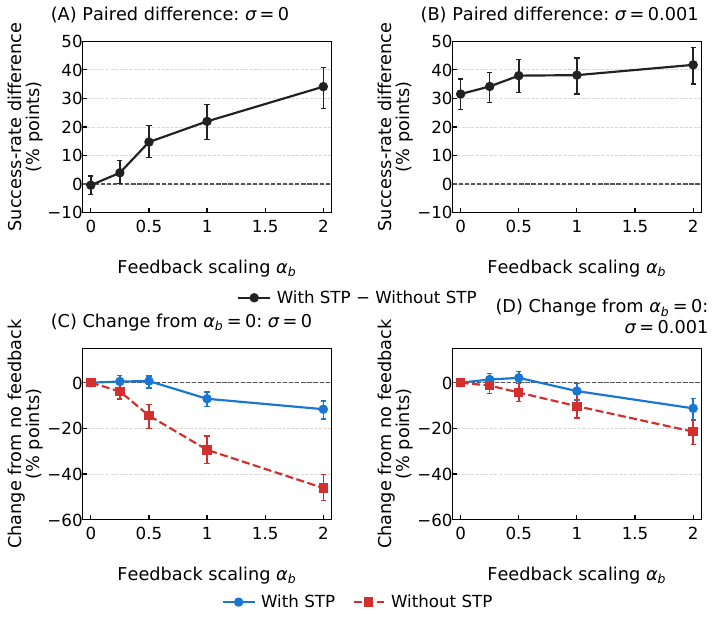}\end{composegraphic}\caption{\textbf{Paired success-rate differences across Q-value feedback strengths.} \textbf{(A)} State noise $\sigma=0$. \textbf{(B)} $\sigma=0.001$. Horizontal axes show feedback scaling $\alpha_b$ on a linear scale; vertical axes show With STP minus Without STP, in percentage points. Points show mean differences across 100 network pairs; bars are 95\% BCa CIs (2{,}000 resamples). Lines connect tested settings. Each feedback strength was fixed during 1{,}900 training and 100 evaluation episodes per network, with $\alpha_r=3.0$ and $\tau=1{,}000$\,ms. Training and evaluation used the same noise level. \textbf{(C)} State noise $\sigma=0$. \textbf{(D)} $\sigma=0.001$. Vertical axes show success-rate changes from $\alpha_b=0$ within each condition, in percentage points. Blue circles denote With STP; red squares denote Without STP. The values at $\alpha_b=0$ are zero by definition. The horizontal axes and the training and evaluation conditions match those in panels (A) and (B).}\label{fig:30}\label{fig:31}\end{composecell}\end{composerow}\end{composepage}\end{figure}\endgroup \section*{Q-value feedback strength}\label{sec:43} We varied the Q-value feedback strength over 5 values, using the settings in \S\ref{sec:12}, to test whether the With-STP advantage depends on it. At $\alpha_r{=}3.0$, the difference without noise widened as feedback scaling increased, with success decreasing more strongly without STP, and the With-STP advantage under state noise persisted across feedback strengths (Fig.~\ref{fig:30}A--C). \par\begingroup\setlength{\emergencystretch}{3em}At $\alpha_b=0,\allowbreak 0.25,\allowbreak 0.5,\allowbreak 1,\allowbreak 2$, mean success rates (With STP, Without STP) without state noise were $(\num[round-mode=places,round-precision=1]{92.66},\num[round-mode=places,round-precision=1]{93.06})\%$, $(\num[round-mode=places,round-precision=1]{93.17},\num[round-mode=places,round-precision=1]{89.22})\%$, $(\num[round-mode=places,round-precision=1]{93.35},\num[round-mode=places,round-precision=1]{78.64})\%$, $(\num[round-mode=places,round-precision=1]{85.61},\num[round-mode=places,round-precision=1]{63.65})\%$, and $(\num[round-mode=places,round-precision=1]{81.04},\num[round-mode=places,round-precision=1]{46.92})\%$, respectively. At $\sigma=0.001$, the corresponding pairs were $(\num[round-mode=places,round-precision=1]{85.58},\num[round-mode=places,round-precision=1]{54.04})\%$, $(\num[round-mode=places,round-precision=1]{86.97},\num[round-mode=places,round-precision=1]{52.78})\%$, $(\num[round-mode=places,round-precision=1]{87.70},\num[round-mode=places,round-precision=1]{49.72})\%$, $(\num[round-mode=places,round-precision=1]{81.90},\num[round-mode=places,round-precision=1]{43.74})\%$, and $(\num[round-mode=places,round-precision=1]{74.34},\num[round-mode=places,round-precision=1]{32.58})\%$. At $\alpha_b=1$ without state noise, success-rate changes from $\alpha_b=0$ were $\ensuremath{-}7.1$ percentage points with STP (95\% BCa CI [$\ensuremath{-}10.5,\ensuremath{-}4.0$]) and $\ensuremath{-}29.4$ without STP ([$\ensuremath{-}35.5,\ensuremath{-}23.3$]).\par\endgroup \begingroup\setlength{\emergencystretch}{3em}Supplementary Figure~\ref{fig:24:partb}C--E also reports comparisons of networks trained and evaluated without state noise or Q-value feedback. Success was higher with STP at the recurrent scalings selected at $\sigma{=}0.001$, whereas the 95\% BCa CI of the paired difference included zero at the same recurrent scaling and when the scalings were re-selected at $\sigma{=}0$ (Supplementary Fig.~\ref{fig:24:partb}C, E). \par\endgroup  \par

\par  \par   \begingroup\begin{figure}[tp]\begin{composepage}[fig:32|1|pg_fig_32_1]\ComposePart{a}\begin{composerow}\begin{composecell}[121.254000000]\begin{composetable}\begingroup\fontsize{8}{9}\selectfont\setlength{\tabcolsep}{2pt}%
\begin{tabular}{llll@{\hspace{3mm}}l@{\hspace{2mm}}l@{\hspace{2mm}}l@{\hspace{3mm}}l}\toprule
\shortstack[l]{Setting} & \shortstack[l]{Variant\\{}mean\\{}(\%)} & \shortstack[l]{$b$} & \shortstack[l]{$\tau_g$\\{}(ms)} & \shortstack[l]{(A) Variant\\{}$-$ reference} & \shortstack[l]{(B) With STP\\{}$-$ variant} & \shortstack[l]{(C) Variant\\{}$-$ Without\\{}STP} & \shortstack[l]{(D)\\{}Mean\\{}scaling}\\\midrule
\multicolumn{8}{l}{\textbf{(A, B, C, D) Success and realized scaling}}\\
\rule{0pt}{4mm}Half G & $30.59$ & $0.3037$ & $500$ & \begin{tabular}[t]{@{}l@{}}$7.09$\\{}$[1.74, 12.56]$\end{tabular} & \begin{tabular}[t]{@{}l@{}}$51.31$\\{}$[44.44, 56.97]$\end{tabular} & \begin{tabular}[t]{@{}l@{}}$-13.15$\\{}$[-19.71, -6.02]$\end{tabular} & $1.1510$\\ %
\rule{0pt}{4mm}Half L & $23.33$ & $0.3037$ & $500$ & \begin{tabular}[t]{@{}l@{}}$7.00$\\{}$[3.17, 10.90]$\end{tabular} & \begin{tabular}[t]{@{}l@{}}$58.57$\\{}$[53.25, 63.10]$\end{tabular} & \begin{tabular}[t]{@{}l@{}}$-20.41$\\{}$[-26.60, -14.50]$\end{tabular} & $1.1520$\\ %
\rule{0pt}{4mm}Double G & $29.63$ & $1.2148$ & $500$ & \begin{tabular}[t]{@{}l@{}}$6.13$\\{}$[0.75, 11.97]$\end{tabular} & \begin{tabular}[t]{@{}l@{}}$52.27$\\{}$[45.89, 57.54]$\end{tabular} & \begin{tabular}[t]{@{}l@{}}$-14.11$\\{}$[-21.20, -7.19]$\end{tabular} & $1.6010$\\ %
\rule{0pt}{4mm}Double L & $17.06$ & $1.2148$ & $500$ & \begin{tabular}[t]{@{}l@{}}$0.73$\\{}$[-3.34, 5.15]$\end{tabular} & \begin{tabular}[t]{@{}l@{}}$64.84$\\{}$[58.76, 69.38]$\end{tabular} & \begin{tabular}[t]{@{}l@{}}$-26.68$\\{}$[-32.72, -19.96]$\end{tabular} & $1.6010$\\ %
\rule{0pt}{4mm}Short G & $51.71$ & $0.6057$ & $100$ & \begin{tabular}[t]{@{}l@{}}$28.21$\\{}$[22.08, 34.44]$\end{tabular} & \begin{tabular}[t]{@{}l@{}}$30.19$\\{}$[23.54, 35.55]$\end{tabular} & \begin{tabular}[t]{@{}l@{}}$7.97$\\{}$[0.23, 15.06]$\end{tabular} & $1.3010$\\ %
\rule{0pt}{4mm}Short L & $32.71$ & $0.6057$ & $100$ & \begin{tabular}[t]{@{}l@{}}$16.38$\\{}$[11.48, 21.13]$\end{tabular} & \begin{tabular}[t]{@{}l@{}}$49.19$\\{}$[42.71, 54.51]$\end{tabular} & \begin{tabular}[t]{@{}l@{}}$-11.03$\\{}$[-17.70, -4.48]$\end{tabular} & $1.3000$\\ %
\bottomrule\end{tabular}
\par\vspace{1mm}\noindent\begin{minipage}{120.24671371246714mm}\fontsize{8}{9}\selectfont
n = $100$\par
resamples = $2{,}000$\par
Reference mean: G $23.50$; L $16.33$\par
With STP mean = $81.90$\par
Without STP mean = $43.74$\par
G, Population-mean scaling; L, Own-unit scaling. Differences and 95\% BCa CIs are paired across networks.\par
\end{minipage}
\endgroup
\end{composetable}\caption{\textbf{The STP success advantage persisted across changes to the single-variable rules' coefficient and relaxation time at the same recurrent scaling.} \textbf{(A--C)} Success-rate differences; within each metric the comparisons are variant minus original rule, With STP minus variant, and variant minus Without STP. Rows list mean paired differences and 95\% BCa CIs (2{,}000 resamples; 100 network pairs). Rows identify halving or doubling of coefficient $b$ at 500\,ms, or 100-ms relaxation with recalibration of $b$. Population-mean scaling is labeled G in the rows and Own-unit scaling is labeled L. \textbf{(D)} Mean relative scaling $\langle s_j/2\rangle$ realized in each trained condition.}\ComposePartLabel{a}\label{fig:32}\end{composecell}\end{composerow}\end{composepage}\end{figure}\endgroup \begingroup\begin{figure}[tp]\begin{composepage}[fig:32|2|pg_fig_32_2]\ComposePart{b}\begin{composerow}\begin{composecell}[121.254000000]\begin{composetable}\begingroup\fontsize{8}{9}\selectfont\setlength{\tabcolsep}{2pt}%
\begin{tabular}{ll@{\hspace{3mm}}l@{\hspace{2mm}}l@{\hspace{2mm}}l}\toprule
\shortstack[l]{Setting} & \shortstack[l]{Variant\\{}mean (Q)} & \shortstack[l]{(E) Variant $-$\\{}reference} & \shortstack[l]{(F) With STP $-$\\{}variant} & \shortstack[l]{(G) Variant $-$\\{}Without STP}\\\midrule
\multicolumn{5}{l}{\textbf{(E, F, G) Action-value differences (Q units)}}\\
\rule{0pt}{4mm}Half G & $0.0080$ & \begin{tabular}[t]{@{}l@{}}$0.0065$\\{}$[0.0005, 0.0119]$\end{tabular} & \begin{tabular}[t]{@{}l@{}}$0.0378$\\{}$[0.0319, 0.0433]$\end{tabular} & \begin{tabular}[t]{@{}l@{}}$-0.0131$\\{}$[-0.0193, -0.0067]$\end{tabular}\\ %
\rule{0pt}{4mm}Half L & $0.0008$ & \begin{tabular}[t]{@{}l@{}}$0.0043$\\{}$[-0.0016, 0.0096]$\end{tabular} & \begin{tabular}[t]{@{}l@{}}$0.0450$\\{}$[0.0397, 0.0500]$\end{tabular} & \begin{tabular}[t]{@{}l@{}}$-0.0203$\\{}$[-0.0261, -0.0143]$\end{tabular}\\ %
\rule{0pt}{4mm}Double G & $0.0092$ & \begin{tabular}[t]{@{}l@{}}$0.0077$\\{}$[0.0019, 0.0136]$\end{tabular} & \begin{tabular}[t]{@{}l@{}}$0.0366$\\{}$[0.0310, 0.0422]$\end{tabular} & \begin{tabular}[t]{@{}l@{}}$-0.0119$\\{}$[-0.0184, -0.0057]$\end{tabular}\\ %
\rule{0pt}{4mm}Double L & $-0.0059$ & \begin{tabular}[t]{@{}l@{}}$-0.0024$\\{}$[-0.0081, 0.0020]$\end{tabular} & \begin{tabular}[t]{@{}l@{}}$0.0517$\\{}$[0.0471, 0.0567]$\end{tabular} & \begin{tabular}[t]{@{}l@{}}$-0.0270$\\{}$[-0.0334, -0.0212]$\end{tabular}\\ %
\rule{0pt}{4mm}Short G & $0.0282$ & \begin{tabular}[t]{@{}l@{}}$0.0266$\\{}$[0.0210, 0.0323]$\end{tabular} & \begin{tabular}[t]{@{}l@{}}$0.0176$\\{}$[0.0130, 0.0224]$\end{tabular} & \begin{tabular}[t]{@{}l@{}}$0.0071$\\{}$[0.0011, 0.0127]$\end{tabular}\\ %
\rule{0pt}{4mm}Short L & $0.0125$ & \begin{tabular}[t]{@{}l@{}}$0.0160$\\{}$[0.0104, 0.0207]$\end{tabular} & \begin{tabular}[t]{@{}l@{}}$0.0333$\\{}$[0.0280, 0.0385]$\end{tabular} & \begin{tabular}[t]{@{}l@{}}$-0.0086$\\{}$[-0.0151, -0.0026]$\end{tabular}\\ %
\bottomrule\end{tabular}
\par\vspace{1mm}\noindent\begin{minipage}{119.02919607029196mm}\fontsize{8}{9}\selectfont
n = $100$\par
resamples = $2{,}000$\par
Reference mean: G $0.0016$; L $-0.0035$\par
With STP mean = $0.0458$\par
Without STP mean = $0.0211$\par
G, Population-mean scaling; L, Own-unit scaling. Differences and 95\% BCa CIs are paired across networks.\par
\end{minipage}
\endgroup
\end{composetable}\end{composecell}\end{composerow}\vspace{2mm}\begin{composerow}\begin{composecell}[121.254000000]\begin{composetable}\begingroup\fontsize{8}{9}\selectfont\setlength{\tabcolsep}{2pt}%
\begin{tabular}{ll@{\hspace{3mm}}l@{\hspace{2mm}}l@{\hspace{2mm}}l}\toprule
\shortstack[l]{Setting} & \shortstack[l]{Variant\\{}mean\\{}(\%)} & \shortstack[l]{(H) Variant $-$\\{}reference} & \shortstack[l]{(I) With STP $-$\\{}variant} & \shortstack[l]{(J) Variant $-$\\{}Without STP}\\\midrule
\multicolumn{5}{l}{\textbf{(H, I, J) First-move differences (percentage points)}}\\
\rule{0pt}{4mm}Half G & $86.80$ & \begin{tabular}[t]{@{}l@{}}$4.69$\\{}$[1.46, 7.51]$\end{tabular} & \begin{tabular}[t]{@{}l@{}}$12.66$\\{}$[10.45, 15.19]$\end{tabular} & \begin{tabular}[t]{@{}l@{}}$-4.93$\\{}$[-8.12, -1.89]$\end{tabular}\\ %
\rule{0pt}{4mm}Half L & $83.48$ & \begin{tabular}[t]{@{}l@{}}$6.49$\\{}$[3.55, 9.53]$\end{tabular} & \begin{tabular}[t]{@{}l@{}}$15.98$\\{}$[13.70, 18.54]$\end{tabular} & \begin{tabular}[t]{@{}l@{}}$-8.25$\\{}$[-11.09, -5.77]$\end{tabular}\\ %
\rule{0pt}{4mm}Double G & $85.28$ & \begin{tabular}[t]{@{}l@{}}$3.17$\\{}$[0.01, 6.73]$\end{tabular} & \begin{tabular}[t]{@{}l@{}}$14.18$\\{}$[11.73, 16.72]$\end{tabular} & \begin{tabular}[t]{@{}l@{}}$-6.45$\\{}$[-9.41, -3.74]$\end{tabular}\\ %
\rule{0pt}{4mm}Double L & $76.11$ & \begin{tabular}[t]{@{}l@{}}$-0.88$\\{}$[-4.10, 2.15]$\end{tabular} & \begin{tabular}[t]{@{}l@{}}$23.35$\\{}$[20.97, 26.08]$\end{tabular} & \begin{tabular}[t]{@{}l@{}}$-15.62$\\{}$[-19.10, -12.37]$\end{tabular}\\ %
\rule{0pt}{4mm}Short G & $92.85$ & \begin{tabular}[t]{@{}l@{}}$10.74$\\{}$[7.80, 13.69]$\end{tabular} & \begin{tabular}[t]{@{}l@{}}$6.61$\\{}$[4.79, 8.90]$\end{tabular} & \begin{tabular}[t]{@{}l@{}}$1.12$\\{}$[-1.72, 4.00]$\end{tabular}\\ %
\rule{0pt}{4mm}Short L & $86.55$ & \begin{tabular}[t]{@{}l@{}}$9.56$\\{}$[6.83, 11.99]$\end{tabular} & \begin{tabular}[t]{@{}l@{}}$12.91$\\{}$[10.80, 15.46]$\end{tabular} & \begin{tabular}[t]{@{}l@{}}$-5.18$\\{}$[-8.26, -2.31]$\end{tabular}\\ %
\bottomrule\end{tabular}
\par\vspace{1mm}\noindent\begin{minipage}{107.02919607029196mm}\fontsize{8}{9}\selectfont
n = $100$\par
resamples = $2{,}000$\par
Reference mean: G $82.11$; L $76.99$\par
With STP mean = $99.46$\par
Without STP mean = $91.73$\par
G, Population-mean scaling; L, Own-unit scaling. Differences and 95\% BCa CIs are paired across networks.\par
\end{minipage}
\endgroup
\end{composetable}\caption{\textbf{(E--G)} Differences in the action-value difference $D_Q$ averaged over GO2--GO3 within each network; \textbf{(H--J)} differences in first-move accuracy.}\ComposePartLabel{b}\end{composecell}\end{composerow}\end{composepage}\end{figure}\endgroup \begin{figure}[tp]\begin{composepage}[fig:33|1|pg_fig_33_1]\ComposePart{a}\begin{composerow}\begin{composecell}[121.254000000]\begin{composegraphic}\includegraphics{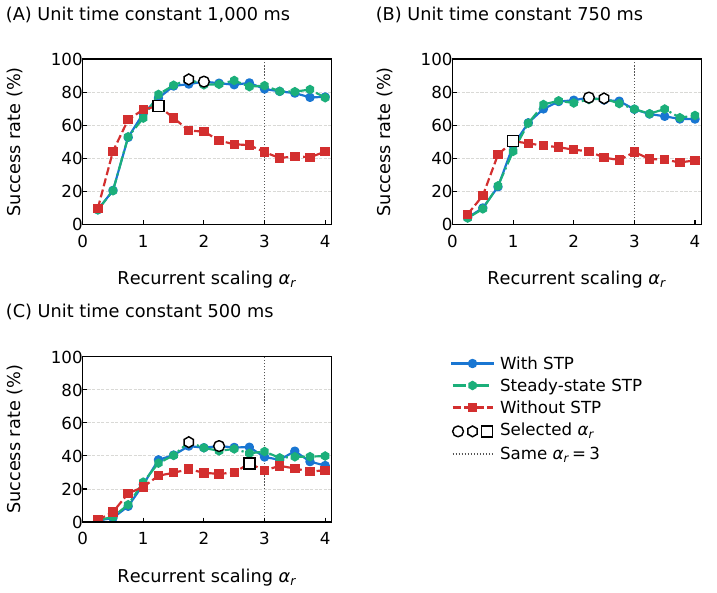}\end{composegraphic}\caption{\textbf{Success rates across recurrent scalings for With STP, steady-state STP and Without STP.} \textbf{(A--C)} Mean success rates for With STP (blue), Steady-state STP (green) and Without STP (red) across the same 16 recurrent scalings at unit time constants of $\tau{=}1{,}000$ (A), 750 (B) and 500\,ms (C), with $n{=}100$ network seeds per condition and scaling. All networks used $\sigma{=}0.001$ during training and evaluation, 1{,}900 training episodes and 100 evaluation episodes. The steady-state STP control (section ``Steady-state STP modulation'') used $U_{\mathrm{SE}}{=}0.1$, $\tau_{\mathrm{R}}{=}100$\,ms and $\tau_{\mathrm{F}}{=}500$\,ms. Large white markers indicate the scalings selected by mean success across all 100 seeds; in (A), $\alpha_r{=}2.00$ with STP, $1.75$ with steady-state STP, and $1.25$ without STP. The vertical dotted line marks the main recurrent scaling, $\alpha_r{=}3.0$. Curves show means; Figs.~\ref{fig:37} and~\ref{fig:38} give the paired differences and their 95\% BCa CIs. Selected $\alpha_r$ for With STP/Steady-state STP/Without STP: (B) $2.25/2.50/1.00$; (C) $2.25/1.75/2.75$.}\ComposePartLabel{a}\label{fig:33}\end{composecell}\end{composerow}\end{composepage}\end{figure}\begin{figure}[tp]\begin{composepage}[fig:33|2|pg_fig_33_2]\ComposePart{b}\begin{composerow}\begin{composecell}[121.254000000]\begin{composegraphic}\includegraphics{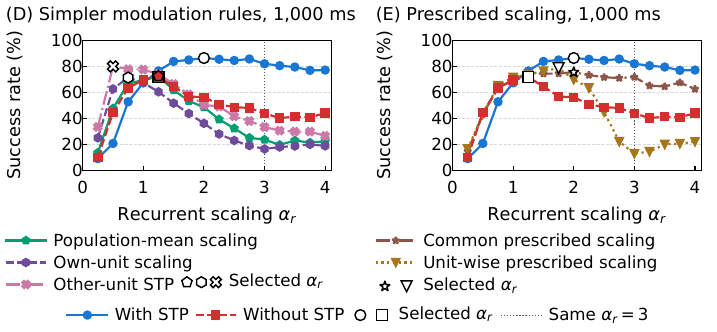}\end{composegraphic}\caption{\textbf{(D)} Population-mean scaling, Own-unit scaling and Other-unit STP across the same scalings at $\tau{=}1{,}000$\,ms, with the With-STP and Without-STP curves of Fig.~\ref{fig:33:parta}A repeated as references; the coefficient $b$ of the two single-variable rules was kept at its value calibrated at $\alpha_r{=}3.0$. For panel (\ref{fig:33:partb}D), selected $\alpha_r$ values were $2.00$ for With STP and $1.25$ for Population-mean scaling. Own-unit scaling used $0.75$ and Other-unit STP used $0.5$. Without STP used $1.25$. \textbf{(E)} Common and Unit-wise prescribed scaling across the same scalings at $\tau{=}1{,}000$\,ms, using the time courses recorded from the With-STP networks at $\alpha_r{=}3.0$, with the With-STP and Without-STP curves of Fig.~\ref{fig:33:parta}A repeated as references. For panel (\ref{fig:33:partb}E), selected $\alpha_r$ values were $2.00$ for With STP, $2.0$ for Common prescribed scaling, $1.75$ for Unit-wise prescribed scaling and $1.25$ for Without STP.}\ComposePartLabel{b}\end{composecell}\end{composerow}\end{composepage}\end{figure}\begin{figure}[tp]\begin{composepage}[fig:34|1|pg_fig_34_1]\begin{composerow}\begin{composecell}[121.254000000]\begin{composegraphic}\includegraphics{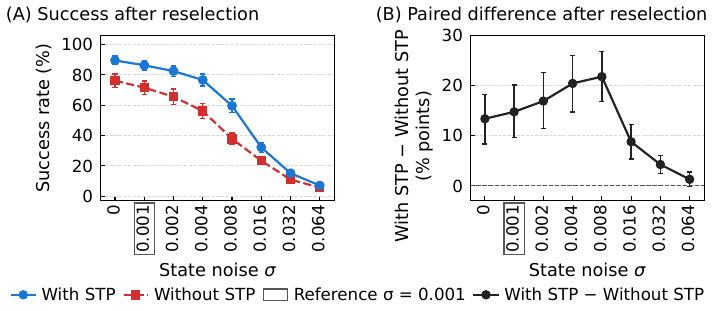}\end{composegraphic}\end{composecell}\end{composerow}\vspace{2mm}\begin{composerow}\begin{composecell}[121.254000000]\begin{composetable}\begingroup\fontsize{8}{9}\selectfont\setlength{\tabcolsep}{2pt}%
\setbox\FinalTabularBox=\hbox{%
\begin{tabular}{@{}lrrrr@{}}\toprule%
\multicolumn{5}{l}{\textbf{(C) Noise-specific recurrent-scaling selection}}\\%
\shortstack{State\\noise $\sigma$} & \shortstack{With STP\\$\alpha_r$} & \shortstack{Without STP\\$\alpha_r$} & \shortstack{With\\grid points} & \shortstack{Without\\grid points}\\\midrule%
\rule{0pt}{4mm}0 & $2.75$ & $1.25$ & $16$ & $16$\\ %
\rule{0pt}{4mm}0.001 & $2.00$ & $1.25$ & $16$ & $16$\\ %
\rule{0pt}{4mm}0.002 & $2.00$ & $1.00$ & $16$ & $16$\\ %
\rule{0pt}{4mm}0.004 & $1.50$ & $0.75$ & $16$ & $16$\\ %
\rule{0pt}{4mm}0.008 & $1.50$ & $0.75$ & $16$ & $16$\\ %
\rule{0pt}{4mm}0.016 & $1.25$ & $0.75$ & $16$ & $16$\\ %
\rule{0pt}{4mm}0.032 & $1.50$ & $1.25$ & $16$ & $16$\\ %
\rule{0pt}{4mm}0.064 & $1.50$ & $1.25$ & $16$ & $16$\\ %
\bottomrule\end{tabular}}%
\vbox{\hsize=121mm\parindent=0pt\box\FinalTabularBox\par%
\noindent\parbox{121mm}{Paired n = $100$.}\par%
}\endgroup%
\end{composetable}\caption{\textbf{With the recurrent scaling re-selected at each noise level, the With-STP condition reached higher success rates from $\sigma{=}0$ to $\sigma{=}0.032$.} \textbf{(A)} Success rate after training, with the mean and the $t$-based 95\% confidence interval across networks; blue circles and solid lines denote With STP, and red squares and dashed lines denote Without STP. The comparison with the selected values held fixed across the noise levels is shown in Fig.~\ref{fig:06}. \textbf{(B)} Paired difference, With STP minus Without STP, in percentage points; black circles and bars show the mean and the 95\% BCa bootstrap confidence interval (2{,}000 resamples; \S\ref{sec:20}), and the dashed gray line marks zero. To ask how the advantage behaves when the recurrent scaling is adapted to the noise, $\alpha_r$ was selected separately for With STP and Without STP at each state-noise level, from the same 16-point grid ($\alpha_r{=}0.25$--$4.0$) by the highest mean success rate; ties were resolved by the smaller $\alpha_r$. Noise had the same level during training and evaluation, at 8 levels from $\sigma{=}0$ to $\sigma{=}0.064$ shown as equally spaced categories, and the main level $\sigma{=}0.001$ is marked as a reference. Each level has $n{=}100$ paired networks (seeds 0--99) at $\tau{=}1{,}000$\,ms, trained for 1{,}900 episodes and evaluated over 100 episodes per network. \textbf{(C)} The two selected values of $\alpha_r$, the number of grid points available at each level for each condition, and the paired network count.}\label{fig:34}\end{composecell}\end{composerow}\end{composepage}\end{figure}\begin{figure}[tp]\begin{composepage}[fig:35|1|pg_fig_35_1]\begin{composerow}\begin{composecell}[121.254000000]\begin{composetable}\begingroup\fontsize{8}{9}\selectfont\setlength{\tabcolsep}{2pt}%
\setbox\FinalTabularBox=\hbox{%
\begin{tabular}{@{}llrrrr@{}}\toprule%
\multicolumn{6}{l}{\textbf{(A) Goal decoding (\%)}}\\%
Setting & Method & \shortstack{With STP\\(\%)} & \shortstack{Without STP\\(\%)} & \shortstack{Difference\\(pp)} & \shortstack{95\% BCa CI\\(pp)}\\\midrule%
\rule{0pt}{4mm}Delay (3--4 s) & same & $99.90$ & $98.99$ & $0.92$ & $[0.73, 1.12]$\\ %
\rule{0pt}{4mm}Delay (3--4 s) & selected & $99.99$ & $100.00$ & $-0.01$ & $[-0.04, 0.00]$\\ %
\rule{0pt}{4mm}Late (8--9 s) & same & $98.50$ & $89.14$ & $9.36$ & $[7.75, 11.19]$\\ %
\rule{0pt}{4mm}Late (8--9 s) & selected & $99.57$ & $98.74$ & $0.83$ & $[0.50, 1.23]$\\ %
\bottomrule\end{tabular}}%
\vbox{\hsize=121mm\parindent=0pt\box\FinalTabularBox\par%
\noindent\parbox{121mm}{Paired n = $100$.}\par%
}\endgroup%
\end{composetable}\end{composecell}\end{composerow}\vspace{2mm}\begin{composerow}\begin{composecell}[121.254000000]\begin{composetable}\begingroup\fontsize{8}{9}\selectfont\setlength{\tabcolsep}{2pt}%
\setbox\FinalTabularBox=\hbox{%
\begin{tabular}{@{}llrrrr@{}}\toprule%
\multicolumn{6}{l}{\textbf{(B) Action-value difference (Q-value units)}}\\%
Setting & Method & \shortstack{With STP\\(Q-value\\units)} & \shortstack{Without STP\\(Q-value\\units)} & \shortstack{Difference\\(Q-value\\units)} & \shortstack{95\% BCa CI\\(Q-value\\units)}\\\midrule%
\rule{0pt}{4mm}GO2 & same & $0.0648$ & $0.0419$ & $0.0228$ & $[0.0175, 0.0281]$\\ %
\rule{0pt}{4mm}GO2 & selected & $0.0636$ & $0.0599$ & $0.0037$ & $[0.0003, 0.0074]$\\ %
\rule{0pt}{4mm}GO3 & same & $0.0268$ & $0.0002$ & $0.0266$ & $[0.0216, 0.0314]$\\ %
\rule{0pt}{4mm}GO3 & selected & $0.0263$ & $0.0140$ & $0.0123$ & $[0.0079, 0.0167]$\\ %
\bottomrule\end{tabular}}%
\vbox{\hsize=121mm\parindent=0pt\box\FinalTabularBox\par%
\noindent\parbox{121mm}{Paired n = $100$.}\par%
}\endgroup%
\end{composetable}\end{composecell}\end{composerow}\vspace{2mm}\begin{composerow}\begin{composecell}[121.254000000]\begin{composetable}\begingroup\fontsize{8}{9}\selectfont\setlength{\tabcolsep}{2pt}%
\setbox\FinalTabularBox=\hbox{%
\begin{tabular}{@{}llrrrr@{}}\toprule%
\multicolumn{6}{l}{\textbf{(C) Scatter ratio}}\\%
Setting & Method & \shortstack{With STP\\(ratio)} & \shortstack{Without STP\\(ratio)} & \shortstack{Difference\\(ratio)} & \shortstack{95\% BCa CI\\(ratio)}\\\midrule%
\rule{0pt}{4mm}Delay (3--4 s) & same & $145.94$ & $9.04$ & $136.89$ & $[91.91, 284.72]$\\ %
\rule{0pt}{4mm}Delay (3--4 s) & selected & $238.40$ & $26.83$ & $211.57$ & $[174.49, 261.39]$\\ %
\bottomrule\end{tabular}}%
\vbox{\hsize=121mm\parindent=0pt\box\FinalTabularBox\par%
\noindent\parbox{121mm}{Paired n = $100$.}\par%
}\endgroup%
\end{composetable}\end{composecell}\end{composerow}\vspace{2mm}\begin{composerow}\begin{composecell}[121.254000000]\begin{composetable}\begingroup\fontsize{8}{9}\selectfont\setlength{\tabcolsep}{2pt}%
\setbox\FinalTabularBox=\hbox{%
\begin{tabular}{@{}llrrrr@{}}\toprule%
\multicolumn{6}{l}{\textbf{(D) Success after feedback removal (\%)}}\\%
Setting & Method & \shortstack{With STP\\(\%)} & \shortstack{Without STP\\(\%)} & \shortstack{Difference\\(pp)} & \shortstack{95\% BCa CI\\(pp)}\\\midrule%
\rule{0pt}{4mm}Feedback removed & same & $74.03$ & $35.10$ & $38.93$ & $[32.73, 44.72]$\\ %
\rule{0pt}{4mm}Feedback removed & selected & $75.17$ & $54.40$ & $20.77$ & $[15.37, 25.74]$\\ %
\bottomrule\end{tabular}}%
\vbox{\hsize=121mm\parindent=0pt\box\FinalTabularBox\par%
\noindent\parbox{121mm}{Paired n = $100$.}\par%
}\endgroup%
\end{composetable}\caption{\textbf{With the recurrent scaling selected for each condition, goals remained highly decodable in both conditions, while action-value differences $D_Q$ were larger with STP.} The comparison at the same recurrent scaling ($\alpha_r{=}3.0$ in both conditions) is shown next to the comparison with the recurrent scaling selected for each condition ($\alpha_r{=}2$ with STP and $\alpha_r{=}1.25$ without STP), which were selected at $\sigma{=}0.001$ and held fixed. Rows list the mean paired difference, With STP minus Without STP, with 95\% BCa bootstrap confidence intervals (2{,}000 resamples; \S\ref{sec:20}). Each comparison uses 100 paired seeds with 100 evaluation episodes per network, at $\tau{=}1{,}000$\,ms and $\sigma{=}0.001$ during training and evaluation after 1{,}900 training episodes, with $U_{\mathrm{SE}}{=}0.1$, $\tau_{\mathrm{R}}{=}100$\,ms and $\tau_{\mathrm{F}}{=}500$\,ms. \textbf{(A)} Goal-decoding accuracy in the delay ($t{=}3$--$4$\,s) and late ($t{=}8$--$9$\,s) windows. Time windows and GO opportunities are identified by rows. Each panel lists both group means, the paired difference with its BCa CI, and $n$. \textbf{(B)} Action-value difference $D_Q$ at GO2 and GO3. \textbf{(C)} Delay-period scatter ratio. \textbf{(D)} Success rate after the Q-value feedback was set to zero from trial onset, in a separate evaluation of the trained networks.}\label{fig:35}\end{composecell}\end{composerow}\end{composepage}\end{figure}\FloatBarrier\section*{Sensitivity of the single-variable rules} We varied the strength and relaxation time of the single-variable rules to test whether their original settings explain their lower success rates relative to With STP (Fig.~\ref{fig:32}).\par At $\alpha_r{=}3.0$, we halved or doubled $b$ at $\tau_g=500$\,ms and recalibrated $b$ at $\tau_g=100$\,ms to the reference mean scaling used for the original controls. After training, mean relative scaling was approximately 1.15 with the halved coefficient, 1.60 with the doubled coefficient, and 1.30 with the shorter relaxation time and recalibrated coefficient; the doubled-coefficient controls exceeded the With-STP mean relative scaling of 1.303. With STP retained a 30.2--64.8 percentage-point success-rate advantage over both rules across these three parameter settings (all 95\% BCa confidence intervals above zero; Fig.~\ref{fig:32}). Population-mean scaling with the 100-ms relaxation time improved success by 28.2 percentage points relative to its original setting (95\% BCa CI [22.1, 34.4]) and exceeded Without STP by 8.0 percentage points (95\% BCa CI [0.2, 15.1]).\par   In Supplementary Fig.~\ref{fig:32}, differences in success rate and first-move accuracy are in percentage points, and $D_Q$ differences are in raw Q-value units. Variant-minus-original comparisons pair each Population-mean (G) or Own-unit (L) scaling variant with the corresponding original control of the same seed, and With STP and Without STP are shared references. Panel D of Fig.~\ref{fig:32:parta} gives the mean relative scaling $\langle s_j/2\rangle$, averaged over $3\leq t<4$\,s with equal weighting of represented goals within networks, as realized in each trained condition; its row labels give the coefficient $b$ and the relaxation time $\tau_g$. The original controls use $b{=}0.6074$ and $\tau_g{=}500$\,ms; for the shorter relaxation time, $b$ was recalibrated on the reference With-STP activity to the same mean scaling as the original controls. The shared training and evaluation settings follow \S\ref{sec:12}. The reference values of With STP, Without STP and the original controls are given in Fig.~\ref{fig:29}.   \begingroup\par\endgroup    
\par     \section*{Steady-state STP modulation}\label{sec:44} We trained a control using the steady-state solutions of Eqs.~\eqref{eq:03}--\eqref{eq:04} at each integration step, without separately evolving $x_j$ and $u_j$. For a fixed effective presynaptic rate $r$, setting both increments to zero gives \[ \begin{aligned} u_{\mathrm{ss}}(r)&=\frac{U_{\mathrm{SE}}(1+\tau_{\mathrm{F}}r)}{1+U_{\mathrm{SE}}\tau_{\mathrm{F}}r},\\ x_{\mathrm{ss}}(r)&=\frac{1}{1+\tau_{\mathrm{R}}r\,u_{\mathrm{ss}}(r)},\\ s_{\mathrm{ss}}(r)&=\frac{2x_{\mathrm{ss}}(r)u_{\mathrm{ss}}(r)}{U_{\mathrm{SE}}}. \end{aligned} \] For unit $j$, we evaluated these expressions at $r=r_j^{\mathrm{STP}}(t)=\bar r_j(t)/\Delta t$ immediately before the recurrent update, and used the recurrent drive \[ I_i^{\mathrm{rec}}(t)=\sum_j W_{ij}^{\mathrm{rec}}\left[s_{\mathrm{ss}}\!\left(r_j^{\mathrm{STP}}(t)\right)\bar r_j(t)-1\right]. \] The original $U_{\mathrm{SE}}{=}0.1$, $\tau_{\mathrm{R}}{=}100$\,ms and $\tau_{\mathrm{F}}{=}500$\,ms determine the curve. The unit integration, Q-value feedback and TD learning follow the same rules as in the With-STP and Without-STP conditions. \par  The steady-state STP control used the training and evaluation protocol and recurrent-scaling grid in \S\ref{sec:12}, at unit time constants $\tau{=}500$, 750 and 1{,}000\,ms with $\sigma{=}0.001$ (Fig.~\ref{fig:33}). Each condition's selected $\alpha_r$ maximized mean success across the same 100 seeds; we also selected on 50 seeds and evaluated the other 50, reversing the halves, at $\tau{=}1{,}000$\,ms. Ties were resolved by selecting the smaller scaling. We computed differences within network-seed pairs and obtained 95\% BCa CIs from 2{,}000 bootstrap resamples, keeping the selected scalings fixed. \par For $\tau{=}1{,}000$\,ms, at the same recurrent scaling and with the recurrent scaling selected for each condition, steady-state STP reached success similar to that of With STP (With STP minus steady-state STP, \ensuremath{-}2.0 [\ensuremath{-}6.0, 1.9] and \ensuremath{-}1.4 [\ensuremath{-}4.9, 1.9] percentage points) and exceeded Without STP (\num[round-mode=places,round-precision=1]{40.18} [\num[round-mode=places,round-precision=1]{34.49}, \num[round-mode=places,round-precision=1]{46.38}] and 16.1 [11.3, 21.0] percentage points; 95\% BCa CIs; Figs.~\ref{fig:37:partb}B, C and~\ref{fig:38}). In the split comparisons at $\tau{=}1{,}000$\,ms, With STP minus steady-state STP was \ensuremath{-}5.5 [\ensuremath{-}12.0, \ensuremath{-}0.7] percentage points with selection on seeds 0--49 and evaluation on seeds 50--99, and \ensuremath{-}3.6 [\ensuremath{-}8.8, 1.3] percentage points in the reverse direction (Fig.~\ref{fig:37:partb}B). \leavevmode{} At $\tau{=}750$ and 500\,ms, With STP minus steady-state STP ranged from \ensuremath{-}3.3 to 0.5 percentage points at the same recurrent scaling and with the recurrent scaling selected for each condition, with every 95\% BCa CI including zero, and steady-state STP exceeded Without STP by 11.4--26.0 percentage points, with every interval above zero (Figs.~\ref{fig:33:parta}B, C and~\ref{fig:37:partb}B, C). The steady-state STP control preserved the success-rate advantage over Without STP in this task at unit time constants of 500, 750 and 1{,}000\,ms with $\sigma{=}0.001$ (Figs.~\ref{fig:33:parta}A--C and~\ref{fig:37:partb}B, C).     \begingroup\begingroup\normalsize\normalbaselines \section*{Simpler activity-dependent rules across recurrent scalings} With the recurrent scaling selected for each condition, success rates for the three simpler activity-dependent rules remained lower than for With STP, extending the comparison beyond their original scaling choice (Fig.~\ref{fig:33:partb}D). Both controls with prescribed time-varying scaling also had lower success rates than With STP when their time courses were kept fixed across the scaling search (Fig.~\ref{fig:33:partb}E).\par\par\endgroup \endgroup  \par     \begingroup\par \needspace{6\baselineskip}{\section*{Internal analyses with the recurrent scaling selected for each condition} We repeated the internal comparisons with the recurrent scaling selected for each condition to test whether their differences depend on using the same scaling in both conditions (Fig.~\ref{fig:35}). Decoding and scatter-ratio entries in Fig.~\ref{fig:35} are window averages of time-resolved measures; decoding accuracy and success rate are in percent and their differences in percentage points, $D_Q$ is in raw Q-value units, and the scatter ratio is dimensionless. Success rates under ordinary evaluation at both recurrent scalings are given in Table~\ref{tab:04} and Fig.~\ref{fig:03}.}\par \endgroup \par  \par \relax \begingroup\begingroup\normalsize\normalbaselines \ComposeKeep{\parindent=15pt \tolerance=200 \emergencystretch=10pt \clubpenalty=150 \parskip=0pt plus 1pt \hsize=\textwidth} \par\endgroup \begingroup\normalsize\normalbaselines \begingroup\begin{figure}[!tp]\begin{composepage}[fig:36|1|pg_fig_36_1]\begin{composerow}\begin{composecell}[121.254000000]\begin{composetable}\begingroup\fontsize{8}{9}\selectfont\setlength{\tabcolsep}{2pt}%
\setbox\FinalTabularBox=\hbox{%
\begin{tabular}{@{}>{\raggedright\arraybackslash}p{31mm}rrrr@{}}\toprule%
\multicolumn{5}{l}{\textbf{(A) Trial outcomes (counts)}}\\%
Condition & Success by GO3 & Success at GO4 & Failure & \\\midrule%
\rule{0pt}{4mm}With STP & $8185$ & $5$ & $1810$ & \\ %
\rule{0pt}{4mm}Without STP & $4335$ & $39$ & $5626$ & \\ %
\midrule%
\multicolumn{5}{l}{\textbf{(B) Goal-directed readout selection (\%)}}\\%
GO & With STP (\%) & Without STP (\%) &  & \\\midrule%
\rule{0pt}{4mm}GO1 & $99.46$ & $91.73$ &  & \\ %
\rule{0pt}{4mm}GO2 & $96.50$ & $78.51$ &  & \\ %
\rule{0pt}{4mm}GO3 & $84.55$ & $54.53$ &  & \\ %
\midrule%
\multicolumn{5}{l}{\textbf{(C) Progress after an error (\%)}}\\%
Setting & With STP (\%) & Without STP (\%) & Difference (pp) & 95\% BCa CI\\\midrule%
\rule{0pt}{4mm}After an error & $61.28$ & $47.61$ & $13.67$ & $[5.54, 21.45]$\\ %
\rule{0pt}{4mm}At GO1 (14 pairs) & $57.09$ & $44.31$ & $12.77$ & $[-11.81, 38.76]$\\ \rule{0pt}{4mm}At GO2 (32 pairs) & $69.97$ & $36.68$ & $33.28$ & $[15.72, 47.94]$\\ \rule{0pt}{4mm}At GO3 (87 pairs) & $58.97$ & $47.42$ & $11.55$ & $[2.77, 20.23]$\\ \rule{0pt}{4mm}Distance-increasing move (87 pairs) & $62.89$ & $48.08$ & $14.81$ & $[6.60, 23.07]$\\ \rule{0pt}{4mm}Invalid move, cursor in place (4 pairs) & $25.00$ & $50.00$ & $-25.00$ & --\\ \bottomrule\end{tabular}}%
\vbox{\hsize=121mm\parindent=0pt\box\FinalTabularBox\par%
\noindent\parbox{121mm}{(C): After an error, 88 paired networks; the row labels give the number of pairs in each stratum. Strata overlap. No interval is reported for the four-pair stratum.}\par%
}\endgroup%
\end{composetable}\caption{\textbf{With STP, more trials ended in success by the third GO opportunity, the readout selected goal-consistent moves more often at GO1--GO3, and progress toward the goal after an erroneous move was higher.} \textbf{(A)} Trial outcomes at the main setting, as the counts of trials that ended in success by the third GO opportunity, that ended in success at the fourth GO opportunity, and that failed. Each condition contains 100 networks with 100 evaluation trials each (10{,}000 trials); rows give the counts of each outcome for With STP and Without STP, and of successes at the fourth GO opportunity. The main setting uses $\alpha_r{=}3.0$ in both conditions, $\tau{=}1{,}000$\,ms, $\sigma{=}0.001$, and the STP parameters $U_{\mathrm{SE}}{=}0.1$, $\tau_{\mathrm{R}}{=}100$\,ms and $\tau_{\mathrm{F}}{=}500$\,ms. \textbf{(B)} The percentage of trials in which the readout selected a move toward the goal, for each of GO1--GO3; all 100 evaluation trials from each of the 100 networks contribute at each GO opportunity. \textbf{(C)} Next-move progress after an erroneous move: the proportion of erroneous moves that were followed by a move toward the goal at the next GO opportunity (\S\ref{sec:23}), averaged over the seed-wise proportions of the 88 networks that have a defined value in both conditions. The table lists the condition means, the paired difference (With STP minus Without STP, in percentage points) with its 95\% BCa confidence interval (2{,}000 resamples; \S\ref{sec:20}) and the number of pairs, for all errors, for each GO opportunity and for each error type.}\label{fig:36}\end{composecell}\end{composerow}\end{composepage}\end{figure}\endgroup \begin{figure}[tp]\begin{composepage}[fig:37|1|pg_fig_37_1]\ComposePart{a}\begin{composerow}\begin{composecell}[121.254000000]\begin{composetable}\begingroup\fontsize{8}{9}\selectfont\setlength{\tabcolsep}{2pt}%
\begin{tabular}{l>{\raggedright\arraybackslash}p{29mm}ll@{\hspace{3mm}}ll@{\hspace{3mm}}lll}\toprule
 & & \multicolumn{2}{l}{With} & \multicolumn{2}{l}{Without} & & &\\ \shortstack[l]{$\tau$\\{}(ms)} & \shortstack[l]{Setting} & $\alpha_r$ & \shortstack[l]{mean\\{}(\%)} & $\alpha_r$ & \shortstack[l]{mean\\{}(\%)} & \shortstack[l]{Difference\\{}(pp)} & \shortstack[l]{95\% BCa CI} & \shortstack[l]{Eval.\\{}$n$}\\\midrule
\multicolumn{9}{l}{\textbf{(A) With STP \ensuremath{-} Without STP}}\\
1,000 & Select A; evaluate B & $2.5$ & $\underline{\mathbf{84.06}}$ & $1.25$ & $74.28$ & $9.78$ & $[3.63, 16.12]$ & $50$\\ %
1,000 & Select B; evaluate A & $2.25$ & $\underline{\mathbf{82.18}}$ & $1.25$ & $68.88$ & $13.30$ & $[5.18, 21.36]$ & $50$\\ %
\bottomrule\end{tabular}
\par\vspace{1mm}\noindent\begin{minipage}{121.15255292652553mm}\fontsize{8}{9}\selectfont
resamples = $2{,}000$\par
\end{minipage}
\endgroup
\end{composetable}\caption{\textbf{Selecting recurrent scaling on separate networks retained the STP success advantage.} \textbf{(A)} Success-rate differences, With STP minus Without STP, when the recurrent scaling was selected on seed set A and evaluated on seed set B, and when it was selected on B and evaluated on A. Seed set A comprises network seeds 0--49 and B comprises seeds 50--99. The split rows use 50 evaluation networks. Each condition's recurrent scaling was selected by mean success on the same 16-point grid ($\alpha_r{=}0.25$--$4.0$ in steps of 0.25); ties selected the smaller value, and the selected scalings were held fixed during evaluation and resampling. Rows list the paired mean differences for the separately selected recurrent scalings with 95\% BCa bootstrap confidence intervals (2{,}000 resamples of evaluation-seed pairs; \S\ref{sec:20}). Differences are in percentage points and follow the subtraction order in each panel heading. Panel (A) lists the evaluation counts, both selected scalings, the group mean success rates (\%) and the differences [95\% BCa CI]. Bold underlined values mark the larger mean in each row. All networks were trained for 1{,}900 episodes and evaluated over 100 episodes at $\sigma{=}0.001$ during training and evaluation, with Q-value feedback present, and at $\tau{=}1{,}000$\,ms. The STP parameters were $U_{\mathrm{SE}}{=}0.1$, $\tau_{\mathrm{R}}{=}100$\,ms and $\tau_{\mathrm{F}}{=}500$\,ms.}\ComposePartLabel{a}\label{fig:37}\end{composecell}\end{composerow}\end{composepage}\end{figure}\begin{figure}[tp]\begin{composepage}[fig:37|2|pg_fig_37_2]\ComposePart{b}\begin{composerow}\begin{composecell}[121.254000000]\begin{composetable}\begingroup\fontsize{8}{9}\selectfont\setlength{\tabcolsep}{2pt}%
\begin{tabular}{l>{\raggedright\arraybackslash}p{29mm}ll@{\hspace{3mm}}ll@{\hspace{3mm}}lll}\toprule
 & & \multicolumn{2}{l}{With} & \multicolumn{2}{l}{Steady} & & &\\ \shortstack[l]{$\tau$\\{}(ms)} & \shortstack[l]{Setting} & $\alpha_r$ & \shortstack[l]{mean\\{}(\%)} & $\alpha_r$ & \shortstack[l]{mean\\{}(\%)} & \shortstack[l]{Difference\\{}(pp)} & \shortstack[l]{95\% BCa CI} & \shortstack[l]{Eval.\\{}$n$}\\\midrule
\multicolumn{9}{l}{\textbf{(B) With STP \ensuremath{-} Steady-state STP}}\\
1,000 & Same \ensuremath{\alpha_r} = 3 & $3.00$ & $81.90$ & $3.00$ & $\underline{\mathbf{83.92}}$ & $-2.02$ & $[-5.96, 1.92]$ & $100$\\ %
1,000 & Select / evaluate: all & $2.00$ & $86.33$ & $1.75$ & $\underline{\mathbf{87.70}}$ & $-1.37$ & $[-4.93, 1.93]$ & $100$\\ %
1,000 & Select A; evaluate B & $2.50$ & $84.06$ & $1.75$ & $\underline{\mathbf{89.60}}$ & $-5.54$ & $[-12.02, -0.68]$ & $50$\\ %
1,000 & Select B; evaluate A & $2.25$ & $82.18$ & $1.75$ & $\underline{\mathbf{85.80}}$ & $-3.62$ & $[-8.80, 1.28]$ & $50$\\ %
750 & Same \ensuremath{\alpha_r} = 3 & $3.00$ & $69.61$ & $3.00$ & $\underline{\mathbf{69.73}}$ & $-0.12$ & $[-5.92, 5.25]$ & $100$\\ %
750 & Select / evaluate: all & $2.25$ & $\underline{\mathbf{76.70}}$ & $2.50$ & $76.19$ & $0.51$ & $[-4.46, 5.68]$ & $100$\\ %
500 & Same \ensuremath{\alpha_r} = 3 & $3.00$ & $39.33$ & $3.00$ & $\underline{\mathbf{42.64}}$ & $-3.31$ & $[-10.27, 3.42]$ & $100$\\ %
500 & Select / evaluate: all & $2.25$ & $45.94$ & $1.75$ & $\underline{\mathbf{48.29}}$ & $-2.35$ & $[-9.30, 4.10]$ & $100$\\ %
\bottomrule\end{tabular}
\par\vspace{1mm}\noindent\begin{minipage}{121.15255292652553mm}\fontsize{8}{9}\selectfont
resamples = $2{,}000$\par
\end{minipage}
\endgroup
\end{composetable}\end{composecell}\end{composerow}\vspace{2mm}\begin{composerow}\begin{composecell}[121.254000000]\begin{composetable}\begingroup\fontsize{8}{9}\selectfont\setlength{\tabcolsep}{2pt}%
\begin{tabular}{l>{\raggedright\arraybackslash}p{29mm}ll@{\hspace{3mm}}ll@{\hspace{3mm}}lll}\toprule
 & & \multicolumn{2}{l}{Steady} & \multicolumn{2}{l}{Without} & & &\\ \shortstack[l]{$\tau$\\{}(ms)} & \shortstack[l]{Setting} & $\alpha_r$ & \shortstack[l]{mean\\{}(\%)} & $\alpha_r$ & \shortstack[l]{mean\\{}(\%)} & \shortstack[l]{Difference\\{}(pp)} & \shortstack[l]{95\% BCa CI} & \shortstack[l]{Eval.\\{}$n$}\\\midrule
\multicolumn{9}{l}{\textbf{(C) Steady-state STP \ensuremath{-} Without STP}}\\
1,000 & Same \ensuremath{\alpha_r} = 3 & $3.00$ & $\underline{\mathbf{83.92}}$ & $3.00$ & $43.74$ & $40.18$ & $[34.49, 46.38]$ & $100$\\ %
1,000 & Select / evaluate: all & $1.75$ & $\underline{\mathbf{87.70}}$ & $1.25$ & $71.58$ & $16.12$ & $[11.29, 21.01]$ & $100$\\ %
1,000 & Select A; evaluate B & $1.75$ & $\underline{\mathbf{89.60}}$ & $1.25$ & $74.28$ & $15.32$ & $[9.26, 22.14]$ & $50$\\ %
1,000 & Select B; evaluate A & $1.75$ & $\underline{\mathbf{85.80}}$ & $1.25$ & $68.88$ & $16.92$ & $[9.55, 23.68]$ & $50$\\ %
750 & Same \ensuremath{\alpha_r} = 3 & $3.00$ & $\underline{\mathbf{69.73}}$ & $3.00$ & $43.77$ & $25.96$ & $[18.35, 33.28]$ & $100$\\ %
750 & Select / evaluate: all & $2.50$ & $\underline{\mathbf{76.19}}$ & $1.00$ & $50.34$ & $25.85$ & $[19.24, 32.36]$ & $100$\\ %
500 & Same \ensuremath{\alpha_r} = 3 & $3.00$ & $\underline{\mathbf{42.64}}$ & $3.00$ & $31.21$ & $11.43$ & $[3.77, 18.35]$ & $100$\\ %
500 & Select / evaluate: all & $1.75$ & $\underline{\mathbf{48.29}}$ & $2.75$ & $35.42$ & $12.87$ & $[5.94, 19.53]$ & $100$\\ %
\bottomrule\end{tabular}
\par\vspace{1mm}\noindent\begin{minipage}{121.15255292652553mm}\fontsize{8}{9}\selectfont
resamples = $2{,}000$\par
\end{minipage}
\endgroup
\end{composetable}\caption{\textbf{(B)} With STP minus Steady-state STP. \textbf{(C)} Steady-state STP minus Without STP. At $\tau{=}1{,}000$\,ms, each panel reports four comparisons. The first two are the same recurrent scaling ($\alpha_r{=}3.0$ in all conditions, as in Fig.~\ref{fig:38}A) and selection and evaluation on all networks. The other two select on A and evaluate on B, or reverse this direction. The remaining rows report the same recurrent scaling and selection on all networks at $\tau{=}750$ and 500\,ms. The all-network rows select and evaluate on the 100 networks (seeds 0--99). With, Steady and Without denote With STP, Steady-state STP and Without STP. For the steady-state STP control, see the section ``Steady-state STP modulation''.}\ComposePartLabel{b}\end{composecell}\end{composerow}\end{composepage}\end{figure}\begin{figure}[tp]\begin{composepage}[fig:37|3|pg_fig_37_3]\ComposePart{c}\begin{composerow}\begin{composecell}[121.254000000]\begin{composetable}\begingroup\fontsize{8}{9}\selectfont\setlength{\tabcolsep}{2pt}%
\begin{tabular}{l>{\raggedright\arraybackslash}p{29mm}ll@{\hspace{3mm}}ll@{\hspace{3mm}}lll}\toprule
 & & \multicolumn{2}{l}{With} & \multicolumn{2}{l}{Control} & & &\\ \shortstack[l]{$\tau$\\{}(ms)} & \shortstack[l]{Setting} & $\alpha_r$ & \shortstack[l]{mean\\{}(\%)} & $\alpha_r$ & \shortstack[l]{mean\\{}(\%)} & \shortstack[l]{Difference\\{}(pp)} & \shortstack[l]{95\% BCa CI} & \shortstack[l]{Eval.\\{}$n$}\\\midrule
\multicolumn{9}{l}{\textbf{(D) With STP \ensuremath{-} each control (selected \ensuremath{\alpha_r})}}\\
1,000 & Without STP & $2.00$ & $\underline{\mathbf{86.33}}$ & $1.25$ & $71.58$ & $14.75$ & $[9.58, 20.08]$ & $100$\\ %
1,000 & Population-mean & $2.00$ & $\underline{\mathbf{86.33}}$ & $1.25$ & $72.38$ & $13.95$ & $[9.20, 18.93]$ & $100$\\ %
1,000 & Own-unit & $2.00$ & $\underline{\mathbf{86.33}}$ & $0.75$ & $71.29$ & $15.04$ & $[9.38, 20.68]$ & $100$\\ %
1,000 & Other-unit STP & $2.00$ & $\underline{\mathbf{86.33}}$ & $0.50$ & $79.89$ & $6.44$ & $[1.59, 11.23]$ & $100$\\ %
1,000 & Common prescribed & $2.00$ & $\underline{\mathbf{86.33}}$ & $2.00$ & $75.45$ & $10.88$ & $[5.80, 16.81]$ & $100$\\ %
1,000 & Unit-wise prescribed & $2.00$ & $\underline{\mathbf{86.33}}$ & $1.75$ & $78.51$ & $7.82$ & $[3.90, 12.49]$ & $100$\\ %
1,000 & Steady-state STP & $2.00$ & $86.33$ & $1.75$ & $\underline{\mathbf{87.70}}$ & $-1.37$ & $[-4.93, 1.93]$ & $100$\\ %
\bottomrule\end{tabular}
\par\vspace{1mm}\noindent\begin{minipage}{121.15255292652553mm}\fontsize{8}{9}\selectfont
resamples = $2{,}000$\par
\end{minipage}
\endgroup
\end{composetable}\end{composecell}\end{composerow}\vspace{2mm}\begin{composerow}\begin{composecell}[121.254000000]\begin{composetable}\begingroup\fontsize{8}{9}\selectfont\setlength{\tabcolsep}{2pt}%
\begin{tabular}{l>{\raggedright\arraybackslash}p{29mm}ll@{\hspace{3mm}}ll@{\hspace{3mm}}lll}\toprule
 & & \multicolumn{2}{l}{Control} & \multicolumn{2}{l}{Without} & & &\\ \shortstack[l]{$\tau$\\{}(ms)} & \shortstack[l]{Setting} & $\alpha_r$ & \shortstack[l]{mean\\{}(\%)} & $\alpha_r$ & \shortstack[l]{mean\\{}(\%)} & \shortstack[l]{Difference\\{}(pp)} & \shortstack[l]{95\% BCa CI} & \shortstack[l]{Eval.\\{}$n$}\\\midrule
\multicolumn{9}{l}{\textbf{(E) Each control \ensuremath{-} Without STP (selected \ensuremath{\alpha_r})}}\\
1,000 & Population-mean & $1.25$ & $\underline{\mathbf{72.38}}$ & $1.25$ & $71.58$ & $0.80$ & $[-4.86, 7.25]$ & $100$\\ %
1,000 & Own-unit & $0.75$ & $71.29$ & $1.25$ & $\underline{\mathbf{71.58}}$ & $-0.29$ & $[-5.67, 5.97]$ & $100$\\ %
1,000 & Other-unit STP & $0.50$ & $\underline{\mathbf{79.89}}$ & $1.25$ & $71.58$ & $8.31$ & $[2.59, 14.09]$ & $100$\\ %
1,000 & Common prescribed & $2.00$ & $\underline{\mathbf{75.45}}$ & $1.25$ & $71.58$ & $3.87$ & $[-2.67, 10.21]$ & $100$\\ %
1,000 & Unit-wise prescribed & $1.75$ & $\underline{\mathbf{78.51}}$ & $1.25$ & $71.58$ & $6.93$ & $[1.19, 12.67]$ & $100$\\ %
1,000 & Steady-state STP & $1.75$ & $\underline{\mathbf{87.70}}$ & $1.25$ & $71.58$ & $16.12$ & $[11.29, 21.01]$ & $100$\\ %
\bottomrule\end{tabular}
\par\vspace{1mm}\noindent\begin{minipage}{121.15255292652553mm}\fontsize{8}{9}\selectfont
resamples = $2{,}000$\par
\end{minipage}
\endgroup
\end{composetable}\caption{\textbf{(D)} With STP minus each separately trained control and  \textbf{(E)} each control minus Without STP at $\tau{=}1{,}000$\,ms, with the recurrent scaling of every condition selected on all 100 networks ($\alpha_r{=}2.00$ with STP; the value for each control is given in the $\alpha_r$ columns); selection and comparison used the same networks. The Without-STP row in (D) repeats the comparison with the recurrent scaling selected for each condition of Table~\ref{tab:04}, and the Steady-state STP rows in (D) and (E) repeat the all-network rows at $\tau{=}1{,}000$\,ms in Fig.~\ref{fig:37:partb}B, C. The Common prescribed scaling and Unit-wise prescribed scaling rows in (D) and (E) use the time courses recorded from the With-STP networks at $\alpha_r{=}3.0$.}\ComposePartLabel{c}\end{composecell}\end{composerow}\end{composepage}\end{figure}\begin{figure}[tp]\begin{composepage}[fig:38|1|pg_fig_38_1]\begin{composerow}\begin{composecell}[121.254000000]\begin{composetable}\begingroup\fontsize{8}{9}\selectfont\setlength{\tabcolsep}{2pt}%
\begin{tabular}{l@{\hspace{3mm}}l@{\hspace{3mm}}l@{\hspace{3mm}}l@{\hspace{3mm}}l}\toprule
\shortstack[l]{With STP} & \shortstack[l]{Steady-state\\{}STP} & \shortstack[l]{Without\\{}STP} & \shortstack[l]{With STP $-$\\{}steady-state [95\%\\{}BCa CI]} & \shortstack[l]{Steady-state $-$\\{}Without STP [95\%\\{}BCa CI]}\\\midrule
\multicolumn{5}{l}{\textbf{(A) Success (\%)}}\\
$81.90$ & $83.92$ & $43.74$ & \begin{tabular}[t]{@{}l@{}}$-2.02$\\{}$[-5.96, 1.92]$\end{tabular} & \begin{tabular}[t]{@{}l@{}}$40.18$\\{}$[34.49, 46.38]$\end{tabular}\\[\medskipamount] %
\multicolumn{5}{l}{\textbf{(B) Action-value difference $D_Q$ (Q units)}}\\
$0.0458$ & $0.0472$ & $0.0211$ & \begin{tabular}[t]{@{}l@{}}$-0.0014$\\{}$[-0.0039, 0.0014]$\end{tabular} & \begin{tabular}[t]{@{}l@{}}$0.0261$\\{}$[0.0214, 0.0309]$\end{tabular}\\[\medskipamount] %
\multicolumn{5}{l}{\textbf{(C) First-move accuracy (\%)}}\\
$99.46$ & $99.37$ & $91.73$ & \begin{tabular}[t]{@{}l@{}}$0.09$\\{}$[-0.46, 0.84]$\end{tabular} & \begin{tabular}[t]{@{}l@{}}$7.64$\\{}$[5.88, 9.72]$\end{tabular}\\ %
\bottomrule\end{tabular}
\par\vspace{1mm}\noindent\begin{minipage}{113.02919607029196mm}\fontsize{8}{9}\selectfont
n = $100$\par
resamples = $2{,}000$\par
\end{minipage}
\endgroup
\end{composetable}\caption{\textbf{Steady-state STP preserved the success and readout advantages over Without STP at the main recurrent scaling.} \textbf{(A)} Success rate,  \textbf{(B)} the goal-directed action-value difference $D_Q$ averaged over GO2--GO3 within each network, and  \textbf{(C)} first-move accuracy, the percentage of evaluation trials in which the first action reduces the Manhattan distance to the cued goal. All three conditions use $\alpha_r{=}3.0$, $\tau{=}1{,}000$\,ms and $\sigma{=}0.001$ during training and evaluation; means are across $n{=}100$ paired networks, each trained for 1{,}900 episodes and evaluated over 100 episodes. The steady-state STP control (section ``Steady-state STP modulation'') used $U_{\mathrm{SE}}{=}0.1$, $\tau_{\mathrm{R}}{=}100$\,ms and $\tau_{\mathrm{F}}{=}500$\,ms. Rows list the paired mean differences with 95\% BCa confidence intervals (2{,}000 resamples of network pairs; \S\ref{sec:20}); the last two columns give With STP minus Steady-state STP and Steady-state STP minus Without STP. Success rate and first-move accuracy are in percent and their differences in percentage points, and $D_Q$ means and differences are in raw Q-value units. The first three columns give the condition means for With STP, Steady-state STP and Without STP.}\label{fig:38}\end{composecell}\end{composerow}\end{composepage}\end{figure}\begin{figure}[tp]\begin{composepage}[fig:39|1|pg_fig_39_1]\begin{composerow}\begin{composecell}[121.254000000]\begin{composetable}\begingroup\fontsize{8}{9}\selectfont\setlength{\tabcolsep}{2pt}%
\begin{tabular}{>{\raggedright\arraybackslash}p{34mm}lllll}\toprule
\shortstack[l]{Schedule} & \shortstack[l]{$\alpha_r$\\{}With} & \shortstack[l]{$\alpha_r$\\{}Without} & \shortstack[l]{With STP\\{}mean $\pm$ SD\\{}(\%)} & \shortstack[l]{Without STP\\{}mean $\pm$ SD\\{}(\%)} & \shortstack[l]{Difference [95\%\\{}BCa CI] (pp)}\\\midrule
\multicolumn{6}{l}{\textbf{(A) Same recurrent scaling; \ensuremath{\sigma} = 0}}\\
Switch after 633 episodes & 3.0 & 3.0 & $76.81 \ensuremath{\pm} 22.87$ & $59.23 \ensuremath{\pm} 26.94$ & $17.58\ [10.21, 24.22]$\\ %
Switch after 1,267 episodes & 3.0 & 3.0 & $84.70 \ensuremath{\pm} 18.86$ & $68.25 \ensuremath{\pm} 25.14$ & $16.45\ [10.57, 21.41]$\\ %
No switch & 3.0 & 3.0 & $82.75 \ensuremath{\pm} 23.16$ & $57.00 \ensuremath{\pm} 28.49$ & $25.75\ [18.81, 32.04]$\\ %
Lower rate from the start & 3.0 & 3.0 & $83.40 \ensuremath{\pm} 21.31$ & $75.11 \ensuremath{\pm} 23.46$ & $8.29\ [1.70, 14.15]$\\ %
\multicolumn{6}{l}{\rule{0pt}{4mm}\textbf{(B) Selected recurrent scaling; \ensuremath{\sigma} = 0}}\\
Switch after 633 episodes & 2 & 1.25 & $75.14 \ensuremath{\pm} 22.90$ & $63.57 \ensuremath{\pm} 26.01$ & $11.57\ [5.52, 17.09]$\\ %
Switch after 1,267 episodes & 2 & 1.25 & $86.86 \ensuremath{\pm} 17.73$ & $72.46 \ensuremath{\pm} 24.26$ & $14.40\ [9.03, 19.62]$\\ %
No switch & 2 & 1.25 & $84.76 \ensuremath{\pm} 17.12$ & $71.14 \ensuremath{\pm} 22.89$ & $13.62\ [8.83, 18.69]$\\ %
Lower rate from the start & 2 & 1.25 & $83.46 \ensuremath{\pm} 18.23$ & $73.19 \ensuremath{\pm} 22.75$ & $10.27\ [5.56, 15.34]$\\ %
\bottomrule\end{tabular}
\par\vspace{1mm}\noindent\begin{minipage}{115.93503528435035mm}\fontsize{8}{9}\selectfont
n = 100\par
resamples = $2{,}000$\par
\end{minipage}
\endgroup
\end{composetable}\end{composecell}\end{composerow}\vspace{2mm}\begin{composerow}\begin{composecell}[121.254000000]\begin{composetable}\begingroup\fontsize{8}{9}\selectfont\setlength{\tabcolsep}{2pt}%
\begin{tabular}{>{\raggedright\arraybackslash}p{34mm}lllll}\toprule
\shortstack[l]{Schedule} & \shortstack[l]{$\alpha_r$\\{}With} & \shortstack[l]{$\alpha_r$\\{}Without} & \shortstack[l]{With STP\\{}mean $\pm$ SD\\{}(\%)} & \shortstack[l]{Without STP\\{}mean $\pm$ SD\\{}(\%)} & \shortstack[l]{Difference [95\%\\{}BCa CI] (pp)}\\\midrule
\multicolumn{6}{l}{\textbf{(C) Same recurrent scaling; \ensuremath{\sigma} = 0.001}}\\
Switch after 633 episodes & 3.0 & 3.0 & $74.76 \ensuremath{\pm} 21.00$ & $44.54 \ensuremath{\pm} 22.30$ & $30.22\ [23.69, 35.68]$\\ %
Switch after 1,267 episodes & 3.0 & 3.0 & $80.99 \ensuremath{\pm} 20.51$ & $50.33 \ensuremath{\pm} 22.89$ & $30.66\ [24.06, 36.74]$\\ %
No switch & 3.0 & 3.0 & $77.89 \ensuremath{\pm} 23.74$ & $41.01 \ensuremath{\pm} 24.42$ & $36.88\ [29.23, 43.19]$\\ %
Lower rate from the start & 3.0 & 3.0 & $77.31 \ensuremath{\pm} 20.17$ & $49.82 \ensuremath{\pm} 21.33$ & $27.49\ [21.54, 32.75]$\\ %
\multicolumn{6}{l}{\rule{0pt}{4mm}\textbf{(D) Selected recurrent scaling; \ensuremath{\sigma} = 0.001}}\\
Switch after 633 episodes & 2 & 1.25 & $72.01 \ensuremath{\pm} 22.20$ & $58.77 \ensuremath{\pm} 22.83$ & $13.24\ [7.35, 18.39]$\\ %
Switch after 1,267 episodes & 2 & 1.25 & $83.65 \ensuremath{\pm} 16.53$ & $69.26 \ensuremath{\pm} 20.24$ & $14.39\ [9.18, 19.20]$\\ %
No switch & 2 & 1.25 & $84.46 \ensuremath{\pm} 16.73$ & $65.05 \ensuremath{\pm} 21.74$ & $19.41\ [14.46, 24.03]$\\ %
Lower rate from the start & 2 & 1.25 & $79.76 \ensuremath{\pm} 17.38$ & $64.01 \ensuremath{\pm} 22.23$ & $15.75\ [11.33, 20.38]$\\ %
\bottomrule\end{tabular}
\par\vspace{1mm}\noindent\begin{minipage}{115.93503528435035mm}\fontsize{8}{9}\selectfont
n = 100\par
resamples = $2{,}000$\par
\end{minipage}
\endgroup
\end{composetable}\caption{\textbf{The With-STP success-rate advantage held across the 4 learning-rate schedules.} Learning-rate schedules were fixed identically in the With-STP and Without-STP networks. Panels (A) and (B) use $\sigma{=}0$, and panels (C) and (D) use state noise $\sigma{=}0.001$, during training and evaluation. Panels (A) and (C) use the same recurrent scaling, $\alpha_r{=}3.0$, in both conditions; panels (B) and (D) hold the selected scalings fixed across schedules and noise levels ($\alpha_r{=}2$ with STP and $\alpha_r{=}1.25$ without STP). The unit time constant is $\tau{=}1{,}000$\,ms, and each network received 1{,}900 training and 100 evaluation episodes. The rows show a switch from the higher to the lower learning rate after 633 total training episodes, a switch after 1{,}267 total training episodes, no switch (the higher rate throughout), and the lower rate from the start. Rows list mean differences in success rate (With STP minus Without STP, in percentage points) with 95\% BCa bootstrap confidence intervals (2{,}000 resamples of the 100 paired networks, seeds 0--99; \S\ref{sec:20}) for the same and the selected recurrent scalings. All 16 intervals lie above zero. The panels list the success rates of both conditions (mean $\pm$ SD across networks, \%), the paired difference [95\% BCa CI], both recurrent scalings and $n$.}\label{fig:39}\end{composecell}\end{composerow}\end{composepage}\end{figure}\begin{figure}[tp]\begin{composepage}[fig:40|1|pg_fig_40_1]\ComposePart{a}\begin{composerow}\begin{composecell}[121.254000000]\begin{composetable}\begingroup\fontsize{8}{9}\selectfont\setlength{\tabcolsep}{2pt}%
\begin{tabular}{@{}>{\raggedright\arraybackslash}p{26mm}>{\raggedright\arraybackslash}p{19mm}>{\raggedright\arraybackslash}p{19mm}>{\raggedright\arraybackslash}p{42mm}@{}}\toprule
\multicolumn{4}{@{}p{110mm}@{}}{\textbf{(A) Success rate (\%)}}\\
\shortstack[l]{STP setting\\$(\tau_{\mathrm{R}},\tau_{\mathrm{F}})$ (ms)} & With STP & Without STP & \shortstack[l]{Difference [95\%\\BCa CI]}\\\midrule
$(100,500)$ & $81.90$ & $43.74$ & $38.16\ [31.55, 44.20]$\\ %
$(150,1500)$ & $75.19$ & $43.74$ & $31.45\ [24.10, 38.83]$\\ %
$(150,1000)$ & $75.39$ & $43.74$ & $31.65\ [24.91, 38.77]$\\ %
\bottomrule\end{tabular}
\endgroup
\end{composetable}\end{composecell}\end{composerow}\vspace{2mm}\begin{composerow}\begin{composecell}[121.254000000]\begin{composetable}\begingroup\fontsize{8}{9}\selectfont\setlength{\tabcolsep}{2pt}%
\begin{tabular}{@{}>{\raggedright\arraybackslash}p{26mm}>{\raggedright\arraybackslash}p{19mm}>{\raggedright\arraybackslash}p{19mm}>{\raggedright\arraybackslash}p{42mm}@{}}\toprule
\shortstack[l]{STP setting\\$(\tau_{\mathrm{R}},\tau_{\mathrm{F}})$ (ms)} & With STP & Without STP & \shortstack[l]{Difference [95\%\\BCa CI]}\\\midrule
\multicolumn{4}{@{}p{110mm}@{}}{\textbf{(B) Goal-decoding accuracy: late (8--9 s) (\%)}}\\
$(100,500)$ & $98.50$ & $89.14$ & $9.36\ [7.75, 11.19]$\\ %
$(150,1500)$ & $99.58$ & $89.14$ & $10.44\ [8.85, 12.24]$\\ %
$(150,1000)$ & $99.48$ & $89.14$ & $10.34\ [8.73, 12.10]$\\\midrule %
\multicolumn{4}{@{}p{110mm}@{}}{\textbf{(C) Goal-decoding accuracy: delay (3--4 s) (\%)}}\\
$(100,500)$ & $99.90$ & $98.99$ & $0.92\ [0.73, 1.12]$\\ %
$(150,1500)$ & $99.99$ & $98.99$ & $1.00\ [0.83, 1.20]$\\ %
$(150,1000)$ & $99.99$ & $98.99$ & $1.01\ [0.83, 1.21]$\\\midrule %
\multicolumn{4}{@{}p{110mm}@{}}{\textbf{(D) Action-value difference $D_Q$: GO2 (Q)}}\\
$(100,500)$ & $0.0648$ & $0.0419$ & $0.0228\ [0.0175, 0.0281]$\\ %
$(150,1500)$ & $0.0636$ & $0.0419$ & $0.0217\ [0.0161, 0.0278]$\\ %
$(150,1000)$ & $0.0640$ & $0.0419$ & $0.0221\ [0.0171, 0.0277]$\\\midrule %
\multicolumn{4}{@{}p{110mm}@{}}{\textbf{(E) Action-value difference $D_Q$: GO3 (Q)}}\\
$(100,500)$ & $0.0268$ & $0.0002$ & $0.0266\ [0.0216, 0.0314]$\\ %
$(150,1500)$ & $0.0185$ & $0.0002$ & $0.0183\ [0.0127, 0.0244]$\\ %
$(150,1000)$ & $0.0177$ & $0.0002$ & $0.0175\ [0.0117, 0.0237]$\\ %
\bottomrule\end{tabular}
\par\vspace{1mm}\noindent\begin{minipage}{118.62335685623357mm}\fontsize{8}{9}\selectfont n = $100$\par resamples = $2{,}000$\par Intervals: 95\% BCa CI.\par \end{minipage} \endgroup
\end{composetable}\caption{\textbf{With STP had higher success at all three tested STP parameter settings.} \textbf{(A)} Success rate. The panels compare the main STP time constants, $(\tau_{\mathrm{R}},\tau_{\mathrm{F}}){=}(100,500)$\,ms, with $(150,1500)$\,ms and $(150,1000)$\,ms, all with $U_{\mathrm{SE}}{=}0.1$, $\tau{=}1{,}000$\,ms, $\alpha_r{=}3.0$ and $\sigma{=}0.001$; the first row within each panel is the main setting. Each row lists the With-STP and Without-STP means and their paired difference (With STP minus Without STP) with its 95\% BCa bootstrap confidence interval (2{,}000 resamples; \S\ref{sec:20}). Success and decoding means are percentages and their differences are percentage points; $D_Q$ is in raw Q-value units. Each setting uses 100 paired independently initialized networks (seeds 0--99). Each panel lists the group means and the estimates with their intervals in the same setting order. \textbf{(B, C)} Goal-decoding accuracy, averaged over the time-resolved accuracies, in the late ($t{=}8$--$9$\,s) and delay ($t{=}3$--$4$\,s) windows. \textbf{(D, E)} Action-value difference $D_Q$ at GO2 and GO3.}\ComposePartLabel{a}\label{fig:40}\end{composecell}\end{composerow}\end{composepage}\end{figure}\begin{figure}[tp]\begin{composepage}[fig:40|2|pg_fig_40_2]\ComposePart{b}\begin{composerow}\begin{composecell}[121.254000000]\begin{composetable}\begingroup\fontsize{8}{9}\selectfont\setlength{\tabcolsep}{2pt}%
\begin{tabular}{@{}>{\raggedright\arraybackslash}p{26mm}>{\raggedright\arraybackslash}p{19mm}>{\raggedright\arraybackslash}p{19mm}>{\raggedright\arraybackslash}p{42mm}@{}}\toprule
\multicolumn{4}{@{}p{110mm}@{}}{\textbf{(F) $\Delta_{\mathrm{sim}}$: late (8--9 s) minus delay (3--4 s) (dimensionless)}}\\
\shortstack[l]{STP setting\\$(\tau_{\mathrm{R}},\tau_{\mathrm{F}})$ (ms)} & \multicolumn{3}{l}{With STP [95\% BCa CI]}\\\midrule
$(100,500)$ & \multicolumn{3}{l}{$0.0348\ [0.0300, 0.0395]$}\\ %
$(150,1500)$ & \multicolumn{3}{l}{$0.0911\ [0.0835, 0.0987]$}\\ %
$(150,1000)$ & \multicolumn{3}{l}{$0.0723\ [0.0654, 0.0790]$}\\\midrule %
\multicolumn{4}{@{}p{110mm}@{}}{\textbf{(G) Relative spectral radius $\rho_{\mathrm{rel}}$: delay (3--4 s) (dimensionless)}}\\
$(100,500)$ & \multicolumn{3}{l}{$1.55\ [1.51, 1.61]$}\\ %
$(150,1500)$ & \multicolumn{3}{l}{$2.56\ [2.50, 2.63]$}\\ %
$(150,1000)$ & \multicolumn{3}{l}{$2.01\ [1.96, 2.05]$}\\\midrule %
\multicolumn{4}{@{}p{110mm}@{}}{\textbf{(H) 95th percentile of eigenvalue magnitudes, delay (3--4 s) (dimensionless)}}\\
\shortstack[l]{STP setting\\$(\tau_{\mathrm{R}},\tau_{\mathrm{F}})$ (ms)} & With STP & Without STP & \shortstack[l]{Difference [95\%\\BCa CI]}\\\midrule
$(100,500)$ & $7.69$ & $5.80$ & $1.88\ [1.79, 1.98]$\\ %
$(150,1500)$ & $12.31$ & $5.80$ & $6.50\ [6.29, 6.73]$\\ %
$(150,1000)$ & $9.85$ & $5.80$ & $4.05\ [3.89, 4.22]$\\\midrule %
\multicolumn{4}{@{}p{110mm}@{}}{\textbf{(I) Next-move progress after an erroneous move (\%)}}\\
$(100,500)$ & $61.28$ & $47.61$ & \shortstack[l]{$13.67\ [5.54, 21.45]$\\$n=88$}\\ %
$(150,1500)$ & $70.07$ & $48.91$ & \shortstack[l]{$21.16\ [13.58, 27.96]$\\$n=92$}\\ %
$(150,1000)$ & $66.80$ & $48.47$ & \shortstack[l]{$18.32\ [10.39, 25.90]$\\$n=93$}\\ %
\bottomrule\end{tabular}
\par\vspace{1mm}\noindent\begin{minipage}{118.62335685623357mm}\fontsize{8}{9}\selectfont
n (F--H) = $100$\par
resamples = $2{,}000$\par
Intervals: 95\% BCa CI. (F) is the With-STP time change; (G) is the absolute With-STP mean.\par
\end{minipage}
\endgroup
\end{composetable}\caption{\textbf{(F)} The With-STP change in the goal-conditioned effective-connectivity separation $\Delta_{\mathrm{sim}}$, late minus delay. \textbf{(G)} The mean With-STP relative effective spectral radius $\rho_{\mathrm{rel}}$ in the delay period. Entries in (\ref{fig:40:partb}F) and (\ref{fig:40:partb}G) are the With-STP time change and the absolute mean, respectively. The reference values are zero for differences and time changes and one for the absolute $\rho_{\mathrm{rel}}$; the Without-STP reference of (\ref{fig:40:partb}F) and (\ref{fig:40:partb}G) is fixed by definition. (H, I): With STP minus Without STP. \textbf{(H)} The condition difference in the 95th percentile of the delay-period eigenvalue magnitudes. $\Delta_{\mathrm{sim}}$, $\rho_{\mathrm{rel}}$ and the eigenvalue magnitudes are dimensionless. \textbf{(I)} Next-move progress after an erroneous move. Panel (I) uses the networks with erroneous moves in both conditions, 88, 92 and 93 pairs in the order of the settings, and its Without-STP means belong to each paired subset. Panel (I) repeats the main-setting next-move progress estimate and confidence interval from Fig.~\ref{fig:36}C.}\ComposePartLabel{b}\end{composecell}\end{composerow}\end{composepage}\end{figure}\section*{Trial outcomes} \label{sec:45} After an erroneous move, too few GO opportunities may remain to reach the goal. We therefore assessed continued goal use by measuring progress toward the goal at the next GO opportunity, separately from eventual trial success. The values at the same recurrent scaling are listed in Fig.~\ref{fig:36}C (\S\ref{sec:23}). With the recurrent scaling selected for each condition, next-move progress averaged 74.2\% with STP and 62.8\% without STP (86 pairs with errors in both conditions; difference 11.4 percentage points, 95\% BCa CI [0.9, 20.4]).\par\par\endgroup\endgroup \par   \begingroup\begingroup\normalsize\normalbaselines \ComposeKeep{\parindent=15pt \tolerance=200 \emergencystretch=10pt \clubpenalty=150 \parskip=0pt plus 1pt \hsize=\textwidth}        \section*{Recurrent-scaling selection and evaluation on separate networks} To evaluate performance after recurrent-scaling selection on separate networks, we selected $\alpha_r$ for each condition from the same 16-point grid by the highest mean success rate across 50 selection networks, choosing the smaller value in a tie. We fixed the selected scalings and evaluated the remaining 50 network pairs, then reversed the selection and evaluation halves (Fig.~\ref{fig:37:parta}A).\par\endgroup\endgroup \par  \par   \par  \begingroup\begingroup\normalsize\normalbaselines \ComposeKeep{\parindent=15pt \tolerance=200 \emergencystretch=10pt \clubpenalty=150 \parskip=0pt plus 1pt \hsize=\textwidth} \leavevmode{} \endgroup \begingroup\normalsize\normalbaselines \begingroup\setlength{\emergencystretch}{3em}With selection and evaluation on the same 100 networks, With STP exceeded the three simpler activity-dependent rules and two controls with prescribed time-varying scaling (Fig.~\ref{fig:37:partc}D, E).\par\endgroup\par\endgroup \endgroup   \begingroup\begingroup\normalsize\normalbaselines \ComposeKeep{\parindent=15pt \tolerance=200 \emergencystretch=10pt \clubpenalty=150 \parskip=0pt plus 1pt \hsize=\textwidth} \section*{Learning-rate schedule controls} To test whether the advantage depends on a success-dependent learning-rate reduction, we compared schedules fixed independently of each network's success history (\S\ref{sec:12}; Fig.~\ref{fig:39}).\par\endgroup\endgroup  \relax \par  \par  \begingroup\begingroup\normalsize\normalbaselines \ComposeFinalParagraphSpace\section*{Additional STP parameter settings} We examined two additional facilitation-dominant pairs of STP time constants to test whether the behavioral and effective-connectivity findings extend beyond the main pair (Fig.~\ref{fig:40}). The two additional settings retained higher success and later $D_Q$ than Without STP at the same recurrent scaling, with mean delay-period relative effective spectral radii above one.\par\par\endgroup \endgroup  \endgroup\par \begingroup\normalsize\normalbaselines\begin{figure}[tp]\begin{composepage}[fig:41|1|pg_fig_41_1]\begin{composerow}\begin{composecell}[121.254000000]\centering\includegraphics{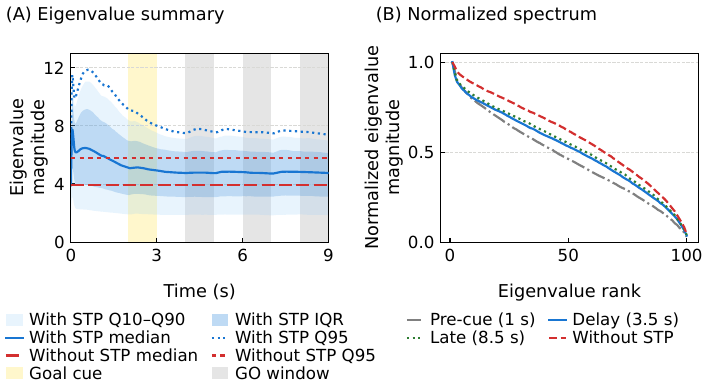}\caption{\textbf{Temporal evolution of the eigenvalue-magnitude distribution of the episode-averaged effective recurrent connectivity $W^{\mathrm{rec}}_{\mathrm{eff}}(t)$.} \textbf{(A)} Fan chart of the time-varying eigenvalue-magnitude distribution. At each time point, percentiles were computed from the $N_m$ eigenvalue magnitudes of episode-averaged $W^{\mathrm{rec}}_{\mathrm{eff}}(t)$ for each seed and then averaged across seeds ($n{=}100$ seeds). The outer band shows Q10--Q90, the darker inner band shows the interquartile range (Q25--Q75), the blue solid line shows the median, and the blue dotted line shows Q95. These bands describe the within-network eigenvalue distribution averaged across networks. In the Without-STP condition, $W^{\mathrm{rec}}_{\mathrm{eff}}(t){=}2W^{\mathrm{rec}}$ is time-invariant; the median and Q95 of its eigenvalue-magnitude distribution are shown as horizontal red reference lines (long-dashed for the median, short-dashed for Q95; bands omitted). Yellow band: goal cue (2--3\,s); gray bands: GO windows (4--5, 6--7, 8--9\,s). As with $\rho_{\mathrm{eff}}$, this analysis is an operational spectral description of $W^{\mathrm{rec}}_{\mathrm{eff}}(t)$ and is not a local Jacobian analysis of the full nonlinear system. \textbf{(B)} Comparison of eigenvalue-magnitude spectra averaged rank-wise across seeds and then normalized by their largest value, $\langle|\lambda_k|\rangle/\langle|\lambda_1|\rangle$, at representative time points before cue presentation ($t{=}1.0$\,s; gray dash-dot), during the delay period ($t{=}3.5$\,s; blue solid), and within the third GO window ($t{=}8.5$\,s; green dotted), together with the Without-STP spectrum (red dashed).}\label{fig:41}\end{composecell}\end{composerow}\end{composepage}\end{figure}\section*{Eigenvalue-magnitude distributions of effective recurrent connectivity} \begingroup\LayoutKeepFourLines To identify which part of the effective recurrent spectrum changes with STP, we examined its eigenvalue-magnitude distribution (Fig.~\ref{fig:41}). For each seed, we computed the average STP scaling vector across $E$ evaluation episodes, $\bar{\mathbf{s}}(t)=\frac{1}{E}\sum_{e=1}^{E}\mathbf{s}_e(t)$, and constructed $W^{\mathrm{rec}}_{\mathrm{eff}}(t)=W^{\mathrm{rec}}\mathrm{diag}(\bar{\mathbf{s}}(t))$. Thus, instead of performing eigendecomposition on each episode-specific matrix $W^{\mathrm{rec}}_{\mathrm{eff},e}(t)$, we built a single $W^{\mathrm{rec}}_{\mathrm{eff}}(t)$ from the averaged $\bar{\mathbf{s}}(t)$. At each time point, the magnitudes of the $N_m$ eigenvalues $\{\lambda_k(t)\}_{k=1}^{N_m}$ were sorted in descending order. We summarized the eigenvalue-magnitude distribution using percentile bands ($Q_{10}$--$Q_{90}$ and $Q_{25}$--$Q_{75}$) and the median $Q_{50}(t)$, with $Q_{95}(t)$ used as an upper-tail summary for predefined-window comparisons. To compare group-averaged \mbox{spectra} across task epochs, we first averaged the rank-wise eigenvalue magnitudes across seeds. Each mean spectrum was then normalized by its largest value, $\langle|\lambda_k|\rangle/\langle|\lambda_1|\rangle$. As with $\rho_{\mathrm{eff}}$, this analysis provides an operational spectral description of the effective recurrent connectivity formed by STP.\par\endgroup  \paragraph{Spread and shape change of the effective spectral structure}\par The quantity $\rho_{\mathrm{rel}}(t)$ in \S\ref{sec:24} summarized global scaling modulation of the effective spectrum. \leavevmode{} In the With-STP condition, the largest eigenvalue magnitude and the central bulk of the distribution (the interquartile range and Q10--Q90) rose after trial onset and varied across task epochs. In the predefined delay window ($t{=}3$--$4$\,s), representative percentiles were $Q_{95}$: With STP $7.69 \pm 0.51$ vs Without STP $5.80 \pm 0.19$ ($d_z{=}3.86$, $p < 0.001$), and $Q_{50}$: With STP $4.82 \pm 0.24$ vs Without STP $3.91 \pm 0.16$ ($d_z{=}3.69$, $p < 0.001$). In the Without-STP condition, by contrast, $W^{\mathrm{rec}}_{\mathrm{eff}}(t){=}2W^{\mathrm{rec}}$ is time-invariant, so the eigenvalue distribution was constant. The normalized group-averaged spectrum differed across the sampled task epochs (Fig.~\ref{fig:41}B). STP modulation affected both the largest eigenvalue magnitude and the broader eigenvalue-magnitude distribution. \par \par\endgroup
\end{document}